\documentclass[12pt,a4paper,oneside,openright]{book}
\usepackage[english]{babel} 
\usepackage{amsmath,amsthm,amssymb,amsfonts,latexsym} 
\usepackage[latin1]{inputenc} 
\usepackage{graphics}
\usepackage[dvips]{graphicx}
\usepackage [Lenny]{fncychap}
\usepackage{epigraph}
\usepackage[avantgarde]{quotchap}

\usepackage{psboxit}
\PScommands
\usepackage{fancyhdr}

\makeatletter
\def\cleardoublepage{\clearpage\if@twoside \ifodd\c@page\else
	\hbox{}
	\vspace*{\fill}
	\thispagestyle{empty}
	\newpage
	\if@twocolumn\hbox{}\newpage\fi\fi\fi}
\makeatother

\usepackage[calcwidth]{titlesec}
\titleformat{\section}[hang]{\sffamily\bfseries}
 {\Large\thesection}{12pt}{\Large}[{\titlerule[0.5pt]}]
\renewcommand\chapterheadstartvskip{\vspace*{-5\baselineskip}}

\begin{document}
	
\begin{titlepage}
\linespread{1.4}
\center{{\Large{Università Cattolica del Sacro Cuore}}\\
\normalsize{Sede di Brescia}\\}
\center{\Large{Facoltà di Scienze~Matematiche,~Fisiche~e~Naturali}}\\
\large{Corso di Laurea Specialistica in Fisica} \vspace{0.5cm}

\begin{figure}[h!]
\center{\includegraphics[width=3.5cm]{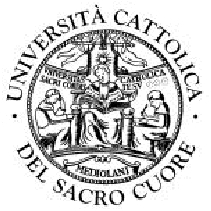}}\\
\end{figure}

\vspace{2.0cm}
\center{\LARGE{\textsc{Coherent quantum transport \\
in a star graph}}}
\vspace{1.5cm}
\normalsize
\flushleft{Supervisor:\\Prof. Fausto Borgonovi}
\flushleft{Assistant supervisor:\\Dott. Giuseppe Luca Celardo}
\hspace*{10.2cm}
Author:\\
\hspace*{10cm}
Angelo Ziletti \\
\hspace*{10cm}

\vspace{0.0cm}
\center{Academic Year 2009/2010}

\end{titlepage}

\newpage

\null \vspace {\stretch {1}}
\begin{flushright}
{\large
A Monica e Camilla}
\end{flushright}
\vspace {\stretch {2}}\null

 	\tableofcontents



\chapter*{Abstract}
	\addcontentsline{toc}{chapter}{Abstract}

The miniaturization of electronic components follows a dizzying pace, and very soon electronic circuits will have to operate in a new regime - the mesoscopic regime - in which a full quantum mechanical treatment is needed.
Furthermore, there are experimental evidences that the photosynthesis process fully exploit the \textit{quantum coherent transport} and it is precisely this fact that enables natural light-harvesting systems to be so efficient. 
In this context, the system considered may have important applications in the information technology and also in the modeling of complex systems, such as light-harvesting systems.

A star graph system which consists of different chains of
 sites coupled with the origin, has been considered
in the tight-binding approximation, within a closed and open formulation.
As far as 
 the closed system is concerned, we  showed  analytically 
that there are two localized states outside the normal Bloch band,
providing  
the localization lengths  as a function of the system parameters.
Regarding the open system, we used
 an approach based on the effective non-Hermitian Hamiltonian in
order to describe the coupling between the internal 
states and the environment.
In this framework, we studied the eigenvalues and the localization
 lengths of the eigenstates by changing the coupling with the external
 world. We found that the degree of opening 
weakly  affects
the localized states 
and we also gave
an analitical expression for the decay rates
in the limit of small coupling.

On the other hand, transport properties are very sensitive to the degree 
of opening of the system:   an analytical estimation of the value 
of the coupling at which the superradiance transition occurs,
 which is also valid for chains of different lengths among them and
 with different couplings with the origin has been found. 
Furthermore, we have shown by numerical simulations that 
(for large number of sites in each chain) the maximum of the integrated transmission coincides with the value of the coupling at which the transition to superradiance occurs.\\
This thesis is organised as follows: in Chap.1 we provide an introduction about the nanoworld and the importance of the system that we have studied; in Chap.2 we present a derivation of the effective Hamiltonian. In Chap.3 we introduce the star graph model and the solution of this system for the closed case (no external coupling). In Chap.4 we study in detail the open system, with particular attention to transport properties and superradiance transition. In Chap.5 we present a preliminary result about the disorder and in Chap.6 we summarize the main results obtained in this thesis.

\label{chap:1}

\begin{savequote}[15pc]
\sffamily
``One day sir, you may tax it.''\\
\qauthor{Michael Faraday's reply to British politician, William Gladstone, when asked of the practical value of electricity (1850)}
\end{savequote}
\chapter{Introduction}

\section{The Nanoworld revolution}
\label{sec:nanoworld}
\textit{Nanotechnology} is the study of manipulating matter on an atomic and molecular scale. Generally, nanotechnology deals with structures sized between 1 to 100 nanometre in at least one dimension, and involves developing materials or devices possessing at least one dimension within that size. Thus, nanometer-scale materials straddle the border between the molecular and the macroscopic. They are small enough to exhibit quantum properties reminiscent of molecules, but large enough for their size and shape to be designed and controlled. The potential of nanoscale materials is almost limitless, but scientists must first overcome two fundamental challenges. The first is physical: how can one control individual atoms in a nanosolid and then assemble them into real-world system? The second is conceptual: how does one attack problems too big to be solved by brute force calculation but too small to be tackled by statistical method?
Overcome these problems will enable for nanoscience and nanotechnology to revolutionize science and technology in ways that will make the world 50 years from now unrecognizable compared with today. In addition, the nanotechnology revolution meets three very specific scientific needs, from three different branches of science. 
The \textit{first} is the rapid advances in molecular biology that have completely changed people's understanding of life over the past 30 years; the \textit{second} is the evolution of chemistry from the study of single atoms and molecules to the fabrication of very large complexes such as quantum dots and proteins.
The \textit{third}, and perhaps the most important need comes from Information Technology (IT): the continuous request to be able to handle increasing volumes of information imposed a rapid miniaturization of electronic components, which began with the birth of the transistor and the microchip.
Already in 1965, Gordon Moore, co-founder and president of Intel, while preparing a speech for a meeting, noticed that the number of transistors that can be placed inexpensively on an integrated circuit doubles approximately every two years: this is the well-know \textit{Moore's law}, that describes a long-term trend in the history of computing hardware. This trend has continued for more than half a century and is expected to continue until 2015 or 2020 or later. This process is really amazing: there are no other branches of industry where advances in technology follow an exponential process, and for so long time. 
\begin{figure}[h!,b,t]
\vspace{0cm}
\center{\includegraphics[width=10cm,angle=0]{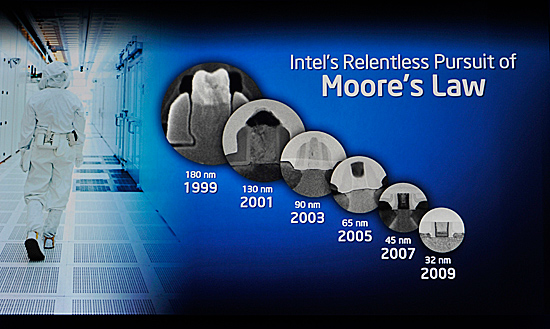}}
\caption{{\footnotesize The Moore's law and the development of Intel processors. The shrinking of devices is impressive; the electronic devices have nearly reached molecular scale (molecules commonly used as building blocks for organic synthesis have a dimension of a few $\mbox{\AA}$ to several dozen \AA) and will achieve very quickly lengths comparable to the atomic scale (Bohr radius of hydrogen has approximately a value of 0.53 \AA). At present, everyone can buy a PC with a processor built with 32nm (320 \AA) manufacturing process technology. Figure taken from the IDF - Intel Developer Forum - which took place from September 22 through September 24th, 2009.
}}
\label{fig:32nm}
\end{figure}
Extrapolation of Moore's law suggests that in the next 20-30 years, electronic circuit elements will shrink to the size of single atoms. Even before this fundamental limit is reached, electronic circuit will have to operate in a \textit{new regime} in which \textit{quantum mechanics} cannot be ignored. New devices that store, process and communicate information in a faster way will have to be invented in order to extend the IT revolution. No matter what device concepts are pursued, smaller devices must eventually embrace a profound change in the way we think about computing. Present devices average the behaviour of a great number
of quantum particles. Make any of these devices small enough and there will be only a few quantum particle in the system. In this limit, the laws of Quantum 
Mechanics manifest themselves most vividly. Thus, researchers need to traverse the threshold and look beyond the limits of classical physics.
And it is in this fervent and extremely important context that the physics of mesoscopic systems - systems in the intermediate size range between microscopic and macroscopic - moves its footsteps.

\section{Mesoscopic physics}
\label{sec:meso}
Mesoscopic physics is a rather young branch of science. It started about 25 years ago and it enjoys the unique combination of being able to deal with and provide answers to fundamental questions of physics while being relevant for applications in the not-too distant future, as explained in the prevoius section. In fact, some of the experimental possibilities in this field have been developed with an eye to reducing the size of electronic components.\\
More formally, \textit{mesoscopic physics} is a sub-discipline of condensed matter physics which deals with materials of an intermediate length scale. The 
length scale 
of such materials can be put
 between the size of few atoms (or a molecule) and of materials 
measuring microns. The lower limit can be also
defined as being the size of individual atoms. At the micron level are bulk materials. Mesoscopic and macroscopic objects have in common that they both contain a large number of atoms. Whereas average properties derived from its constituent materials describe macroscopic objects, as they usually obey the laws of Classical Mechanics, a mesoscopic object, by contrast, is affected by fluctuations around the average, and is subject to Quantum Mechanics. Thus, we are in the so-called \textit{quantum realm}, where quantum mechanical effects become important and not negligible. Typically, this means distances of 100 nanometers or less.
Even if 
this distance seems extremely small, it can be achieved 
quite easily today. Indeed, modern lithographic techniques allow researchers to pattern materials into devices with dimensions down to approximately 30 nanometers, roughly 100 atoms across. These systems are so small that they can be easily simulated with a personal computer, as we will do in Chapter \ref{chap:3} and Chapter \ref{chap:4}.
\begin{figure}[h!,b,t]
\vspace{0cm}
\center{\includegraphics[width=12cm,angle=0]{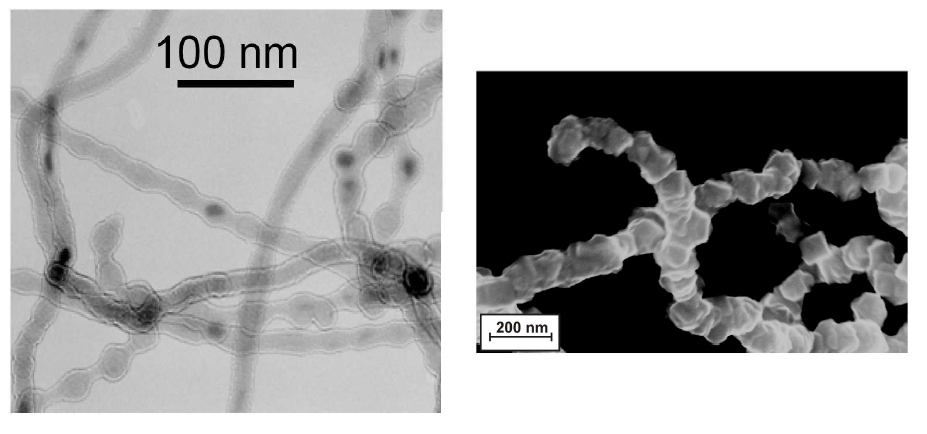}}
\caption{{\footnotesize Electron microscope pictures of nanoparticle chains. Si nanoparticle chain (left) and Fe nanoparticle chain (right). Figures taken from Ref.\cite{2} and Ref.\cite{1} respectively.
}}
\label{fig:nanochain}
\end{figure}
\subsection{Mesoscopic conductors}
\begin{figure}[h!,b,t]
\vspace{0cm}
\center{\includegraphics[width=11.2cm,angle=0]{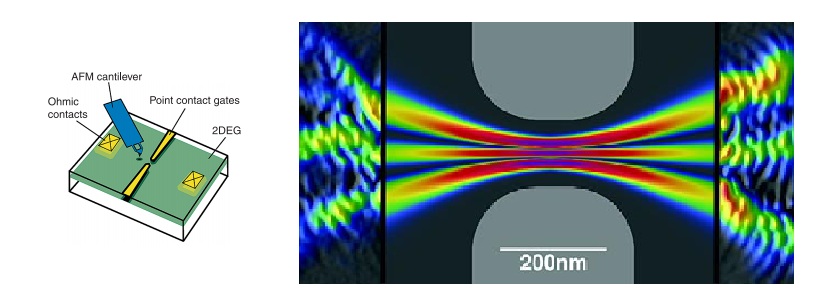}}
\caption{{\footnotesize This image shows the coherent flow of electrons through a Quantum Point Contact (QPC) formed in a two dimensional electron gas (2DEG) inside a GaAs/AlGaAs heterostructure. The gray areas outline the position of electrostatic gates. Scanned probe microscope images on the outside agree well with theoretical simulations inside. The fringes are spaced by half the electron wavelength. Source: Robert Westervelt and Eric Heller at Harvard University, Arthur C. Gossard at UC Santa Barbara, Physics Today 56(12) (2003). 
}}
\label{fig:electron_transp}
\end{figure}

While the  resistance of an electrical element measures its opposition to the passage of an electric current, the \textit{electrical conductance} measures how easily electricity flows along a certain path 
(it is the inverse of the electrical resistance). 
Applying a voltage $V$ across the point contact of a circuit induces a current to flow, the magnitude of this current is given by $I = GV$, where $G$ is the conductance of the contact. It is well-know that the conductance of a conductor is directly proportional to its cross sectional area $W$ and inversely proportional to its length $L$; namely,
\begin{equation}
\label{eq:conduc}
G = \dfrac{\sigma W}{L}
\end{equation}
where the conductivity $\sigma$ is a material property of the sample independent of its dimensions.
How small can we make, \textit{W} and/or \textit{L}, before the
 ohmic behavior breaks down?
This question led to important developments, both theoretical and experimental, in our understanding of the meaning of resistance at the microscopic level. Small conductors whose dimensions are indermediate between the microscopic and the macroscopic are called \textit{mesoscopic.} They are much larger than microscopic objects like atoms, but not large enough to be ``ohmic"\footnote{A notable example of this type of conductors is a Quantum Point Contact (QPC), a narrow constriction between two wide electrically-conducting regions, of a width comparable to the electronic wavelength (from nanometer to micrometer), see Fig.(\ref{fig:electron_transp}).}.
For this type of conductors Equation (\ref{eq:conduc}) is no longer 
valid and two important points should be considered.
Firstly, there is an interface resistance independent of the length \textit{L} of the sample. Secondly, the conductance does not increase linearly with the width \textit{W}. Instead it depends on the number of transverse modes $M$ in the conductor and goes down in discrete steps. These corrections lead 
 to the famous \textit{Laundauer's formula}\footnote{A formal derivation of the Laundauer's formula can be found in Refs. \cite{datta} and \cite{imry}.}:
\begin{equation}
G = \dfrac{2 e^2}{h} M T
\end{equation}
The factor \textit{T} represents the average probability that
 an electron injected at one end of the conductor will transmit to the other end.
Now, a question arises: when a conductor shows an ohmic behavior and when not? 
Do these
characteristic length scales,
 enable us to roughly define a borderline between the
 two regimes?

\subsection{Quantum coherent transport}
A conductor usually shows ohmic behavior if its dimensions are much larger than each of three characteristic length scales: 
\begin{enumerate}
\item the \textit{de Broglie wavelength}, which is related to the kinetic energy of the electrons;
\item the \textit{inelastic mean free path}, which is the distance that an electron travels between two collisions;
\item the \textit{phase-relaxation length}, which is the distance that an electron travels before its initial phase is destroyed.
\end{enumerate}
These length scales change widely from one material to another and are also strongly affected by temperature, magnetic fields 
etc. For this reason, mesoscopic transport phenomena have been observed in conductors having a wide range of dimensions from a few nanometers to hundreds of microns. Now, we want to provide some realistic estimations for the characteristic lengths introduced above. In our examples, we will consider a 2D system because this is a situation that often occurs in experiments; for instance, quantum point contacts are formed in 2-dimensional electron gases (2DEG), e.g. in GaAs/AlGaAs heterostructures\footnote{See for example Ref.\cite{qpc}.}.
\subsubsection*{De Broglie wavelength ($\lambda$)}
It is possible to show\footnote{See Ref.\cite{datta} for more details.}
that the Fermi wavenumber $k_f$ goes up as the square root of the electon density. The corresponding wavelength goes down as the square root of the electron density $n_s$:
\begin{equation}
\lambda_f = \dfrac{2 \pi}{k_f} = \sqrt{\dfrac{2 \pi}{n_s}}
\end{equation}
For an electron density of $5\cdot 10^{11} /cm^2$, the Fermi wavelength is about 35 nm, comparable with characteristic lengths achieved in the industrial production of microprocessors, see Sec.(\ref{sec:nanoworld}). At low temperatures the current is carried mainly by electrons having an energy close to the Fermi energy so that the Fermi wavelength is the relevant length. Other electron with less kinetic energy have longer wavelengths but they do not contribute to the conductance.
\subsubsection*{Inelastic mean free path ($L_m$)}
An electron in a perfect crystal moves as if it were in vacuum but with a different mass (called \textit{effective mass}). Any deviation from perfect crystallinity such as impurities, lattice vibrations (phonons) or other electrons leads to ``collisions" that scatter the electron from one state to another thereby 
changing its momentum. The mean free path, $L_m$, is the distance that an electron travels before its initial momentum is destroyed; that is,
\begin{equation}
L_m = v_f \tau_m
\end{equation}
where $\tau_m$  is the momentum relaxation time and $v_f$ is the Fermi velocity.
The Fermi velocity is given by
\begin{align}
v_f = \dfrac{\hbar k_f}{m} = \dfrac{\hbar}{m} \sqrt{2 \pi n_s} = 3 \cdot 10^7 cm/s && \mbox{if } &&n_s = 5 \cdot 10^{11} /cm^2
\end{align}
Assuming a momentum relaxation time $100 ps$ we obtain a mean free path $L_m = 30 \mu m$. 
\subsubsection*{Phase-relaxation length($L_\phi$)}
The most relevant length scale  quantifying a mesoscopic system is 
 the phase
coherence length $L_\phi$, the length scale over which the \textit{carriers preserve their phase information}.
This phase coherence length is, on the other hand, highly sensitive to 
temperature and sharply decreases with the rise of temperature. 
One way to visualize the distruction of phase is in term of an 
ideal
 experiment
 involving interference. For example, suppose we split a beam of 
electrons into two paths and then recombine them. 
In a perfect crystal, the two paths would be identical resulting 
in constructive interference. However, if we introduce impurities and 
defects randomly into each arm, the two paths
 are then no longer identical and  the interference may  be destructive. 
Moreover, there are other important situations that affect the phase-relaxation time $t_\phi$ such as the effect of a dynamic scatter like lattice vibrations (phonons) or electron-electron interactions.
Therefore, to be in the mesoscopic regime we need low  temperatur
 (on the order of liquid He,  $ \sim 4K$) in order 
to neglect  phase randomization processes caused by phonons.

In this thesis, we consider the \textit{coherent quantum transport}, which means that the results obtained have to be applied to samples of a length $L$ such that the electron's phase coherence length $L_\phi$ (the typical distance the electron travels without losing phase coherence) is larger than or comparable to the system's size $L$.
Phase coherence is affected by the coupling of the electron to its environment, and phase-breaking processes
involve a change in the state of the environment. In
most cases phase coherence is lost in inelastic scatterings, e.g., with other electrons or phonons, but spin-flip
scattering from magnetic impurities can also contribute to phase decoherence. Elastic scatterings of the electron,
e.g., from impurities, usually preserve phase coherence and are characterized by the \textit{elastic} mean free path. $L_\phi$ increases rapidly with decreasing temperature, and for $L \sim 1 \mu m$, an open system typically becomes mesoscopic below $\sim 100 mK$. In a mesoscopic sample, the description of transport in terms of local conductivity breaks down, and the whole sample must be treated as a single, coherent object\footnote{See Ref.\cite{qdot} from more details.}.\\
Thus, systems satisfying the condition $L \ll L_\phi$ have to be treated quantum mechanically, in contrast with  the macroscopic objects where usually the laws of classical mechanics are used. We want to point out that these systems are not so far from the size already achieved by the manufacturers of electronic components; since
 2009 everyone can buy a personal computer equipped with microprocessors with 32-nm technology\footnote{Moreover, in September 2009, Intel disclosed that it had the world's first working 22nm silicon technology. Specifically, Intel disclosed that it had functional SRAM test chips with 364 Mbits, a whopping 2.9 billion transistors per chip. These SRAMs had the 
smallest cell in working circuits to date, at 0.092 square microns. The 22nm process is on track for production in the $2^{nd}$ half of 2011, two years after start of 32nm high volume production.}. 
Therefore, the assumption $L \ll L_\phi$, which are at the basis of our study of \textit{coherent quantum transport} is fully justified and are not as far as can be imagined at first glance by important applications in everyday life. Furthermore, several spectacular effects appear as a consequence of quantum phase coherence of the
electronic wave functions in mesoscopic systems like one-dimensional (1D) quantum wires, quantum dots where electrons are fully confined, two-dimensional (2D) electron gases in heterostructures, etc. For example, we may mention Aharonov-Bohm effect, Universal Conductance Fluctuations, Persistent Current, Anderson Localization and a new and very interesting phenomenon which is the \textit{coherent quantum transport in photosynthesis}. Let us analyze this last phenomenon in more detail\footnote{See Ref.\cite{qpath} and the references therein.}.
\begin{figure}[h!,b,t]
\vspace{0cm}
\center{\includegraphics[width=8cm,angle=0]{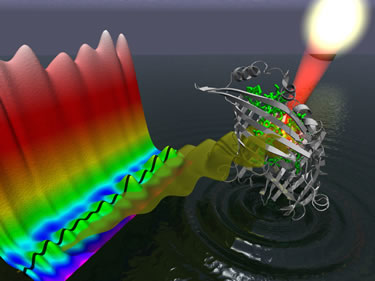}}
\caption{{\footnotesize Like ripples on water, energy absorbed from sunlight moves through the photosynthetic complex (gray-and-green molecular structure) with a wavelike motion. In this artist's rendering, that motion creates interference signals that are analyzed in a spectrum (left). (Image of Greg Engel, Lawrence Berkeley National Laboratory, Physical Biociences Division)
}}
\label{fig:photosyn}
\end{figure}
\subsubsection*{Coherent quantum transport and photosynthesis}
In organisms ranging from blue algae to giant sequoias, complicated assemblies of molecules of the pigment chlorophyll absorb sunlight's photons and channel their energy to enable the plants to turn water and carbon dioxide into oxygen and sugars. The efficiency of photosynthesis, as this process is called, has long astounded scientists. Virtually every photon absorbed by chlorophyll initiates a photosynthetic reaction. Plants use up to 95 percent of the light that strikes them, whereas commercial solar panels use less than 30 percent. But how do plants achieve this amazing result? \\
Before answering this crucial question, let us look at the mechanism of photosynthesis. Photosynthesis is initiated by the excitation, through incident light, of electron in pigment molecules - chromophores - such as chlorophyll. This electronic excitation moves downhill from energy level to energy level through the chromophores before being trapped in a reaction center, where its remaining energy is used to initiate the production of energy-rich carbohydrates.
The mechanism of energy transfert through chromophore complexes has generally been assumed to involve incoherent hopping, thus doing a random walk with a general downhill direction. But at each hopping, the excitation might dissipate as waste heat, so scientists did not understand how the process could be so efficient.
The solution was experimentally discovered by Engel\textit{ et al.}\footnote{See Ref.\cite{qpath2} for more details.}; they found that groups of chlorophyll molecules spend a surprisingly long time in a \textit{superposition of states}.
In the experiment, the team froze (77 kelvin) chlorophyll complexes from blue algae and shot them with sequences of ultrashort laser pulses, each lasting just 40 femtoseconds. Three pulses excited the molecules, and a fourth pulse detected interference patterns.
The complexes stayed in a superposition of states for more than 600 femtoseconds after receiving the pulses. In other words, the electronic excitation that transfers the energy downhill does not simply hop incoherently from state to state, but samples two or more states simultaneously.\\ Thus, the photosynthesis process fully exploit (also at room temperature!) the \textit{quantum coherent transport} and it is precisely this fact that enables natural light-harvesting systems to be so efficient.
Concluding, we can easily understand how the quantum coherent transport \textit{could provide a clean solution to mankind's energy requirements}. Nature already knows to do that, now it's up to us.

\section{The importance of the star graph system}
\begin{figure}[h!,b,t]
\vspace{0cm}
\center{\includegraphics[width=10cm,angle=0]{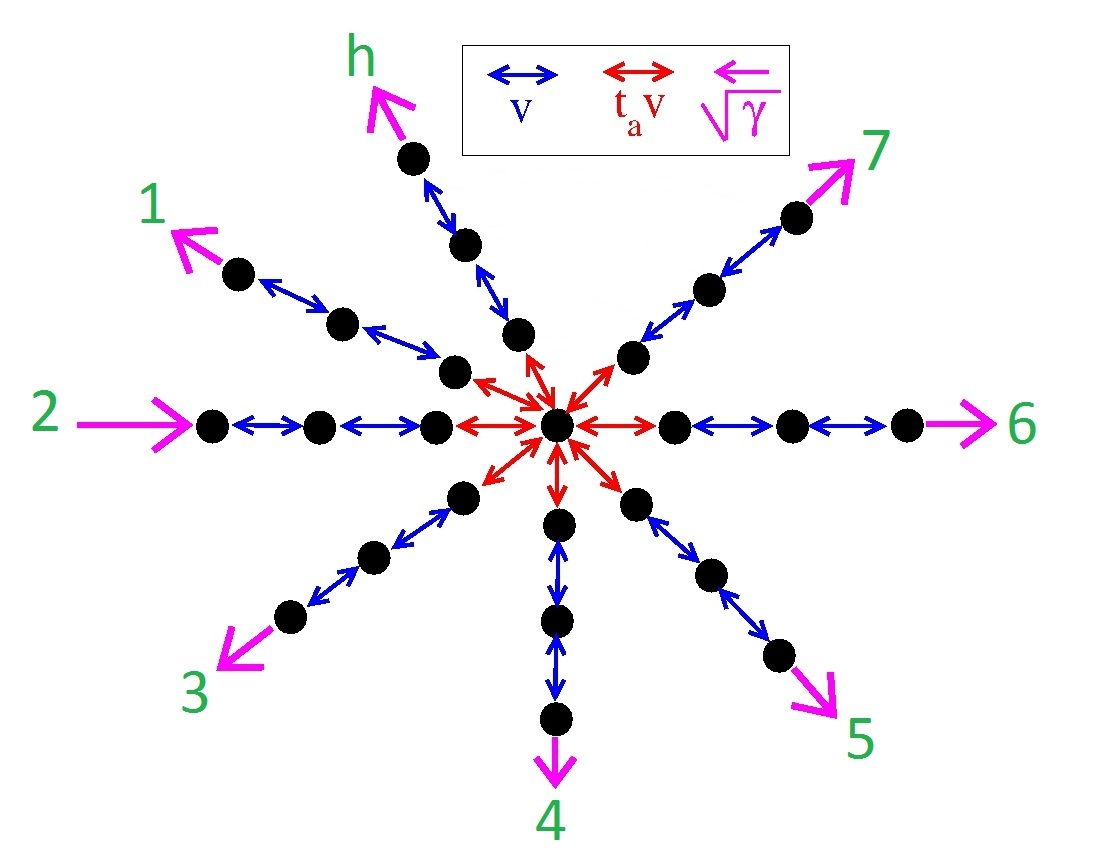}}
\caption{{\footnotesize The star graph system.
}}
\label{fig:stargraph}
\end{figure}
The miniaturization of electronic components follows an impressive pace, and very soon electronic circuits will have to operate in a new regime - the mesoscopic regime - in which a full quantum mechanical treatment is needed, in contrast to the macroscopic objects where usually the laws of classical mechanics are used.
Furthermore, there are experimental evidences that the photosynthesis process fully exploit the \textit{quantum coherent transport} and it is precisely this fact that enables natural light-harvesting systems to be so efficient. 
In this context, the system considered may have important applications in the information technology and also in the modeling of complex systems, such as light-harvesting systems.\\
Up to now, in the literature were considered only very simple systems, such as 1d, 2d and 3d chains; in this thesis, we go beyond because we consider a more complex system, but still analytically treatable, which is closest to the complex systems existing in nature than chains considered so far.\\
More precisely, in this thesis, we will study in details what we named \textit{star graph system}, a system which consist of $h$ chains of sites with a common vertex at the point \textit{O}, the origin. The only coupling between the chains comes through the origin.
Furthermore, when we will considerer an open system, the last site of each chain will be coupled to the external world.
In Fig.(\ref{fig:stargraph}), we show a representation of this system. In this thesis, we will present the results both for the \textit{closed} and the \textit{open system}; we emphasize the fact that all theoretical and numerical results  related to the above system contained in this thesis are brand new and unpublished. In fact, it is the first time that the star graph system has been 
studied in such detail; up to now, such a detailed study of a similar system had been done only for a one-dimensional chain, see  Ref.\cite{lucak}.
Now, we want to share with the reader the importance of the star graph system, and also the coherent quantum transport, in such a system. Here, we present some crucial points:
\begin{itemize}
\item in the mesoscopic regime, electronic transport cannot be investigated by using the conventional Boltzmann transport equation since at this length scale \textit{quantum phase coherence} plays an important role and a full quantum mechanical treatment is needed.
\item The rapid miniaturization of electronic devices will require understanding the mechanism of coherent quantum  transport, see Sec.(\ref{sec:nanoworld}) and Sec.(\ref{sec:meso}); furthermore, if we will build a quantum computer, we will have to know in detail the quantum coherent transport, in order to fully exploit the potential of quantum computing such as the quantum parallelism. In this context, the star graph system could be used as quantum wires to connect different components of a quantum computer.
\item The notion of a closed physical system is always an idealization because the interaction with the outside world, including the measuring apparatus, cannot be removed completely; furthermore, the transport properties depend strongly on the degree of openess of the system. Therefore, our studies on the star graph open system are of great importance in understanding the \textit{interaction} of our quantum system \textit{with the environment}. The effective non-Hermitian Hamiltonian approach to open quantum system used in this thesis has been shown to be a very effective tool to control the effect of the opening of a quantum system. A clear example of non trivial effect of openess on transport properties is the \textit{phenomenon of superradiance}\footnote{See more detail in Chap.\ref{chap:2}, Sec.(\ref{sec:supertrans}).}, recentely shown to occurs in a paradigmatic model of quantum coherent transport\footnote{See Refs.\cite{lucak} and \cite{lucak2}.}.
\item In this thesis, we will demonstrate the existence of localized states in branching regions; we will show that these states are very little affected by the opening of the system and thus have a rather large decay time, see Chap.\ref{chap:4}, Eq.(\ref{eq:tau_loc5}). Accordingly, these localized states, \textit{distinctive feature of the star graph system}, could be used as memory cells.
\end{itemize}
Concluding, we can state that the star graph system studied 
 has lots of new items that could be utilized
 in many different areas, such as condensed matter, nanotechnology, 
molecular biology, information technology and quantum computation.

\begin{figure}[h!,b,t]
\vspace{0cm}
\center{\includegraphics[width=11cm,angle=0]{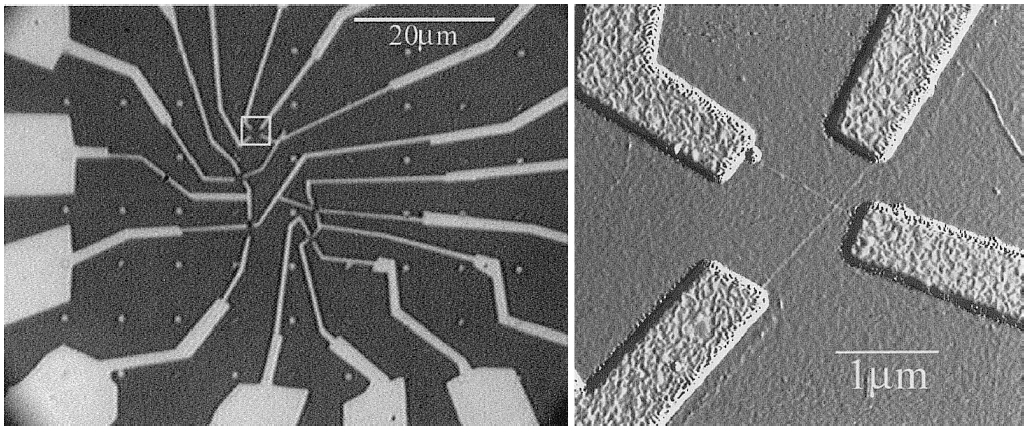}}
\caption{{\footnotesize (a) Optical micrograph showing Cr/Au electrical leads to five crossed SWNT (single-walled carbon nanotubes) devices. (b) Tapping mode AFM image (amplitude signal) of the area indicated by the white box in (a). Figures and caption taken from Ref.\cite{transpnano}.
}}
\label{fig:img_cross}
\end{figure}

\renewcommand\chapterheadstartvskip{\vspace*{0\baselineskip}}





\begin{savequote}[15pc]
\sffamily
There must be no barriers for freedom of inquiry. There is no place for dogma in science. The scientist is free, and must be free to ask any question, to doubt any assertion, to seek for any evidence, to correct any errors.
\qauthor{Robert Oppenheimer - American physicist}
\end{savequote}
\chapter{Open quantum systems: \\the effective Hamiltonian approach}
\label{chap:2}
Transport properties depend strongly on the degree of openess of the system. In important applications, the effect of the opening is large and cannot be treated perturbatively. In a typical situation, we have a discrete quantum system coupled to an external environment characterized by a continuum of states. Elimination of the continuum leads to an effective non-Hermitian Hamiltonian. Analysis of the complex eigenvalues of the effective Hamiltonian reveals a general phenomenon, namely, the \textit{segregation of decay widths} (corresponding to the imaginary parts of the complex eigenvalues). Generically, at weak coupling, all internal states are similarly affected by the opening and acquire small decay widths, resulting in narrow transmission resonances. As the coupling increases and reaches a critical value, the resonances overlap, and a sharp restructuring of the system occurs. Beyond this critical value, a few resonances become \textit{short-lived} states, leaving all other (\textit{long-lived}) states effectively decoupled from the environment. This general phenomenon is referred to as the \textit{super-radiance transition}, due to its analogy with Dicke super-radiance in quantum optics\footnote{See Ref.\cite{dicke}.}. In this chapter, we will present a derivation of the general form of the effective non-Hermitian Hamiltonian starting from standard scattering theory.

\section{The Lippmann-Schwinger equation}
We assume that the Hamiltonian can be written as
\begin{equation}
H = H_0 + V
\end{equation}
where $H_0$ stands for the kinetic-operator
\begin{equation}
H_0 = \dfrac{\textbf{p}}{2m}
\end{equation}
In the absence of a scatterer, $V$ would be zero, and an energy eigenstate would be just a free particle state $\vert \phi \rangle$. The presence of $V$ causes the energy eigenstate to be different from a free-particle state. However, 
in presence of elastic scattering 
we are interested in a solution of
 the full-Hamiltonian Schr\"{o}dinger equation with the same energy eigenvalue. The free particle state $\vert \phi \rangle$ is the energy eigenket of $H_0$:
\begin{equation}
\label{eq:s_h0}
H_0 \vert \phi \rangle = E \vert \phi \rangle
\end{equation}
where $E$ is the energy of the system.
The basic Schr\"{o}dinger equation to be  solved is
\begin{equation}
\label{eq:s_solve}
\left ( H_0 + V \right ) \vert \psi \rangle = E \vert \psi \rangle
\end{equation}
We look for a solution to Eq.(\ref{eq:s_solve}) such that, as $V\rightarrow 0$, we have $\vert \psi \rangle \rightarrow \vert \phi \rangle$, where $\vert \phi \rangle$ is the solution to the free-particle Schr\"{o}dinger equation (\ref{eq:s_h0}) with the same energy eigenvalue.\\
The solution of Eq.(\ref{eq:s_solve}) is:
\begin{equation}
\label{eq:lse}
\vert \psi ^{\left (\pm \right )} \rangle = \vert \phi \rangle + \dfrac{1}{E - H_0 \pm i \varepsilon} V \vert \psi ^{\left (\pm \right )} \rangle
\end{equation}
where $\varepsilon$ is a small real parameter.
This is known as the \textit{Lippmann-Schwinger equation}, named after Bernard A. Lippmann and Julian Schwinger.\\
Now we introduce the Green's functions of the system.
Defining $z = E + i0$\footnote{
This is a shortened form which stands for
\begin{equation*}
z = \lim_{\varepsilon \rightarrow 0^+} \left ( E + i\varepsilon \right )
\end{equation*}.}, the function 
\begin{align}
G\left ( z \right ) = \dfrac{1}{z - H}
\end{align}
is called Green's function (or resolvent, or propagator) 
of $H$ and fulfills the equation
\begin{equation}
\left ( E - H \right ) G = 1
\end{equation}
Similarly, the function 
\begin{equation}
g\left ( z \right ) = \dfrac{1}{z - H_0}
\end{equation}
is the Green's function of the operator $H_0$ and satisfies 
\begin{equation}
\left ( E - H_0 \right ) g = 1
\end{equation}
Now, we want to find out a relation between the two Green's functions. Consider the following equation
\begin{equation}
\label{eq:propag_1}
g^{-1}(z) - G^{-1}(z) = ( z - H_0) - (z - H) = -H_0 + H = V.
\end{equation}
Multiplying Eq.(\ref{eq:propag_1}) to the left by $g(z)$ and to
the right by $G(z)$ gives
\begin{align}
\begin{split}
g(z) \left [ g^{-1} (z) - G^{-1}(z) \right ] G(z) 
= G(z) - g(z) = g(z) V G(z),
\end{split}
\end{align}
from which follows
\begin{equation}
\label{eq:dyson_a}
G(z) = g(z) + g(z)V G(z)
\end{equation}
This is the \textit{Dyson equation}, an equation giving
 the relation between the two \textit{propagators} $g(E)$ and $G(E)$.\\
On the other hand, multiplying Eq.(\ref{eq:propag_1}) to
 the left by $G(z)$ and the right by $g(z)$ we obtain
\begin{equation}
\label{eq:dyson_b}
G(z) = g(z) + G(z)V g(z)
\end{equation}
which is another formulation of the Dyson equation, equivalent to Eq.(\ref{eq:dyson_a}).\\
Coming back to Eq.(\ref{eq:dyson_a}), multiplying to the right by $V$ 
 we obtain
\begin{equation}
G V = g V + G V g V
\end{equation}
which can be written as
\begin{equation}
\label{eq:1_GV}
\left ( 1 + G V \right ) \left ( 1 - gV \right ) = 1
\end{equation}
Now, the Lippman-Schwinger equation (\ref{eq:lse}) reads
\begin{equation}
\left ( 1 - g V \right ) \vert \psi ^{\left (+ \right )}\rangle = \vert \phi \rangle 
\end{equation}
that, multiplying to the left by $\left ( 1 + GV \right )$ and taking into account Eq.(\ref{eq:1_GV}), gives 
\begin{equation}
\label{eq:psi_1_gv}
\vert \psi^{\left ( + \right )} \rangle = \left ( 1 + G V \right )\vert \phi \rangle
\end{equation}
It is convenient to introduce a matrix $\mathcal{T}$ such that
\begin{equation}
\label{eq:trans_m}
\mathcal{T}^{a b} \doteq \langle \phi_b \vert \mathcal{T} \vert \phi_a \rangle \doteq \langle \phi_b \vert V \vert \psi_a^{\left (+\right )}\rangle
\end{equation}
where $\vert \phi_{a,b} \rangle$ and $\vert \psi_{a,b} \rangle$ stand for the eigenkets of the system in the state \textit{a} and \textit{b}, respectively. 
$\mathcal{T}$ will be called the \textit{transition matrix}, and the element $\mathcal{T}^{ab}$ the \textit{amplitude for the transition} $a \rightarrow b$.\\
Multiplying Eq.(\ref{eq:psi_1_gv}) to the left by $V$ we get
\begin{equation}
V \vert \psi^{\left ( + \right )} \rangle = \left ( V + V G V \right )\vert \phi \rangle
\end{equation}
that is, using the definition of the transition matrix 
(\ref{eq:trans_m}),
\begin{equation}
\langle \phi_b \vert \mathcal{T} \vert \phi_a \rangle = \langle \phi_b \vert \left ( V +V G V \right ) \vert \phi_a \rangle
\end{equation}
that implies
\begin{equation}
\label{eq:v_vgv}
\boxed{
\mathcal{T} = V + V G V = V + V \dfrac{1}{E - H + i0} V
}
\end{equation}

\section{The effective Hamiltonian}
Consider a discrete quantum system described by intrinsic basis states coupled to a continuum of states.
First of all, let us divide the Hilbert space of the full Hermitian Hamiltonian $H$ of the system into two mutually orthogonal subspace $S_P$ and $S_Q$ with the use of the projection operators $P$ and $Q$, respectively. The subspace $S_P$ involves the internal states $\left  \lbrace \vert n \rangle \right \rbrace$; $n=1..N$ and the subspace $S_Q$ is composed by the continuum channel states $\left \lbrace \vert c; E \rangle \right \rbrace$ where $c=1..M$ is a discrete quantum number labeling $M$ channels and $E$ is a continuum quantum number representing the energy.\\
Keeping in mind the division of the Hilbert space of $H$, the operator $H$ can be written in the form
\begin{equation}
H =
\left( \begin{array}{cc}
H_{PP} & H_{PQ} \\
H_{QP} & H_{QQ} 
\end{array} \right)
\end{equation}
where we used the notation $H_{XY} \doteq XHY$.
Now, splitting the operator $H$, we obtain
\begin{equation}
\label{eq:h-h0-v}
H = H_0 + V = \left( \begin{array}{cc}
H_{PP} & 0 \\
0 & H_{QQ} 
\end{array} \right) +
\left( \begin{array}{cc}
0 & H_{PQ} \\
H_{QP} & 0 
\end{array} \right)
\end{equation}
Application of the Dyson equation (\ref{eq:dyson_a}) gives
\begin{align}
\begin{split}
\left( \begin{array}{cc}
G_{PP} & G_{PQ} \\
G_{QP} & G_{QQ} 
\end{array} \right) =&
\left( \begin{array}{cc}
g_{PP} & 0 \\
0 & g_{QQ} 
\end{array} \right) + \\
+&\left( \begin{array}{cc}
g_{PP} & 0 \\
0 & g_{QQ} 
\end{array} \right)
\left( \begin{array}{cc}
0 & H_{PQ} \\
H_{QP} & 0 
\end{array} \right)
\left( \begin{array}{cc}
G_{PP} & G_{PQ} \\
G_{QP} & G_{QQ} 
\end{array} \right)
\end{split}
\end{align}
where $g_{PP}(z) = (z- H_{PP})^{-1}$ and $g_{QQ} = (z- H_{QQ})^{-1}$. Performing the algebrical matrix operations indicated in the above expression, it follows
\begin{align}
G_{PP}(z) &= g_{PP}(z) + g_{PP}(z)H_{PQ}G_{QP}(z)\\
G_{QP}(z) &= g_{QQ}(z)H_{QP}G_{PP}(z)
\end{align}
(and similar expressions for exchange of the subscripts P and Q). With straightforward properties of matrix multiplication we have
\begin{equation}
\label{eq:G_aa}
G_{PP}(E) = \frac{1}{E - H_{PP}-H_{PQ}\dfrac{1}{E - H_{QQ}+i0}H_{QP}}
\end{equation}
Until now we have performed exact algebraic transformations.\\
The interpretation of equation (\ref{eq:G_aa}) is immediate. The Green's function in the space $S_P$ is determined by an \textit{effective Hamiltonian} which is given by
\begin{equation}
\label{eq:eff_h}
\boxed{
\mathcal{H}(E) = H_{PP} + H_{PQ}\dfrac{1}{E - H_{QQ}+i0}H_{QP}
}
\end{equation}
The physical meaning of the effective Hamiltonian is self-explanatory: besides $H_{PP}$, it contains the effect of an excursion from $S_P$ to $S_Q$, a propagation within $S_Q$, and an excursion back to $S_P$.\\
Suppose that the splitting of the original space spanned by $H$ is such that in the subspace $S_Q$ the Green's function is known. We can \textit{eliminate} (or as it is commonly said in the literature \textit{decimate}) all the states of the subspace $S_Q$, considering within the subspace $S_P$ the effective Hamiltonian
\begin{equation}
\mathcal{H}(E) = H_{PP} + \Sigma_{PP}(E)
\end{equation}
with
\begin{equation}
\Sigma_{PP}(E) = H_{PQ}\dfrac{1}{E - H_{QQ}+i0}H_{QP}
\end{equation}
The operator $\Sigma_{PP}(E)$, whose origin is linked to the elimination of the subspace $S_Q$, is called \textit{self-energy operator}; the self-energy operator $\Sigma_{PP}(E)$, added to $H_{PP}$, gives the effective Hamiltonian on the preserved subspace $S_P$, now formally decoupled from the subspace $S_Q$.\\
Coming back to Eq.(\ref{eq:eff_h}), we note that the effective Hamiltonian is in term of the projection operators. These are the explicit structures of the projection operators 
\begin{align}
P &= \sum_{n=1}^N \vert n \rangle \langle n \vert \\
Q &= \sum_c \int dE^{\prime} \vert c, E^\prime \rangle \langle c, E^\prime \rangle
\end{align}
so that
\begin{equation}
P + Q =1
\end{equation}
and
\begin{equation}
PQ = QP = 0
\end{equation}
In order to put Eq.(\ref{eq:eff_h}) in a more transparent form we substitute into it the definitions of the projection operators and, after a straightforward calculation, we obtain
\begin{equation}
\mathcal{H}(E) = H_{PP} + \sum_c \sum_{r,s} \int d E^\prime \dfrac{\vert r \rangle \langle r \vert H \vert c, E^\prime \rangle
\langle c, E^\prime \vert H \vert s \rangle \langle s \vert}{E - E^\prime +i0} 
\end{equation}
Multiplying from the left by $\langle m \vert$ and from the right by $\vert n \rangle$ which are two internal states, we get
\begin{equation}
\label{eq:ham_eff}
\langle m \vert \mathcal{H} (E)\vert n \rangle = \langle m \vert H \vert n \rangle + \sum_c \int d E^\prime \dfrac{A_m^c (E^\prime) A_n^c (E^\prime)^*}{E - E^\prime +i0} 
\end{equation}
where we have defined 
\begin{equation}
A_n^c (E^\prime) \doteq \langle n \vert H \vert c,E^\prime \rangle
\end{equation}
which is the \textit{transition amplitude} between the intrinsic states and the continuum. 
The integral in Eq.(\ref{eq:ham_eff}) can be further decomposed into its Hermitian part (principal value) and the remaining-non Hermitian part by using the Sokhotski-Plemelj formula\footnote{
Let $f$ be a complex-valued function and let $a$ and $b$ be real constants with $a < 0 < b$. Then
\begin{equation*}
\lim_{\varepsilon \rightarrow 0^+} \int_a^b \dfrac{f(x)}{x \pm i \varepsilon} dx = \mathcal{PV}\int_a^b \dfrac{f(x)}{x} \mp i \pi f(0)
\end{equation*}
where $\mathcal{PV}$ denotes the Cauchy principal value.
}:
\begin{equation}
\begin{split}
\sum_c \int d E^\prime \dfrac{A_m^c (E^\prime) A_n^c (E^\prime)^*}{E - E^\prime +i0}  = \sum_c \mathcal{PV}& \int d E^\prime \dfrac{A_m^c (E^\prime) A_n^c (E^\prime)^*}{E - E^\prime}\\ - &i\pi \sum_{c(open)} A_m^c(E) A_n^c(E)^* 
\end{split}
\end{equation}
where the notation $\mathcal{PV}$ refers to the Cauchy principal value. Then, the effective Hamiltonian for the intrinsic system, which fully takes into account its opening to the outside, can be written as
\begin{equation}
\label{eq:ham_hpw}
\mathcal{H} (E) = H_{PP} + \Delta(E) - \dfrac{i}{2} W (E)
\end{equation}
with
\begin{equation}
W_{nm} = 2 \pi \sum_{c(open)} A_n^c(E)A_m^c(E)^*
\end{equation}
where the sum is limited to the open channels, and
\begin{equation}
\Delta_{nm} = \sum_c \mathcal{PV} \int dE^\prime \dfrac{A_n^c(E^\prime)A_m^c (E^\prime)^*}{E-E^\prime}
\end{equation}
Assuming $\Delta \left ( E \right )$ and $W \left ( E \right )$ are smooth function of the energy, their energy dependence can be neglected if the region of interest is concentrated in a small energy window. Therefore, the amplitudes $A_n^c$ can be taken as energy-independent parameters.
Taking into account all these considerations, Eq.(\ref{eq:ham_hpw}) becomes
\begin{align}
\label{eq:2sempl_h}
\mathcal{H} = H_{PP} - \dfrac{i}{2}W && W_{mn}=\sum_{c=1}^M A_m^c A_n^{c*}
\end{align}
Under time-reversal invariance, both $H_{PP}$ and $W$ are real symmetric matrices, then the coupling amplitudes $A_n^c$ between intrinsic states $\vert n \rangle$ and channels $c$ can be taken as real.

\section{The scattering matrix}
The effective Hamiltonian (\ref{eq:ham_hpw}) determines the scattering matrix of the system.
First of all, we rewrite Eq.(\ref{eq:h-h0-v}) in a different, but equivalent way:
\begin{equation}
H = H_{PP} + H_{QQ} + H_{PQ} + H_{QP}
\end{equation}
where we used the notation $H_{XY} \doteq XHY$ previously introduced.\\
The first two terms $H_w \doteq H_{PP} + H_{QQ}$ are the part of the full Hamiltonian $H$ acting within the respective subspaces $P$ and $Q$. Similarly, the last two terms $V \doteq H_{PQ} + H_{QP}$ act across the two subspaces. Our full Hamiltonian can now be written as $H = H_w + V$.\\
From standard scattering theory, we know that the scattering matrix is defined as 
\begin{equation}
\mathcal{S} = 1 - i \mathcal{T}
\end{equation}
where $\mathcal{T}$ is the transition matrix already defined in Eq.(\ref{eq:trans_m}). \\
Using Eq.(\ref{eq:v_vgv}), we obtain the following relation for the transition matrix:
\begin{equation}
\mathcal{T} = V + V G V = V + V \dfrac{1}{E - H  +i0} V
\end{equation}
where $G$ is the  Green's function of the full Hamiltonian $H$.\\
The transition amplitude for the process from channel $a$ to $b$ is
\begin{align}
\begin{split}
\mathcal{T}^{ab}\left (E \right ) =& \langle b,E \vert \mathcal{T} \vert a, E \rangle \\
=& \langle b,E \vert V \vert a, E \rangle + \langle b,E \vert V G V \vert a, E \rangle
\end{split}
\end{align}
Now, we insert $V = H_{PQ} + H_{QP}$ and $P = \sum_n \vert n \rangle \langle n \vert$ into the above expression and by exploiting the orthogonality of the two subspace $S_P$ and $S_Q$, we find
\begin{equation}
\label{eq:t_a_b}
\mathcal{T}^{ab} \left (E \right ) = \langle b,E \vert H_{QP} P G P H_{PQ} \vert a,E \rangle
\end{equation}
We note that in the equation above  the propagator $PGP$
appears
 which is equal to $G_{PP}$. The latter
can be seen as a projection of the full propagator $G$ into the 
internal subspace, after the elimination of  channel variables.\\
Using both Eq.(\ref{eq:G_aa}) and (\ref{eq:eff_h}), the equation (\ref{eq:t_a_b}) reads
\begin{align}
\label{eq:trasm}
\begin{split}
\mathcal{T}^{ab} \left (E \right ) &= \sum_{n,m=1}^N\langle b,E \vert H \vert n \rangle \langle n \vert G_{PP} \vert m \rangle \langle m \vert H \vert a, E \rangle \\
&= \sum_{n,m=1}^N A_n^{b*}\left (E\right ) \left ( \dfrac{1}{E - \mathcal{H}\left (E\right )}\right )_{nm} A_m^a \left ( E \right )
\end{split}
\end{align}
The propagator $\left (E - \mathcal{H} \right )^{-1}$ in the scattering amplitude $\mathcal{T}^{ab}$ does not depend on a specific reaction and contains the full effective Hamiltonian (\ref{eq:ham_hpw}) with the same amplitudes $A_n^c$ as those determining the entrance and exit channel in Eq.(\ref{eq:trasm}). This guarantees the unitarity of the $\mathcal{S}$-matrix since the virtual processes of evolution of the open system to and from the continuum channels are included in all orders in the propagator.
From the transition amplitude $\mathcal{T}$, we define the \textit{transmission} $\mathsf{T}^{ab}(E)$ which gives us the transmission from channel $a$ to $b$:
\begin{equation}
\label{eq:2t_ab}
\mathsf{T}^{ab}(E) = \vert \mathcal{T}^{ab}(E) \vert^2
\end{equation}
Now, we want to write $\mathcal{T}^{ab}$ in a different way in order to show that the eigenvalues of $\mathcal{H}$ coincide with the poles of the scattering matrix $\mathcal{S} = 1 - i \mathcal{T}$. First of all, we diagonalize the effective non-Hermitian Hamiltonian $\mathcal{H}$. Its eigenfunctions $\vert r \rangle$ and $\langle \tilde{r} \vert$ form a bi-orthogonal complete set
\begin{align}
\mathcal{H} \vert r \rangle = \mathcal{E}_r \vert r \rangle && \langle \tilde{r} \vert \mathcal{H} = \langle \tilde{r} \vert \mathcal{E}_r^*
\end{align}
and its eigenvalues are complex energies
\begin{equation}
\mathcal{E}_r = E_r - \dfrac{i}{2}\Gamma_r
\end{equation}
corresponding to resonances centered at $E_r$ with widths $\Gamma_r$ which determine the lifetime of the resonances, $\tau_r \sim \hbar/\Gamma_r$. The decay amplitudes $A_n^a$ are transformed according to
\begin{align}
\mathcal{A}_r^a = \sum_n A_n^a \langle n \vert r \rangle && \mathcal{\tilde{A}}_r^b = \sum_m \langle \tilde{r} \vert m \rangle A_m^b
\end{align}
and the transmission amplitudes are given by
\begin{equation}
\label{eq:2pole}
\mathcal{T}^{ab}\left (E \right ) = \sum_{r=1}^N \mathcal{A}_r^a \dfrac{1}{E - \mathcal{E}_r}\mathcal{\tilde{A}}_r^b
\end{equation}
The complex eigenvalues $\mathcal{E}_r$ of $\mathcal{H}$ thus
coincide with the pole of the scattering matrix $\mathcal{S}$. From this consideration, it is clear that the properties of the complex eigenvalues of the effective Hamiltonian $\mathcal{H}$ are extremely important for understanding the transport properties of the system.

\section{Transition to superradiance}
\label{sec:supertrans}
In order to understand what is the transition to superradiance, we can consider a simplified version of Eq.(\ref{eq:2sempl_h}):
\begin{equation}
\label{eq:h_eff_short}
\mathcal{H} = H_0 - \frac{i}{2}\gamma W
\end{equation}
where
\begin{equation}
\label{eq:w_part}
W_{ij} = \sum_{c(open)}A_i^c A_j^{c*}
\end{equation}
and where $\gamma$ is a parameter that controls the coupling strength with the external world (which we assume to be of the same order of magnitude for all the intrinsic states), and the basis states $\vert i \rangle$ are chosen to be the eigenstates of $H_0$, with eigenvalues $E_0^i$.\\
As long as $\gamma$ is small $(\gamma \ll 1)$, the second term of the Hamiltonian $\mathcal{H}$ in Eq.(\ref{eq:h_eff_short}) can be considered as a small perturbation of $H_0$. This condition is always fulfilled if the average width $\langle \Gamma \rangle$ is much smaller than the average distance $\bar{D}$ between neighboring resonance states. In this case, the nondiagonal matrix elements of $\mathcal{H}$ are small and the individual resonances are isolated. So, we have that the first-order complex eigenvalues of $\mathcal{H}$ are
\begin{align}
\mathcal{E}_i = E_0^i - \frac{i}{2}\gamma W_{ii}
\end{align}
In the opposite case of large $\gamma$ ($\gamma \gg 1$), the matrix $W$ determines the behavior of the system. $H_0$ can be viewed as a perturbation acting on $W$. From Eq.(\ref{eq:w_part}) it is evident that the rank of $W$ is $M$, the number of open channels, and, hence, $M$ is also the rank of $\mathcal{H}$. This implies that $W$ has only $M$ nonzero eigenvalues for $M<N$. Thus, only $M$ states will have a decay width in the limit of large coupling, while all others will have zero width to first order. Therefore, as the coupling increases, all widths initially increase linearly with $\gamma$, but at large coupling only $M$ of the widths continue to increase, while the remaining $N-M$ widths approach zero. 
The
 $N-M$ states  almost decoupled from the continuum of decay channels
 become \textit{long-lived} (\textit{trapped}) while $M$ states take almost the whole coupling strength and become \textit{short-lived} (\textit{super-radiant}). This general phenomenon is referred to as the \textit{super-radiance transition}, due to
its analogy with Dicke super-radiance in quantum optics, see Ref.\cite{dicke}.
Therefore, it is clear that a transition between these two regimes may take place at a critical value of $\gamma$. 
Roughly, the transition occurs when 
\begin{equation}
\label{eq:2crit_crit}
\gamma/D \approx 1
\end{equation}
where $D$ is the mean level spacing of $H_0$. Note that the qualitative criterion $\gamma/D \approx 1$ for the transition to superradiance is valid in the case of uniform density of states and negligible energy shift; when the density of states is not uniform, the transition to superradiance occurs as a hierarchical process, see Ref. \cite{hier}.

\renewcommand\chapterheadstartvskip{\vspace*{-5\baselineskip}}





\label{chapter:the_model}

\begin{savequote}[15pc]
\sffamily
A good idea has a way of becoming simpler and solving problems other than for which it was intended.
\qauthor{Robert Tarjan - computer scientist}
\end{savequote}

\chapter{Star graph: \\the closed tight-binding model}
\label{chap:3}
\section{Overview of tight-binding method}
The \textit{tight binding method}, suggested by Bloch in 1928, is an approach to the calculation of electronic band structure using an approximate set of wave functions based upon superposition of wave functions for \textit{isolated atoms} located at each atomic site; these wave functions can be physically interpreted as atomic orbitals.  Orbitals of neighboring (or not neighboring) sites are connected by what is referred to as a \textit{hopping matrix element} or an \textit{overlap integral}. The tight-binding approximation deals with the case in which the overlap of atomic wave functions is enough to require corrections to the picture of isolated atoms, but not so much as to render the atomic description completely irrelevant. This method, when not applied in oversimplified forms, describe electron propagation in any type of crystal (metal, semiconductors and insulators). Indeed, the tight binding model has a long history and has been applied in many ways and with many different purposes and different outcomes. Parts of the model can be filled in or extended by other kinds of calculations and models like the nearly-free electron model. The model itself, or parts of it, can serve as the basis for other calculations. In the study of conductive polymers, organic semiconductors and molecular electronics for example, tight binding like models are applied in which the role of the atoms in the original concept is replaced by the molecular orbitals of conjugated systems and where the inter atomic matrix elements are replaced by inter or intra molecular hopping and tunneling parameters. \\
For example, we imagine construction of a crystal from a hypothetical periodic one-dimensional sequence of $R$ equal atoms. In the case of negligible interaction among atoms, the same atomic orbitals centered in the different lattice sites would have the same energy; in the presence of interaction this $R$ fold degeneracy is removed and evolves into an \textit{energy band}.
\begin{figure}[h!]
\vspace{0cm}
\center{\includegraphics[width=10cm,angle=0]{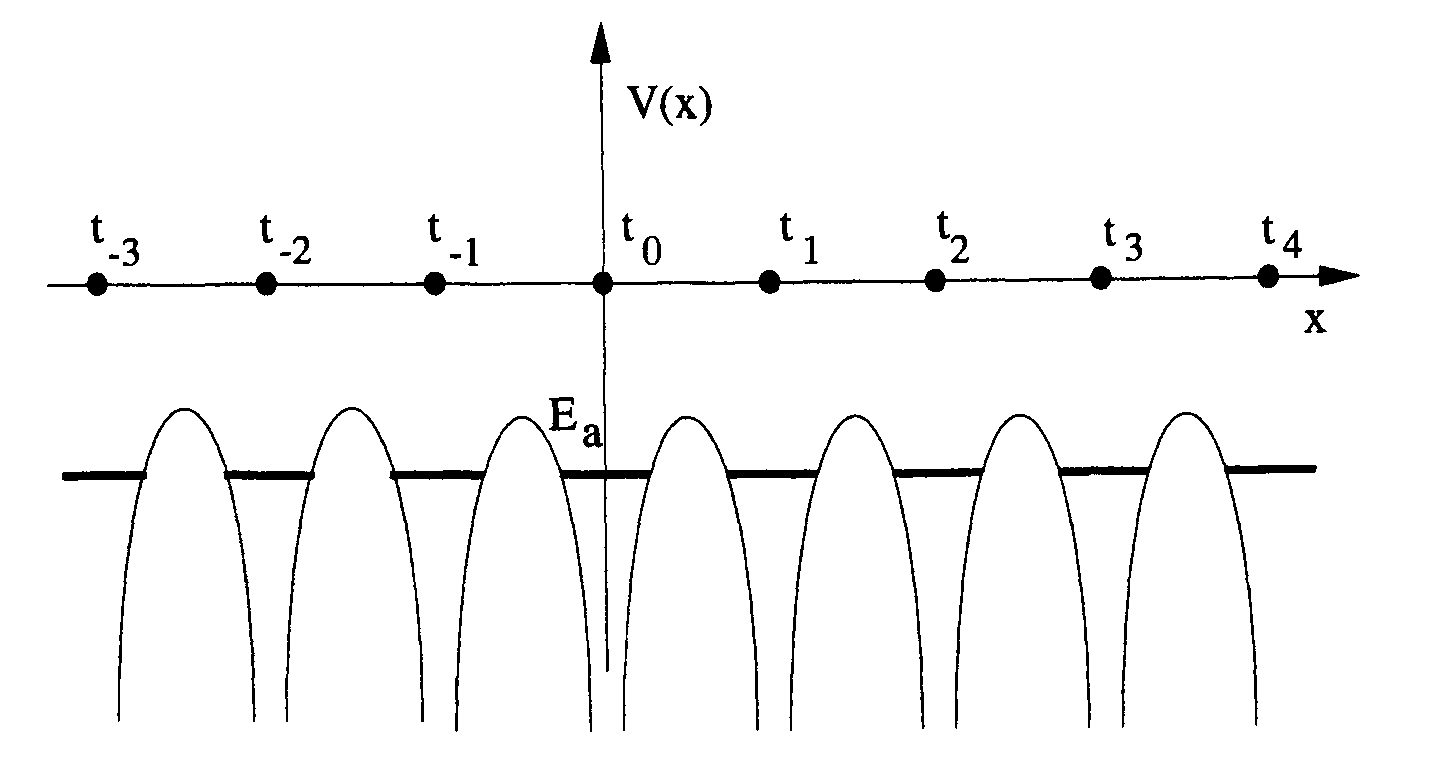}}
\caption{{\small
Schematic representation of the crystal potential as a superposition of atomic-like potentials, centered at the lattice sites $t_n$. In the tight-binding approximation, the interaction between nearby atomic-like orbitals of energy $E_a$ leads to the formation of energy bands. Figure taken from \cite{grosso}. 
}}
\label{fig:img_pastori}
\end{figure}

\section{The model}
\label{sec:3model}
\begin{figure}[h!,b,t]
\vspace{0cm}
\center{\includegraphics[width=10cm,angle=0]{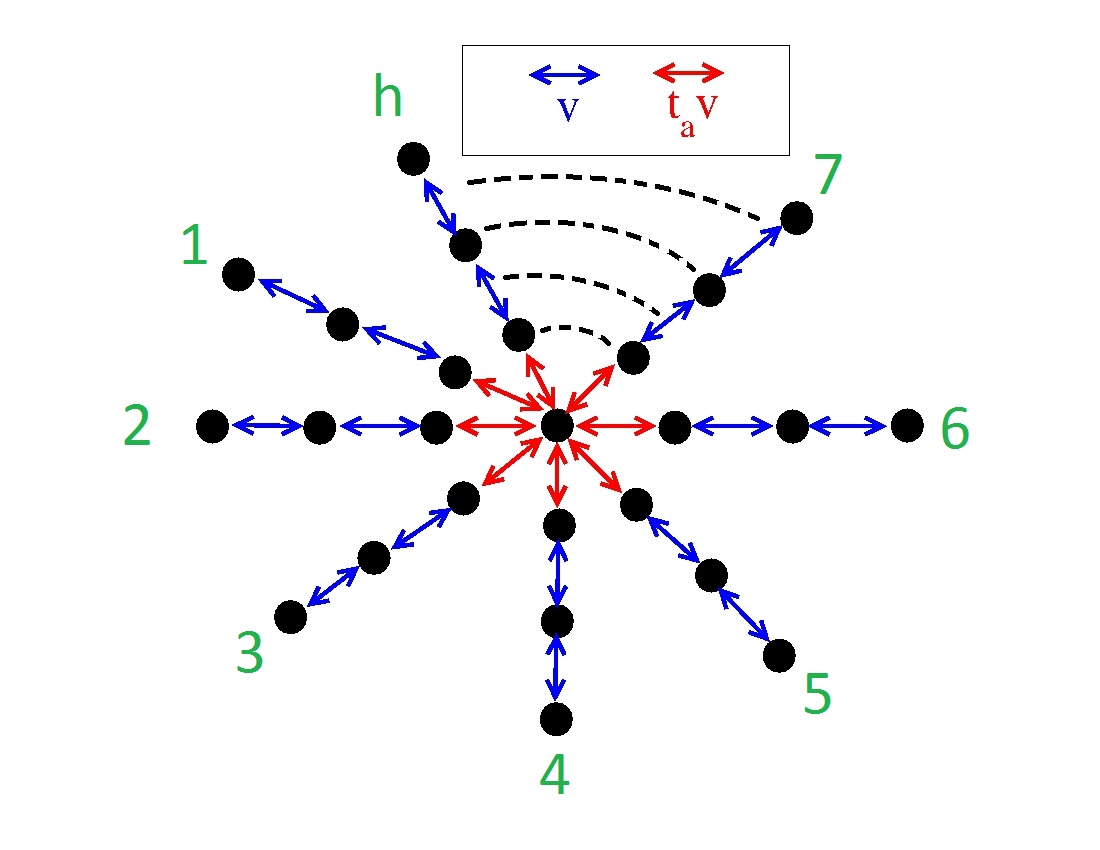}}
\caption{{\footnotesize The star graph system (closed system).
}}
\label{fig:4stargraph_closed}
\end{figure}

In this section, we will introduce the tight-binding model that describes the system we have considered; we will introduce the \textit{general model} (
 the more general model that we will solve in this thesis) and the \textit{simplified model} that will be solved in this chapter.
\subsection{General model}
We consider \textit{h} chains of sites of lengths $N_a$, $a=1, \ldots, h.$; the chains have a common vertex at the point $O$. In each chain, a particle can propagate with the hopping amplitude $v_a$ to closest neighbors\footnote{
For simplicity, fully justified by the localized nature of atomic orbitals, the hopping integrals involving second or further apart neighbours are assumed negligible.}. The channels are coupled through the origin with hopping amplitudes $t_av_a$. 
We name this system $S_h(N_1, \ldots, N_h)$ \textit{system}; indeed, this is a \textit{star graph}.\\
The total wave function of a particle in such a closed system is
\begin{equation}
\label{eq:3psi_form_gen}
\vert \psi\rangle = \sum_{a=1}^h \sum_{n=1}^{N_a}C_n^a \vert a; n\rangle + C_0 \vert 0 \rangle.
\end{equation}
We assume that the numeration of the sites in each chain starts, $n=1$, at the site closest to the origin, while the site $\vert a; N_a\rangle$ will be later coupled to the external world. The level energy in each site equals $\varepsilon_a$.\\
Substituting (\ref{eq:3psi_form_gen}) in the stationary Schr\"{o}dinger equation with energy \textit{E}, we obtain the following equations for the state (\ref{eq:3psi_form_gen}):
\begin{align}
\label{eq:obv_eq1}
v_a C_2 + t_a v_a C_0 = (E - \varepsilon_a)C_1^a \\
\label{eq:obv_eq2}
v_a ( C_{n+1}^a + C_{n-1}^a ) = (E - \varepsilon_a)C_n^a && n=2,\ldots, N_a
\end{align}
while the central amplitude $C_0$ satisfies
\begin{equation}
\label{eq:central_ampl_gen}
\sum_{a=1}^h t_av_a C_1^a = (E - \varepsilon_0)C_0,
\end{equation}
where $\varepsilon_0$ is a local level at the center. The boundary conditions at the outer ends of a closed system are $C_{N_a+1}^a = 0$. The only coupling between the chains comes through eq. (\ref{eq:central_ampl_gen}).
Furthermore, when we will consider
 an open system, the last site of each chain will be coupled to the external world with coupling amplitudes $\sqrt{\gamma_1}, \ldots, \sqrt{\gamma_h}$. 

\subsection{Simplified model}
\label{eq:s_model}
In this chapter, we will consider the case of \textit{h chains} all of identical length $N_a \equiv N$ and identical hopping amplitudes $v_a \equiv v$ among the $h$ chains.
We name this system $S_h(N)$ \textit{system}.
In this case, we have $hN+1$ sites, including the vertex. \\
Taking into account all these considerations, the tight-binding Hamiltonian $\mathsf{H}$ with nearest neighbour interaction can be written as
\begin{align}
\label{eq:3_hamil}
\mathsf{H} = \mathsf{H}_{dc} + \mathsf{H}_{c}
\end{align}
where
\begin{align}
\begin{split}
\label{eq:3h_bloch}
\mathsf{H}_{dc} &= \varepsilon_0 \vert 0 \rangle \langle 0 \vert + \sum_{a=1}^h \sum_{n=1}^N \varepsilon_a \vert a; n \rangle \langle a; n \vert \\ &+\sum_{a=1}^h \sum_{n=1}^{N-1} v \vert a; n \rangle \langle a; n + 1 \vert + 
\sum_{a=1}^h \sum_{n=2}^{N} v \vert a; n \rangle \langle a; n - 1\vert
\end{split}
\end{align}
and
\begin{equation}
\label{eq:3h_co}
\mathsf{H}_{c} = \sum_{a=1}^h t_av \left( \vert 0 \rangle \langle a; 1 \vert + \vert a; 1 \rangle \langle 0 \vert \right)
\end{equation}
It is clear from Eqs.(\ref{eq:3h_bloch}) and (\ref{eq:3h_co}) that $\mathsf{H}_{dc}$ gives us the tight-binding model for $h$ \textit{decoupled chains} of sites (plus the origin) whereas $\mathsf{H}_{c}$ represents the coupling between the origin and the first site of each chain.\\
Moreover, we set the level energy in each site (also the origin) equal to $\varepsilon_0 = \varepsilon_a = 0$.
Thus, the Hamiltonian $\mathsf{H}$ becomes
\begin{align}
\begin{split}
\mathsf{H} =& \sum_{a=1}^h \sum_{n=1}^{N-1} v \vert a; n \rangle \langle a; n + 1 \vert + 
\sum_{a=1}^h \sum_{n=2}^{N} v \vert a; n \rangle \langle a; n - 1\vert \\  &+ \sum_{a=1}^h t_av \left( \vert 0 \rangle \langle a; 1 \vert + \vert a; 1 \rangle \langle 0 \vert \right)
\end{split}
\end{align}
Furthermore, starting from Eq.(\ref{eq:3psi_form_gen}), we have that the total wave function of a particle in such a closed system is
\begin{equation}
\label{eq:psi_form}
\vert \psi\rangle = \sum_{a=1}^h \sum_{n=1}^{N}C_n^a \vert a; n\rangle + C_0 \vert 0 \rangle.
\end{equation}
Taking into account Eqs.(\ref{eq:obv_eq1}), (\ref{eq:obv_eq2}) and (\ref{eq:central_ampl_gen}), the amplitudes of the stationary state (\ref{eq:psi_form}) with energy \textit{E} satisfy the following equations,
\begin{align}
v C_2 + t_av C_0 = E C_1^a \\
v ( C_{n+1}^a + C_{n-1}^a ) = E C_n^a && n=2,\ldots, N
\end{align}
while the central amplitude $C_0$ satisfies
\begin{equation}
\label{eq:central_ampl}
\sum_{a=1}^h t_av C_1^a = E C_0,
\end{equation}

\section{Solution of the model for a closed system}
\label{sec:3sol_closed}
In this section, we will solve the simplified model presented in Paragraph(\ref{eq:s_model}) for a closed system.
The equations that we have to solve are
\begin{equation}
\label{eq:system_rip}
 \left\{ \begin{array}{rl}
 \displaystyle{\sum_{a=1}^h t_a v C_1^a = E C_0} \\
  \displaystyle{v C_2^a + t_a v  C_0  = E C_1^a}  & \hspace*{15pt} \textrm{\textit{a}=1,\dots,
  \textit{h}}\\
 \displaystyle{v \left( C_{n+1}^a + C_{n-1}^a \right ) = E C_n^a}  & \hspace*{15pt} \textrm{\textit{a}=1,\dots, \textit{h}; \textit{n}=2,\ldots, \textit{N}-1}\\
 \displaystyle{v C_{N-1}^a = E C_N^a} & \hspace*{15pt} \textrm{\textit{a}=1,\ldots, \textit{h}}
\end{array} \right.
\end{equation}
One can easily notice that 
we have  $hN+1$ equations, as expected for $hN+1$ sites.\\
The system (\ref{eq:system_rip}) allows us to calculate \textit{numerically} the eigenvalues of the system; it is enough, in fact, diagonalize the matrix that represents the system in the basis
 of the unknowns $\left \lbrace C_0,C_n^a \right \rbrace$. \\
A numerical solution of the system (\ref{eq:system_rip}) gives us \textit{two eigenvalues outside of the
 normal Bloch band}, while the others $hN-1$ are enclosed within the Bloch band.
These two eigenvalues outside of the
 normal Bloch band are a \textit{distinctive feature} of our \textit{star graph system}; in the linear 1d-chains considered so far in the literature, in fact, all the eigenvalues were confined within the Bloch band.
\begin{figure}[h!]
\vspace{0cm}
\center{\includegraphics[width=10cm,angle=-90]{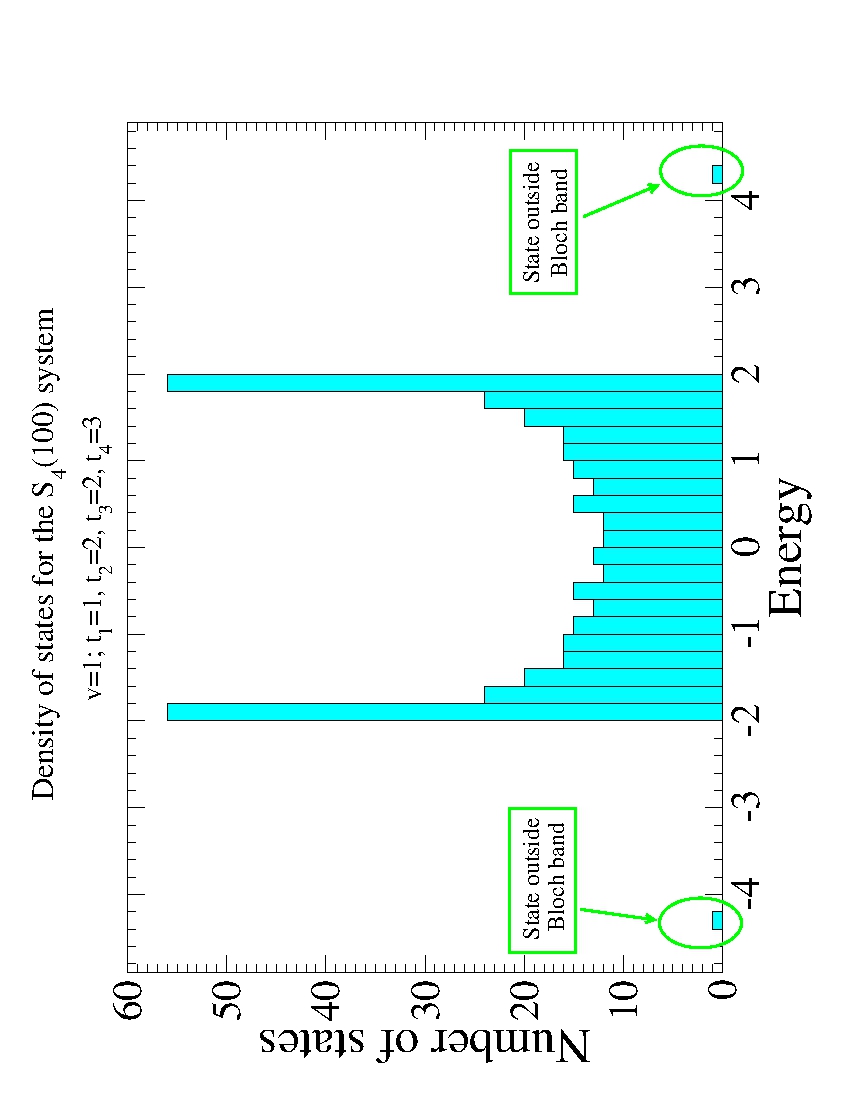}}
\caption{{\small
We plot the number of states as a function of the energy of the system for the $S_{4}(100)$ system with $t_1=1, t_2=2, t_3=2, t_4=3$ and $v=1$. We can see that the states are all within the band of Bloch, [-2,+2] in this case, except two, which are indicated with green circles. The quantities plotted are dimensionless. 
}}
\label{fig:Dos_h_4}
\end{figure}
In Fig.(\ref{fig:Dos_h_4}) we show the density of states (DOS) for the $S_{4}(100)$ system - 4 chains each of 100 sites plus the origin - with $t_1=1, t_2=2, t_3=2, t_4=3$ and $v=1$; it is clear from Fig.(\ref{fig:Dos_h_4}) that we have only two states outside of the  normal Bloch band.
This result is not due to the particular choice of parameters, but it is a general feature of the star graph system due to the coupling of the different chains with the origin, i.e.
 to the term $\mathsf{H}_{c}$ of 
Eq.~(\ref{eq:3h_co}). 
Indeed, if we set $\mathsf{H}_{c} = 0$, all the states are confined within the Bloch band.\\
A numerical study of the eigenvectors of the star graph system leads us to find that the \textit{two states corresponding to the} \textit{eigenvalues outside the Bloch band are localized at the origin}, where the \textit{h} branches cross, while all the others are extended, Bloch-like states. \\
In the following paragraphs, we will analytically calculate the eigenvalues of the states outside of the
 normal Bloch band (see  Paragraph \ref{par:eigen_ls}) and we will show that these states are localized at the origin (see Paragraph \ref{par:eigenv_ls}).
\subsection{Eigenvalues}
\label{par:eigen_ls}
As we said before, we numerically find that the solution of the system (\ref{eq:system_rip}) gives us \textit{two eigenvalues outside of the
 normal Bloch band}, while the others $hN-1$ are enclosed within the Bloch band.
\subsubsection{Eigenvalues outside of the normal Bloch band}
Now, we want to compute analytically the two eigenvalues outside of 
the  normal Bloch band; we are particularly interested in those states because we will demonstrate that they are the only localized states of the system $S_h(N)$.\\
First of all, we want to have a linear system where the unknowns are \textit{a-independent}, in order to have a $(N+1) \times (N+1)$ tridiagonal system which is easier to solve than the system (\ref{eq:system_rip}).
Starting from Eq.(\ref{eq:system_rip}), we multiply and divide the left-hand side of the first equation by $t_av$ and we divide all the other equations by $t_av$; after that, we get
\begin{equation}
\label{eq:system_4_diff_v_t}
 \left\{ \begin{split}
\sum_{a=1}^h \left (t_av\right) ^2 \left ( \dfrac{C_1^a}{t_av}\right ) &= E C_0\\
v\left ( \dfrac{C_2^a}{t_av}\right ) + C_0 &= E \left ( \dfrac{C_1^a}{t_av}\right )  && \hspace*{15pt} \textrm{\textit{a}=1,\ldots,\textit{h}}\\
v \left( \dfrac{C_{n+1}^a}{t_av} \right ) +v \left ( \dfrac{C_{n-1}^a}{t_av}  \right ) &= E \left (\dfrac{C_n^a}{t_av} \right )  && \hspace*{15pt} \textrm{\textit{a}=1,\ldots,\textit{h}; \textit{n}=2,\ldots, \textit{N}-1}\\
v \left ( \dfrac{C_{N-1}^a}{t_av}\right ) &= E\left ( \dfrac{C_N^a}{t_av} \right ) && \hspace*{15pt} \textrm{\textit{a}=1,\ldots,\textit{h}}
\end{split} \right.
\end{equation}
It is easy to show\footnote{
We first prove this statement for the particular case of $N=3$. In this case, we have
\begin{equation}
\label{eq:system_rip4}
 \left\{ \begin{array}{rl}
 \displaystyle{\sum_{a=1}^h t_a v C_1^a = E C_0} \\
  \displaystyle{v C_2^a + t_a v  C_0  = E C_1^a} \\
 \displaystyle{v \left( C_{3}^a + C_{1}^a \right ) = E C_2^a} \\
 \displaystyle{v C_{2}^a = E C_3^a} 
\end{array} \right.
\end{equation}
Replacing backward from the last equation, the second equation becomes $\frac{C_1^a}{t_a v} \left ( E - \frac{v^2 E}{E^2 - v^2} \right ) = C_0$. Considering the fact that $E$, $v$ and $C_0$ are \textit{a}-independent, we find that also $\frac{C_1^a}{t_a v}$ is a-independent. In the general case, we find $\frac{C_1^a}{t_a v} f_N(E,v) = C_0$ where $f_N$ is a function dependent only on \textit{E} and \textit{v} for the case of \textit{N} sites; again, the quantity $\frac{C_1^a}{t_a v}$ is \textit{a}-independent. Please note that this proof is not valid if $E=0$ or $C_0=0$.
}
 that all $\frac{C_i^a}{t_av}$ for $i=1,\ldots, N$ are \textit{a-independent} if $E$ and $C_0$ are different form zero. We are not interested in these cases because they do not lead us to states outside of the normal Bloch band.\\
So, defining
\begin{align}
\hat{C}_i \doteq \dfrac{C_i^a}{t_av} && i=1,\ldots, N
\end{align}
and
\begin{align}
\sum_{a=1}^h \left (t_a \right) ^2 \doteq T_{tot}
\end{align}
the system (\ref{eq:system_4_diff_v_t}) becomes
\begin{equation}
\label{eq:system_4_diff_v_t_semp}
 \left\{ \begin{split}
T_{tot} v^2 \hat{C}_1 - E C_0 &= 0\\
v \hat{C}_2 + C_0 - E \hat{C}_1 &= 0\\
v \hat{C}_{n+1}+v \hat{C}_{n-1} - E \hat{C}_{n}&= 0 && \hspace*{15pt} \textrm{\textit{n}=2,\ldots, \textit{N}-1}\\
v \hat{C}_{N-1} - E\hat{C}_N &= 0
\end{split} \right.
\end{equation}
This is the tridiagonal system (where the unknowns are \textit{a}-independent) that we wanted to achieve and now we want to solve.
This is an homogeneous system of $N+1$ linear equations.\\
Assigning a null value  to each  variable we get
the null solution
which is obviously not normalizable.\\
Furthermore, we know from linear algebra that a square homogeneous linear 
system has solutions (other than the trivial solution) if and only if
  the determinant of the matrix of the coefficients of the system is zero,
namely,
\begin{equation}
\label{eq:detA}
\det \mathcal{M} \doteq \det \left(
\begin{array}{ccccccc}
-E & T_{tot}v^2 & 0 &0 &0 &\cdots & 0 \\
1 & -E & v &0 &0 &\cdots & 0 \\
0 & v & -E & v &0 &\cdots & 0 \\
\vdots & \ddots & \ddots & \ddots & \ddots & \ddots & \vdots\\
0 & \cdots & 0 & v &-E &v & 0 \\
0 & \cdots & 0 & 0 &v &-E & v \\
0 & \cdots & 0 & 0 &0 &v & -E \\
\end{array}
\right) = 0
\end{equation}
This is a $(N+1)\times(N+1)$ determinant that describes the $S_h (N)$ system and allows us to calculate the energy levels of the aforementioned system.
The problem, now, is to find the determinant of the matrix $\mathcal{M}$; this matrix is tridiagonal, then we can calculate quite easily the determinant. \\
Applying Eq.(\ref{eq:det_trid})\footnote{See Appendix B.} to Eq.(\ref{eq:detA}), we get
\begin{equation}
\label{eq:det_A_11}
\begin{split}
\det \mathcal{M} = \left[\left(
\begin{array}{cc}
-E & -v^2 \\
1 & 0 \\
\end{array}
\right)^{N-1}
\left(
\begin{array}{cc}
-E & -T_{tot}v^2 \\
1 & 0 \\
\end{array}
\right)
\left(
\begin{array}{cc}
-E & 0 \\
1 & 0 \\
\end{array}
\right)
\right]_{11}
\end{split}
\end{equation}
Now, we have to calculate a power of a square matrix:
\begin{equation}
\mathcal{B}^{N-1} =
 \left(
 \begin{array}{cc}
-E & -v^2 \\
1 & 0 \\
\end{array}
\right)^{N-1} .
\end{equation}
A powerful method is to use the eigenvalue decomposition of the matrix $\mathcal{B}$.\\
It is straightforward to find that the eigenvectors of $\mathcal{B}$ are
\begin{align}
v^{(-)} =  
\left( \begin{array}{c}
E^{(-)}\\
1 \\
\end{array}
\right) &&
v^{(+)}=  
\left( \begin{array}{c}
E^{(+)}\\
1 \\
\end{array}
\right)
\end{align}
where
\begin{equation}
E^{(\pm)} = \dfrac{-E \pm \sqrt{E^2-4v^2}}{2}
\end{equation}
Now, let $P$ be the matrix with these eigenvectors as its columns:
\begin{equation}
\mathcal{P} = 
 \left(
 \begin{array}{cc}
E^{(-)} & E^{(+)} \\
1 & 1 \\
\end{array}
\right)
\end{equation}
We can also compute the inverse\footnote{
The inversion of $2\times2$ matrices can be done as follows:
\begin{equation}
\mathcal{A}^{-1} = 
 \left(
 \begin{array}{cc}
a & b \\
c & d \\
\end{array}\right)^{-1}
= \dfrac{1}{ad-bc}
 \left(
 \begin{array}{cc}
d & -b \\
-c & a \\
\end{array}\right)
\end{equation}
}
 $\mathcal{P}^{-1}$ of the matrix $\mathcal{P}$
\begin{equation}
\mathcal{P}^{-1} =
\dfrac{1}{E^{(-)}-E^{(+)}} 
 \left(
 \begin{array}{cc}
1 & -E^{(+)} \\
-1 & E^{(-)} \\
\end{array}
\right)
\end{equation}
Thus, the matrix $\mathcal{B}$ can be written as follows:
\begin{equation}
\mathcal{B} = \mathcal{P} \mathcal{D} \mathcal{P}^{-1} 
\end{equation}
where $\mathcal{D}$ is a diagonal matrix with the eigenvalues of $\mathcal{B}$ on the main diagonal:
\begin{equation}
\mathcal{D} =
\left(
 \begin{array}{cc}
E^{(-)} & 0 \\
0 & E^{(+)} \\
\end{array}
\right)
\end{equation}
Computing the power of the matrix $\mathcal{B}$:
\begin{equation}
\begin{split}
\mathcal{B}^{N-1} &= \left(\mathcal{P}\mathcal{D}\mathcal{P}^{-1}\right)^{N-1} = 
\left( \mathcal{P}\mathcal{D}\mathcal{P}^{-1}\right)\cdot \left( \mathcal{P}\mathcal{D}\mathcal{P}^{-1}\right)
\ldots \left( \mathcal{P}\mathcal{D}\mathcal{P}^{-1}\right)\\
&= \left( \mathcal{P}\mathcal{D} \right) \left( \mathcal{P}^{-1} \mathcal{P} \right) \mathcal{D} \left( \mathcal{P}^{-1} \mathcal{P} \right)\ldots \left( \mathcal{P}^{-1} \mathcal{P} \right) \mathcal{D} \mathcal{P}^{-1} \\
&= \mathcal{P} \mathcal{D}^{N-1} \mathcal{P}^{-1}
\end{split}
\end{equation}
Please note that the power of the matrix $\mathcal{D}$ is easy to calculate since it  involves the powers of a diagonal matrix only.\\

We are ready to replace these relations in Eq.(\ref{eq:det_A_11}); after some calculation we get
\begin{equation}
\label{eq:polyn_ttot}
\begin{split}
&\det \mathcal{M} = 0 \\
&\Rightarrow \dfrac{\left[ \left( E^{(-)} \right)^{N} \left(
E^2 + EE^{(+)} - T_{tot}v^2 \right) - \left( E^{(+)} \right)^{N} \left(
E^2 + EE^{(-)} - T_{tot}v^2 \right) \right]}{E^{(-)} - E^{(+)}} = 0
\end{split}
\end{equation}
This is a polynomial equation that allows us, at least in principle, to find the eigenvalues outside of the normal Bloch band for any $N$\footnote{We remind the reader that $N$ is the number of sites for each chain, then the solution for $N$ is the solution of the system that we named $S_h(N)$ system.}; in fact, we just need to find the largest and the smallest root of this polynomial.\\
From the computational point of view, however, is very difficult to provide an analytical solution with increasing degree of the polynomial to be solved; in fact, finding roots of certain functions, especially polynomials, frequently requires the use of standard numerical methods (such as Newton's method, Bisection method, Secant method, etc.).\\
Here, we report the analytical solutions for some small \textit{N} and for large \textit{N}.

\subsubsection*{Small N}
Where possible, we found analytically the roots of the polynomial of 
Equation (\ref{eq:polyn_ttot}); these are the results.\\
For $N=1$: 
\begin{equation}
E = \pm \left (\sqrt{T_{tot}}\right )v
\end{equation}
For $N=2$:
\begin{equation}
E = \pm \left ( \sqrt{1+T_{tot}}\right )v 
\end{equation}
For $N=3$:
\begin{equation}
E = \pm \left ( \dfrac{\sqrt{2+T_{tot}+\sqrt{4+T_{tot}^2}}}{\sqrt{2}}\right )v 
\end{equation}
For $N=4$:
\begin{equation}
E = \pm \left ( \sqrt{\dfrac{3+T_{tot}+\sqrt{T_{tot}^2-2T_{tot}+5}}{2}}\right )v 
\end{equation}
After these values of N, it is very difficult to find an analytical solution because of the complexity of the polynomial; however, is always possible to calculate a numerical solution with the above-mentioned numerical methods.
\subsubsection*{Large N}
\begin{figure}[h!]
\vspace{0cm}
\center{\includegraphics[width=10cm,angle=-90]{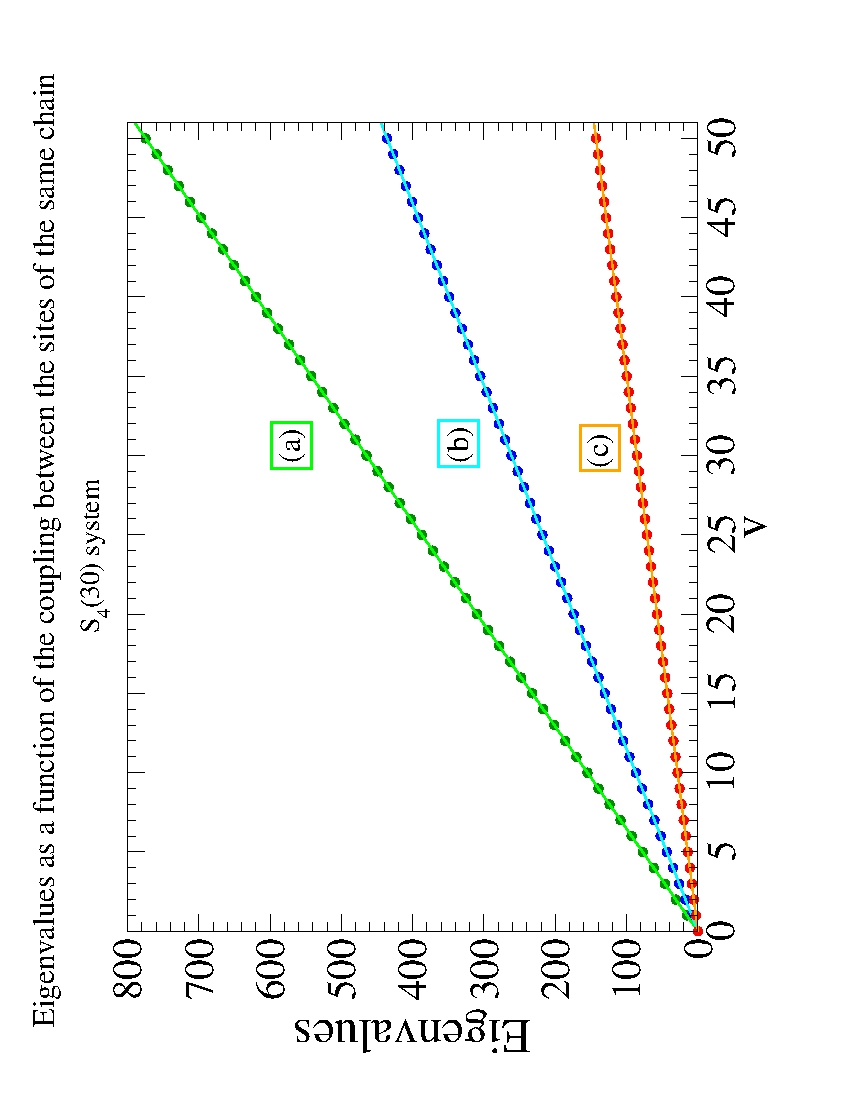}}
\caption{{\small
We plot one of the two eigenvalues outside of the
 normal Bloch band of the $S_4(30)$ system as a function of $v$ with different $t_a$. In particular:\newline
Data set (a): $t_1$ = 15, $t_2$ = 3, $t_3$ = 2, $t_4$ =1 $\Rightarrow T_{tot} = 239$\newline
Data set (b): $t_1$ = 7,\hspace*{3pt} $t_2$ = 5, $t_3$ = 4, $t_4$ = 2 $\Rightarrow T_{tot} = 93$\newline
Data set (c): $t_1$ = 2,\hspace*{3pt} $t_2$ = 1, $t_3$ = 1, $t_4$ = 1 $\Rightarrow T_{tot} = 7$\newline
The circles refer to numerical data, while the lines represent theoretical data obtained by using Eq.(\ref{eq:eigenvalues_4_diff_t}) for different values of $T_{tot}$. The quantities plotted are dimensionless. 
}}
\label{fig:diff_v_t7-93-239}
\end{figure}

\begin{figure}[h!]
\vspace{0cm}
\center{\includegraphics[width=10cm,angle=-90]{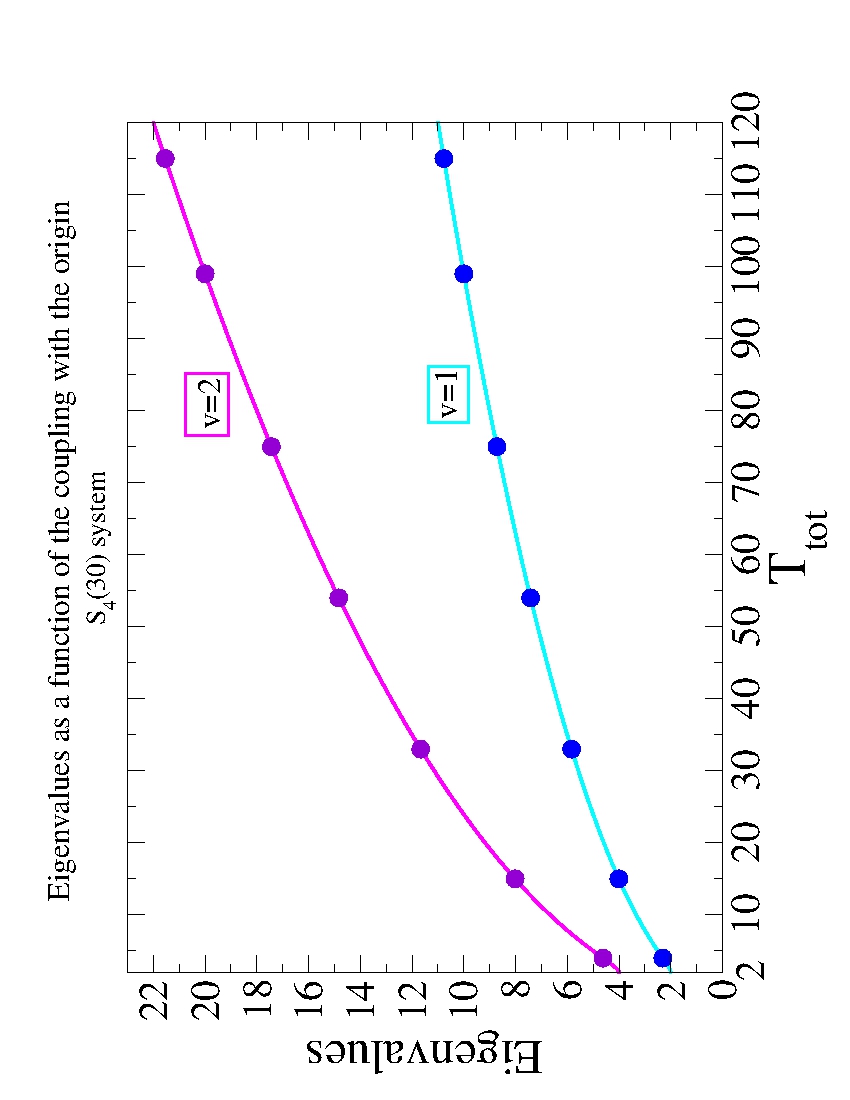}}
\caption{{\small
We plot one of the two eigenvalues outside of the
 normal Bloch band of the $S_4(30)$ system as a function of $T_{tot}$ for two systems with different $v$ ($v=1$ and $v=2$). The circles refer to numerical data, while the lines represent theoretical data obtained by using Eq.(\ref{eq:eigenvalues_4_diff_t}) with $v=1$ and $v=2$. The quantities plotted are dimensionless. 
}}
\label{fig:eig_vfix_t}
\end{figure}

Here, we find the analytical solution for large \textit{N}; 
this  limit 
is very interesting because it appears very frequently 
in experimental situations.\\
In this limit, from Eq.~ (\ref{eq:polyn_ttot}), after some calculations and remembering the definitions of $E^{(-)}$  and $E^{(+)}$, we find that the 
equations to be solved are the following:
\begin{align}
E^2 + E\sqrt{E^2-4v^2} - 2T_{tot}v^2 = 0\\
E^2 - E\sqrt{E^2-4v^2} - 2T_{tot}v^2 = 0
\end{align}
or, 
\begin{align}
\label{eq:eigenvalues_4_diff_t}
\boxed{
E = \pm \left ( \dfrac{T_{tot}}{\sqrt{T_{tot}-1}} \right )v} && \mbox{where} && T_{tot} = \sum_{a=1}^h \left (t_a \right) ^2
\end{align}
\\
This relation gives the asymptotic eigenvalues for the system $S_h$ 
as a function of $T_{tot}$ and $v$; we remind the reader that $v$ is the coupling between the sites of the same chain. 
In the limit of large $T_{tot}$, for the two states outside Bloch band we have $E \approx \pm \sqrt{T_{tot}}v$, while all the others are \textit{always} within the Bloch band [-2v,2v]. This is very important because \textit{one can control} (by using $T_{tot}$ and thus also $h$) \textit{the energy distance} between the states outside Bloch band and the others confined in this band.\\
In Fig.(\ref{fig:diff_v_t7-93-239}) and Fig.(\ref{fig:eig_vfix_t}) we show the results obtained with the $S_4(30)$ system; in these graphs, we compare numerical results (circles) with theoretical predictions (lines) given by the 
Equation (\ref{eq:eigenvalues_4_diff_t}).
We plot only the positive eigenvalues outside of the normal Bloch band; the negative ones are symmetric with respect to the horizontal axis.\\
In Fig.(\ref{fig:diff_v_t7-93-239}), we plot the eigenvalues as a function of $v$ for different $T_{tot}$; we obtain three different straight lines with slope $\frac{T_{tot}}{\sqrt{T_{tot}-1}}$, according to Eq.(\ref{eq:eigenvalues_4_diff_t}).
In Fig.(\ref{fig:eig_vfix_t}), we plot the eigenvalues as a 
function of $T_{tot}$ for two different values of $v$.
In both figures, we can see the perfect agreement between the theoretical and numerical data; this is an evidence of the goodness of our analytical model.
Please also note that the theoretical data used for comparison are those of the asymptotic regime; this means that for the $S_4(30)$ system we can consider the asymptotic regime already reached.\\
We conclude stressing the fact that all the eigenvalues that we found here (for both small and large N) are a \textit{distinctive feature of our crossing configuration system} because they are outside of the
 normal Bloch band; in the linear 1d-chains considered so far in the literature, in fact, all the eigenvalues were confined within the Bloch band.

\subsection{Eigenvectors}
\label{par:eigenv_ls}
A numerical study of the eigenvectors of the star graph system tells us that there are two types of eigenvectors: the two eigenvectors corresponding to the two eigenvalues outside of the
 normal Bloch band are localized at the origin, where the branches cross, while the other are extended, Bloch-like states.\\
In this section, we will prove analitically that the two states outside of 
the  normal Bloch band are localized and we will provide an analytical expression of their localization lengths in the asymptotic case.

\subsubsection{Eigenvectors of the states outside of the normal Bloch band}
\begin{figure}[h!]
\vspace{0cm}
\center{\includegraphics[width=10cm,angle=-90]{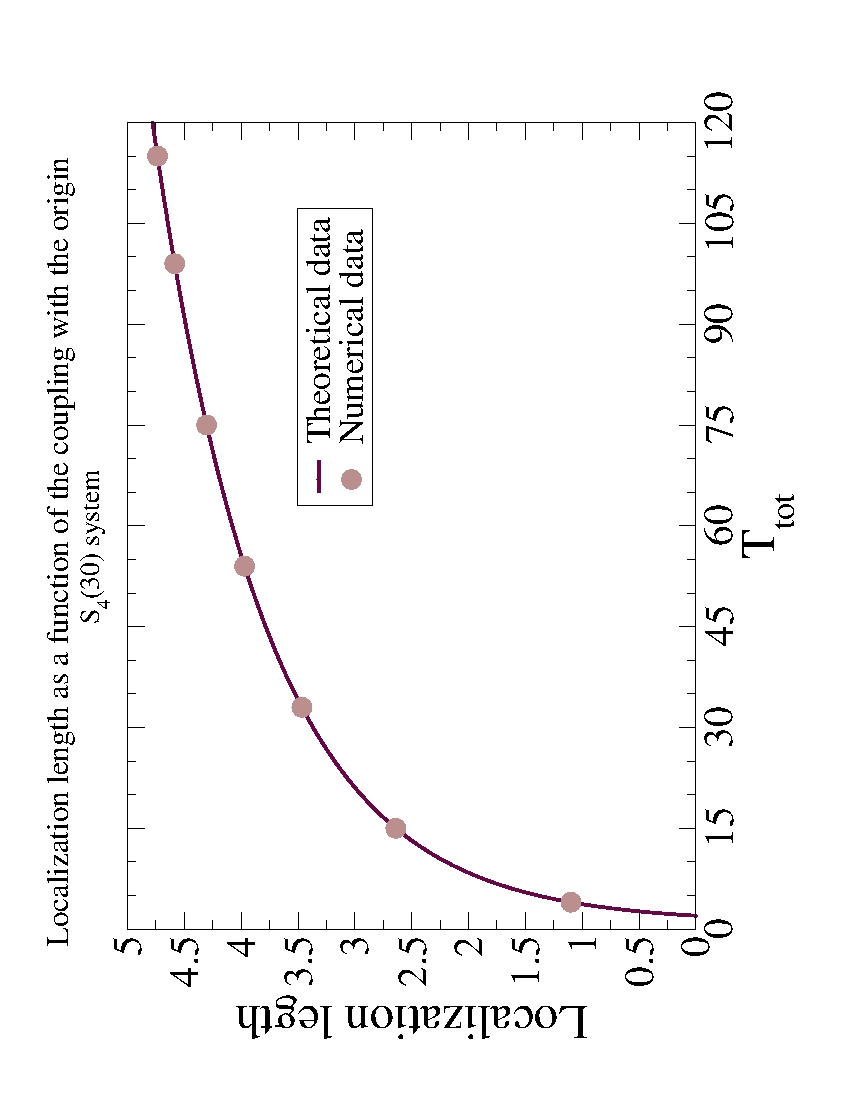}}
\caption{{\small
We show the localization length of the $S_4(30)$ system as a function of $T_{tot}$. The circles refer to numerical data, while the line represents theoretical data obtained by using Equation (\ref{eq:decay_t_4}). The quantities plotted are dimensionless.
}}
\label{fig:loc_h4_t}
\end{figure}

Now, we want to calculate the eigenvectors outside of the
 normal Bloch band. 
The linear system that we have to solve is Eq.(\ref{eq:system_4_diff_v_t_semp}) which is given here for convenience:
\begin{equation}
\label{eq:system_semp}
 \left\{ \begin{split}
T_{tot} v^2 \hat{C}_1  &= E C_0\\
v \hat{C}_2 + C_0 &= E \hat{C}_1\\
v \hat{C}_{n+1}+v \hat{C}_{n-1} &= E \hat{C}_{n} && \hspace*{15pt} \textrm{n=2,\ldots, N-1}\\
v \hat{C}_{N-1} &= E\hat{C}_N 
\end{split} \right.
\end{equation}
where
\begin{align}
\hat{C}_i \doteq \dfrac{C_i^a}{t_av} && i=1,\ldots, N
\end{align}
and\footnote{Here, we are considering the asymptotic regime and then we use the asymptotic expression for the eigenvalues; however, if one wants to consider a case with small \textit{N}, one can get the result for that system simply using the appropriate eigenvalue calculated from Eq.(\ref{eq:polyn_ttot}) instead of Eq.(\ref{eq:eigenvalue_asymp}).}
\begin{equation}
\label{eq:eigenvalue_asymp}
E = \pm \left ( \dfrac{T_{tot}}{\sqrt{T_{tot}-1}} \right )v
\end{equation}
For convenience, which will be clear later, we make the following transformation:
\begin{align}
\label{eq:transf}
\hat{C}_i \longrightarrow \tilde{C}_{N-i} && \mbox{with } i=0,\ldots, N
\end{align}
Thus, the system (\ref{eq:system_semp}) becomes
\begin{equation}
\label{eq:system_eigenvector_T}
 \left\{ \begin{split}
v\tilde{C}_1 &= E\tilde{C}_0 \\
v\tilde{C}_{0} + v\tilde{C}_2 &= E\tilde{C}_1\\
v\tilde{C}_{1} + v\tilde{C}_3 &= E\tilde{C}_2\\
\vdots \hspace*{30pt} \vdots &\hspace*{30pt} \vdots \hspace*{8pt} \\
v\tilde{C}_{N-3} + v\tilde{C}_{N-1} &= E\tilde{C}_{N-2}\\
v\tilde{C}_{N-2} + \tilde{C}_{N} &= E\tilde{C}_{N-1}\\
T_{tot}v^2\tilde{C}_{N-1} &= E\tilde{C}_N 
\end{split} \right.
\end{equation}
From the second row of the above equation, we obtain that
\begin{equation}
\tilde{C}_2 = \dfrac{E}{v}\tilde{C}_1 - \tilde{C}_0
\end{equation}
and from the third row 
\begin{equation}
\tilde{C}_3 = \dfrac{E}{v}\tilde{C}_2 - \tilde{C}_1.
\end{equation}
It is clear that the process can be iterated, and this leads us to recognize the following \textit{recurrence relation}\footnote{For a brief overview of recurrence relations see Appendix A. In addition, please note that the recurrence relation (\ref{eq:rec_rel_1}) is valid until the third from last 
equation in (\ref{eq:system_eigenvector_T}).}:
\begin{align}
\label{eq:rec_rel_1}
\tilde{C}_k = \dfrac{E}{v} \tilde{C}_{k-1} - \tilde{C}_{k-2} && \mbox{with } k=2,\ldots, N-1
\end{align}
This is a linear homogeneous recurrence relation with constant coefficients of order 2. \\
First of all, we determine the characteristic polynomial $p(t)$ of this recurrence relation:
\begin{equation}
p(t) = t^2 - \dfrac{E}{v}t + 1 
\end{equation}
The roots of this polynomial are two, each of multiplicity 1:
\begin{equation}
\label{eq:e_pm}
\mathcal{E}_{\pm} = \dfrac{E \pm \sqrt{E^2 -4v^2} }{2v}
\end{equation}
Therefore, we can write the general solution to our recurrence relation:
\begin{align}
\label{eq:gen_rr}
\tilde{C}_k = A \left ( \mathcal{E}_{-} \right )^k + B \left ( \mathcal{E}_{+} \right )^k && \mbox{with } k=0, \ldots, N-1
\end{align}
which gives us a \textit{closed-form expression} for $\tilde{C}_k$.
This is the most general solution; the two constants $A$ and $B$ can be chosen based on two given initial conditions $\tilde{C}_0$ and $\tilde{C}_1$ to produce a specific solution.\\
We choose the following initial conditions for our recurrence relation:
\begin{align}
\label{eq:initial_cond}
\tilde{C}_0 = 1 &&
\tilde{C}_1 = \dfrac{E}{v}
\end{align}
Obviously, these conditions do not return normalized eigenvectors, but we
are  interested in  the ratios between the eigenvectors and those with conditions (\ref{eq:initial_cond}) are preserved.\\
Thus, the system that we have to solve to find $A$ and $B$ is
\begin{equation}
\label{eq:initial_cond_system}
 \left\{ \begin{split}
\tilde{C}_0 &= A + B = 1 \\
\tilde{C}_1 &= A \left ( \mathcal{E}_{-} \right ) + B \left ( \mathcal{E}_{+} \right ) = \dfrac{E}{v}\\
\end{split} \right.
\end{equation}
that gives,
\begin{align}
\label{eq:i_cond_final}
A = \dfrac{1}{\mathcal{E}_{-} - \mathcal{E}_{+}} \left ( \dfrac{E}{v} - \mathcal{E}_{+} \right ) &&
B = \dfrac{1}{\mathcal{E}_{-} - \mathcal{E}_{+}} \left ( -\dfrac{E}{v} + \mathcal{E}_{-} \right ).
\end{align}
Putting Equation (\ref{eq:i_cond_final}) into (\ref{eq:gen_rr}) and after some algebraic manipulation, we get
\begin{align}
\tilde{C}_k \propto \dfrac{v (-1)^k}{\sqrt{E^2-4v^2}}\left [ \left ( \mathcal{E}_{-}\right )^{k+1} - \left ( \mathcal{E}_{+}\right )^{k+1} \right ] && k=0,\ldots, N-1
\end{align}
Recalling the transformation (\ref{eq:transf}), we have
\begin{align}
\hat{C}_{N-k} \propto \dfrac{v (-1)^{k}}{\sqrt{E^2-4v^2}}\left [ \left ( \mathcal{E}_{-}\right )^{k+1} - \left ( \mathcal{E}_{+}\right )^{k+1} \right ] && k=0,\ldots, N-1
\end{align}
or equivalently
\begin{align}
\hat{C}_{i} \propto \dfrac{v (-1)^{N-i}}{\sqrt{E^2-4v^2}}\left [ \left ( \mathcal{E}_{-}\right )^{N-i+1} - \left ( \mathcal{E}_{+}\right )^{N-i+1} \right ] && i=1,\ldots, N
\end{align}
Now, we can compute the square of the absolute value of the ratio between the eigenvectors
\begin{align}
\label{eq:ratio_gen}
\left \vert \dfrac{\hat{C}_{i+1}}{\hat{C}_i} \right \vert^2 = \left \vert \dfrac{C^a_{i+1}}{C^a_i} \right \vert^2 = \left \vert \dfrac{\left ( \mathcal{E}_{-}\right )^{N-i} - \left ( \mathcal{E}_{+}\right )^{N-i}}{\left ( \mathcal{E}_{-}\right )^{N-i+1} - \left ( \mathcal{E}_{+}\right )^{N-i+1}} \right |^2 && i=1,\ldots, N-1
\end{align}
Up to now, we have not yet used the fact that we are considering \textit{h} chains that are coupled by the following equation:
\begin{equation}
\label{eq:constrain}
T_{tot}v^2\hat{C}_1 = EC_0
\end{equation}
where 
\begin{equation}
\label{eq:eigen_4}
E = \pm \left ( \dfrac{T_{tot}}{\sqrt{T_{tot}-1}} \right )v
\end{equation}
One can check, by using Equation (\ref{eq:ratio_gen}), taking into account (\ref{eq:eigen_4}) and the second row of (\ref{eq:system_4_diff_v_t_semp}), that the Equation (\ref{eq:constrain}) is automatically verified; therefore, we can say that the information on the coupling between the chains is contained only in the eigenvalues of the system.
In addition, we note that if we substitute the Bloch eigenvalues into Eq.(\ref{eq:ratio_gen}), we get  the proper ratios between the Bloch eigenvectors; this is due to the fact that Eq.(\ref{eq:ratio_gen}) is valid also for chains not coupled to the origin.\\
Coming back to Equation (\ref{eq:ratio_gen}), we want to compute explicitly $\mathcal{E}_\pm$ for each eigenvalue; so, we must distinguish two cases, one for each eigenvalue.\\
Considering $E= \left ( \dfrac{T_{tot}}{\sqrt{T_{tot}-1}} \right )v$ and using the definitions of $\mathcal{E}_\pm$, Eq.(\ref{eq:e_pm}), we get
\begin{align}
\label{eq:eps_p}
\mathcal{E}_+ = \sqrt{T_{tot}-1} && \mathcal{E}_- = \dfrac{1}{\sqrt{T_{tot}-1}} && \mbox{with }T_{tot}\geq 2
\end{align}
On the other hand, with $E= -\left ( \dfrac{T_{tot}}{\sqrt{T_{tot}-1}} \right )v$, we find
\begin{align}
\label{eq:eps_m}
\mathcal{E}_+ = \dfrac{1}{\sqrt{T_{tot}-1}} && \mathcal{E}_- = \sqrt{T_{tot}-1}  && \mbox{with }T_{tot}\geq 2
\end{align}
Hence, substituting Eqs.(\ref{eq:eps_p}) and (\ref{eq:eps_m}) into Eq.(\ref{eq:ratio_gen}), we obtain, regardless of the considered eigenvalue, the following relation:
\begin{align}
\label{eq:ratio_gen2}
\left \vert \dfrac{C^a_{i+1}}{C^a_i} \right \vert^2 =
\left ( \dfrac{1}{T_{tot}-1} \right ) \mathcal{B}\left (T_{tot},N,i \right )  
&& i=1,\ldots, N-1
\end{align}
where we have defined
\begin{equation}
\label{eq:boundary}
\mathcal{B}\left (T_{tot},N,i \right ) \doteq \left \vert \dfrac{1 - \left (T_{tot}-1\right )^{-\left ( N-i \right )}}{1-  \left (T_{tot}-1\right )^{-\left ( N+1-i \right )}} \right \vert^2
\end{equation}
This means that the ratio between two subsequent eigenvectors depends on \textit{i}; however, this dependence is confined to the term $\mathcal{B}$ which is a boundary effect different from 1 only for $i$ very close to $N$.
\begin{figure}[t]
\vspace{0cm}
\center{\includegraphics[width=10cm,angle=-90]{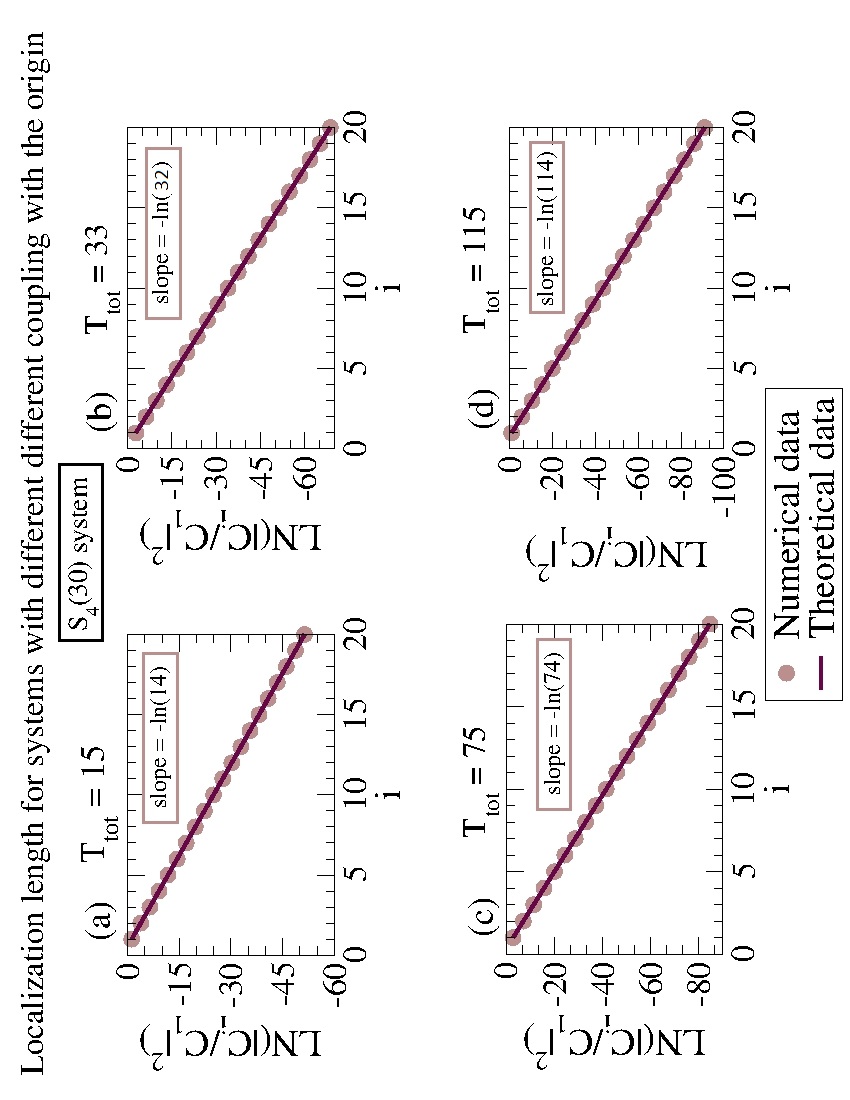}}
\caption{{\small
In the simulation, we used a finite system consisting of 121 sites - $S_4(30)$ system. Theoretical data were calcolated using Equation (\ref{eq:decay_t_4}). One can see the perfect correspondence between numerical results and theoretical predictions. We plotted only the first 20 value because the others were lower that the machine zero. The quantities plotted are dimensionless.
}}
\label{fig:loc_30_diff_t}
\end{figure}
Neglecting the boundary effect\footnote{However, these effects will be crucial in determining the lifetime of localized states, see Chapter \ref{chap:4}, Eq.(\ref{eq:tau_loc5}).}, namely setting $\mathcal{B} \equiv 1$, from Eq.(\ref{eq:ratio_gen2}) we get:
\begin{align}
\label{eq:states_4_t}
\boxed{\left \vert \dfrac{C_i^a}{C_1^a} \right \vert^2 = e^{-\xi\left (i -1 \right )}} && i=1,\ldots, N
\end{align}
where \textit{i} is the distance from the origin, where the \textit{h} branches cross and the \textit{localization length} $\xi$ is equal to
\begin{align}
\label{eq:l_states_4_t}
\boxed{\xi= \ln(T_{tot}-1)} && T_{tot}\geq 2
\end{align}
Therefore, the result indicates that \textit{the two states outside of the
 normal Bloch band are localized with localization length} $\xi$, see Fig.(\ref{fig:loc_30_diff_t}) and Fig.(\ref{fig:loc_h4_t}).
This result is a \textit{distinctive feature of our crossing configuration system}.
From Eq.(\ref{eq:l_states_4_t}), it is clear that the higher is $T_{tot}$, the higher is $\xi$, which means that the two localized states are more localized.
Please also note that this result is \textit{v-independent}.\\
It is also interesting compute the ratio between $C_0$ and $C_1^a$:
\begin{align}
\label{eq:3ratio_t1_t0}
\left \vert \dfrac{C_1^a}{C_0} \right \vert^2 = \dfrac{t_a^2}{T_{tot}-1} && T_{tot} \geq 2
\end{align}
This relation is, in general, different from one chain to another because it depends on the coupling $t_a$ with the origin of the chain considered; the higher is the coupling with the origin, the higher is this ratio, which means that the probability of being in that chain is higher. Thus, since the localization length is independent of the chain, the $C_i^a$ give us \textit{h} different sequences, where, however, the ratio of two successive terms is the same in all \textit{h} sequences (i.e. chains). 
Namely,
\begin{align}
\left \vert \dfrac{C_i^a}{C_0} \right \vert^2 =
\left \vert \dfrac{C_i^a}{C_1^a} \cdot \dfrac{C_1^a}{C_0} \right \vert^2 =
\left \vert \dfrac{C_1^a}{C_0} \right \vert^2 \cdot \left \vert \dfrac{C_i^a}{C_1^a} \right \vert^2    
\end{align}
that gives us
\begin{align}
\label{eq:decay_t_4}
\boxed{\left \vert \dfrac{C_i^a}{C_0} \right \vert^2 = \mathcal{A}\left (t_a,T_{tot}\right )  e^{-\xi \left (i-1 \right )}} && \mbox{with } i=1, \ldots, N; && T_{tot} \geq 2
\end{align}
where
\begin{align}
\label{eq:decay_r_t_4}
\boxed{\xi= \ln(T_{tot}-1)} && T_{tot}\geq 2
\end{align}
and
\begin{align}
\label{eq:decay_a}
\boxed{\mathcal{A}\left (t_a,T_{tot}\right ) = \dfrac{t_a^2}{T_{tot}-1}}  && T_{tot}\geq 2
\end{align}

\section{An explanatory example}
In this Section, we  present an example, in order to illustrate in a 
particular case what has been done in the previously. 
Let us consider $h$ chains coupled with the origin with the same hopping amplitudes $v$.\\
Namely,
\begin{align}
\label{eq:tas}
t_a \doteq 1 && \mbox{with } a=1,\ldots,h
\end{align}
For example, a physical system that can be represented by this example is the crossing of $h$ \textit{identical} quantum wires.
From Eq.(\ref{eq:tas}), we have
\begin{align}
\label{eq:t_tot_s}
T_{tot} = \sum_{a=1}^h \left (t_a \right) ^2 = h
\end{align}
In this paragraph, we will find the eigenvalues and the eigenvectors in the asymptotic regime.
\begin{figure}[h!]
\vspace{0cm}
\center{\includegraphics[width=10cm,angle=-90]{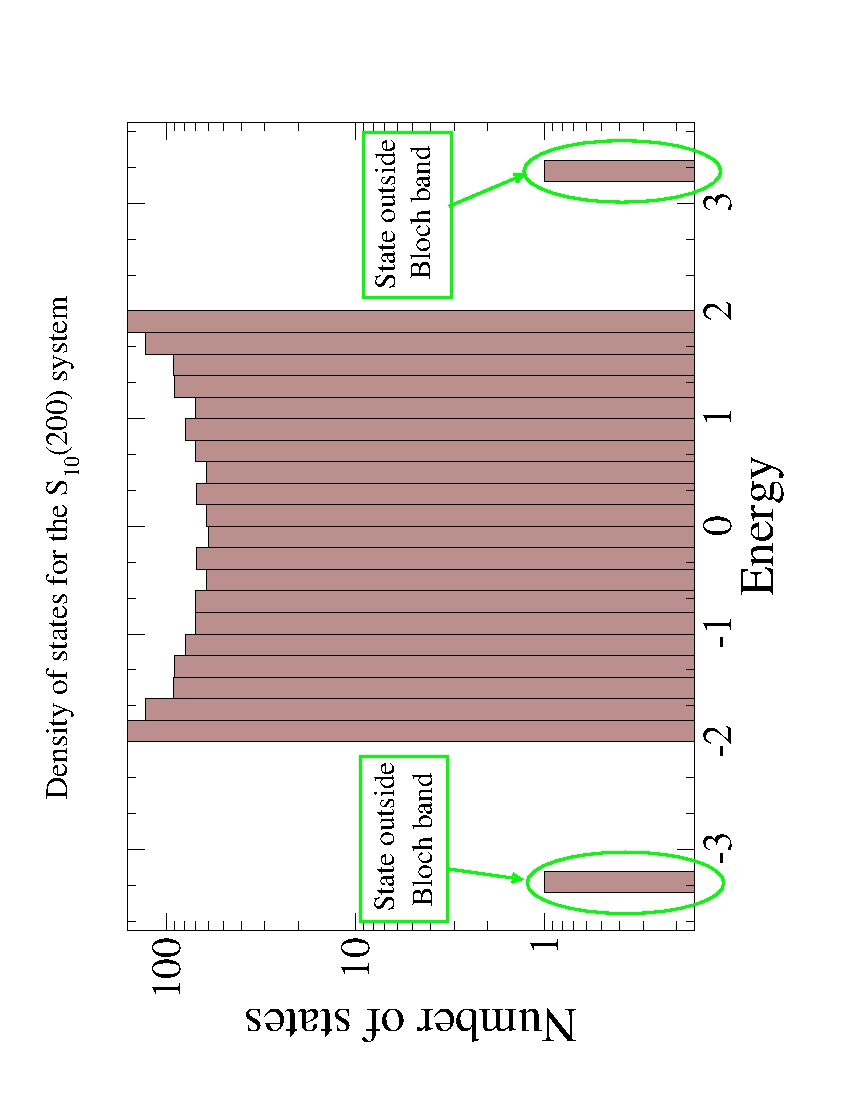}}
\caption{{\small
We plot the number of states as a function of the energy of the system for the $S_{10}(200)$ system with $t_a=1$ $\forall a$ and $v=1$. We can see that the states are all within the band of Bloch, [-2,+2] in this case, except two, which are indicated with green circles. The quantities plotted are dimensionless. 
}}
\label{fig:Dos_h_10}
\end{figure}

\subsubsection{Eigenvalues - asymptotic regime}
A numerical solution of the system gives us two eigenvalues outside of the
 normal Bloch band, which is $[-2v,2v]$, while the others are within the Bloch band; the density of states in a particular case is shown in Fig.(\ref{fig:Dos_h_10}).
Now, we will find an analitical expression for these two eiegnvalues, starting from what has been done in the previous section.\\
Substituting Eq.(\ref{eq:t_tot_s}) into Eq.(\ref{eq:eigenvalues_4_diff_t}) we get
\begin{align}
\label{eq:eigen_h}
\boxed{E = \pm \dfrac{h v}{\sqrt{h-1}}} && \mbox{with } h \geq 2
\end{align}
The value $h=1$ is meaningless; in fact, this would mean that there is a branch to the left of the origin, while on the right there is nothing.\\
So, the first value that $h$ can assume is $h=2$; this means that we have two branches that cross in the origin, so we have a 1-d chain. For this particular case, we know\footnote{Consider a 1-d chain with nearest neighbour interactions composed by a finite number of sites from $n=1,\ldots, N$ with zero boundary conditions at the ends, $n=0$ and $n = N+1$, where the chain terminates. The dispersion relation take the form
\begin{align}
E (q) = \varepsilon_q + 2 v \cos \left (\dfrac{\pi q}{N+1}\right ) && q = 1,2,\ldots, N
\end{align}
In our treatment, we set $\epsilon_q = 0$. Furthermore, we are interested in an infinite chain. It can be easily shown that for such a system the dispersion relation is
\begin{align}
\label{eq:e_cont}
E(\mu) = 2 v \cos \left ( k a \right ) && \mbox{where }-\dfrac{\pi}{a} \leq k \leq \dfrac{\pi}{a}
\end{align}
where $a$ is the lattice spacing and $k$ is a continuous quantum number.
From Eq.(\ref{eq:e_cont}), it is immediate to realize that, in the case of an infinite chain, the largest and smallest eigenvalue are $+2v$ and $-2v$, respectively.
To ensure the convergence to the desired continuum limit of a quantum wire, one has to let the lattice spacing $a$ go to zero while increasing $N$ such that their product (the length of the sample) remains constant.
For more details, see Ref.\cite{suren}.} from the literature that the largest and the smallest eigenvalue of a  1-d chain are respectively $+2v$ and $-2v$; this is the same result that we have from Equation (\ref{eq:eigen_h}) setting $h=2$.
In the limit of large $h$, we have, for the two localized states, $E \approx \pm \sqrt{h} v$, while all the others are \textit{always} confined in the Bloch band [-2v,2v]. Then, one can control, in a straightforward way, 
the energy distance between the two localized states and all the others by using the parameter $h$, i.e. the number of branches.
\subsubsection{Eigenvectors - asymptotic regime}
\begin{figure}[h!]
\vspace{0cm}
\center{\includegraphics[width=10cm,angle=-90]{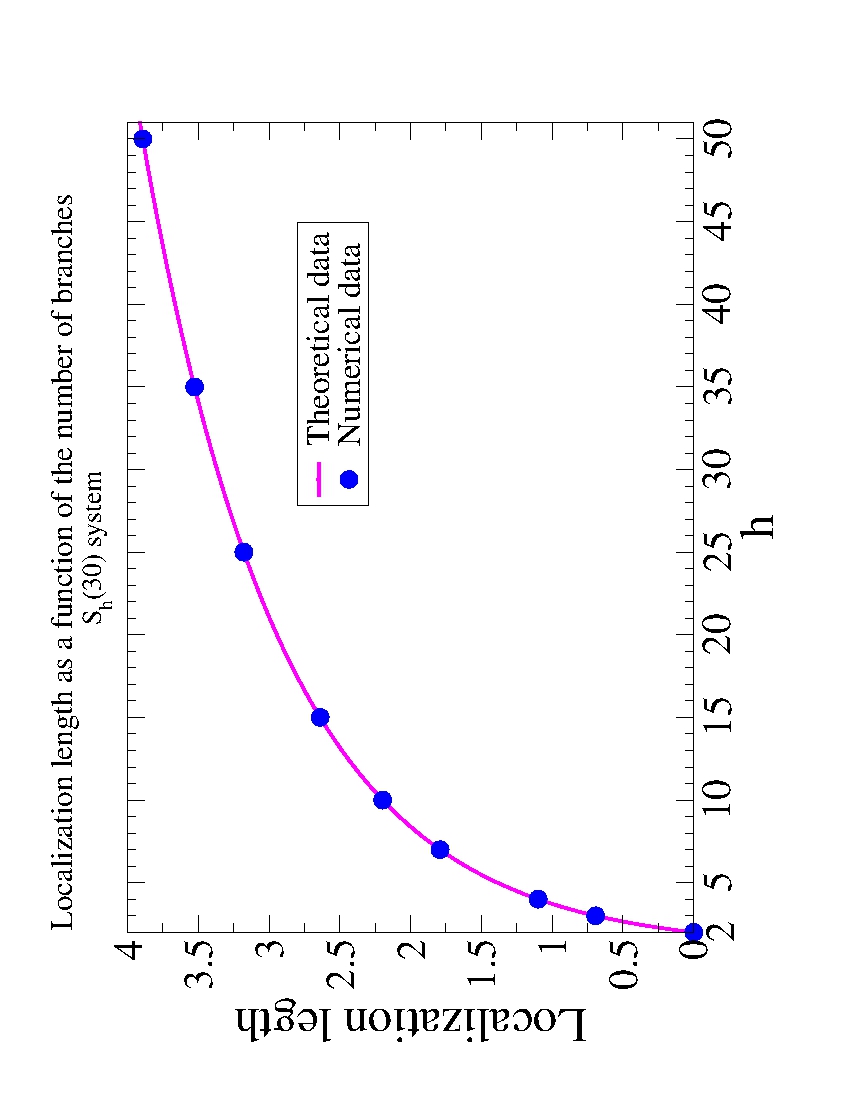}}
\caption{{\small
We plot the localization length of $S_h(30)$ system as a function of $h$; we set $t_a=1$ $\forall a$ and $v=1$. The line represents theoretical predictions obtained using Eq.(\ref{eq:decay_r_t_4n}) while the circles represent numerical data; we can see the perfect agreement between numerical and theoretical data. The quantities plotted are dimensionless.
}}
\label{fig:loc_h_riass}
\end{figure}

A numerical solution of the system leads us to find two classes of eigenvectors: the two eigenvectors corresponding to eigenvalues outside Bloch band are localized in the origin while all the others are extended, Bloch-like states.
Taking into account what have been done in the previous section, we will compute the localization lentghs of the two localized states.
From Eqs.(\ref{eq:decay_t_4}), (\ref{eq:decay_r_t_4}) and (\ref{eq:decay_a}), considering $t_a=1$ $\forall a$, we get:
\begin{align}
\label{eq:decay_t_4n}
\boxed{\left \vert \dfrac{C_i^a}{C_0} \right \vert^2 \doteq \left \vert \dfrac{C_i}{C_0} \right \vert^2 =  e^{-\xi i}} && \mbox{with } i=0, \ldots, N; a=1,\ldots, h 
\end{align}
where
\begin{align}
\label{eq:decay_r_t_4n}
\boxed{\xi= \ln(h-1)} && h\geq 2
\end{align}
and where we have defined $C_i^a \doteq C_i$ $\forall a$ according to the symmetry of the problem.
Therefore, we have $h$ identical sequences, \textit{a}-independent.\\
In figure (\ref{fig:loc_h_riass}), we plot the localization length of $S_h(30)$ systems as a function of the number of branches; we set $t_a=1$ $\forall a$ and $v=1$. The line represents theoretical data obtained using Eq.(\ref{eq:decay_r_t_4n}) while the circles stand for numerical data; we can see the perfect agreement between the two series of data. Please note that theoretical data refer to the asymptotic regime; this implies that the system in question has already been 
achieved, with excellent approximation, this regime.
\newline
\section{An alternative derivation of the asymptotic regime}
In this section we will provide an alternative solution for the $S_h$ system, which is valid only for the asymptotic regime and where we will \textit{neglect boundary conditions}.
We start from Eq.(\ref{eq:system_rip}) which is reported here for convenience:
\begin{equation}
\label{eq:system_rip3}
 \left\{ \begin{array}{rl}
 \displaystyle{\sum_{a=1}^h t_a v C_1^a = E C_0} \\
  \displaystyle{v C_2^a + t_a v  C_0  = E C_1^a}  & \hspace*{15pt} \textrm{\textit{a}=1,\dots,
  \textit{h}}\\
 \displaystyle{v \left( C_{n+1}^a + C_{n-1}^a \right ) = E C_n^a}  & \hspace*{15pt} \textrm{\textit{a}=1,\dots, \textit{h}; \textit{n}=1,\ldots, \textit{N}-1}\\
 \displaystyle{v C_{N-1}^a = E C_N^a} & \hspace*{15pt} \textrm{\textit{a}=1,\ldots, \textit{h}}
\end{array} \right.
\end{equation}
Now, in order to have a system whose unknowns are \textit{a}\textit{-independent}, we define
\begin{align}
\bar{C_i} \doteq \dfrac{C_i^a}{t_a} && i=1,\ldots, N
\end{align}
and
\begin{align}
\sum_{a=1}^h \left (t_a \right) ^2 \doteq T_{tot}
\end{align}
Thus, the system (\ref{eq:system_rip3}) becomes\footnote{Please note that the $\bar{C}_i$ are \textit{a}-independent, see pag.\pageref{eq:system_rip4}.}
\begin{equation}
\label{eq:system_bar}
 \left\{ \begin{split}
T_{tot} v^2 \bar{C}_1 &= E C_0  \\
v \bar{C}_2 + vC_0 &= E \bar{C}_1 \\
v \bar{C}_{n+1}+v \bar{C}_{n-1} &= E \bar{C}_{n}&& \hspace*{15pt} \textrm{\textit{n}=2,\ldots, \textit{N}-1}\\
v \bar{C}_{N-1} &= E\bar{C}_N 
\end{split} \right.
\end{equation}
To solve this system we use the following ``educated guess", or \textit{ansatz}\footnote{
In physics and mathematics, an ansatz is an educated guess that is verified later by its results. An ansatz is the establishment of the starting equation(s) describing a mathematical or physical problem.
}:
\begin{align}
\label{eq:educ_guess}
\bar{C}_j = e^{-kj} && \mbox{with }j=0,\ldots, N
\end{align}
fully justified by the fact that we are looking for states localized at the origin, where the branches cross.
Substituting Eq.(\ref{eq:educ_guess}) into system (\ref{eq:system_bar}), we get
\begin{equation}
\label{eq:system_sos}
 \left\{ \begin{split}
T_{tot} v e^{-k} &= E   \\
v e^{-2k} + v &= E e^{-k} \\
v e^{-k}+ v e^{k} &= E && \hspace*{15pt} \textrm{\textit{n}=2,\ldots, \textit{N}-1}\\
v e^{k} &= E 
\end{split} \right.
\end{equation}
A solution\footnote{Solving the system we neglected boundary conditions.} of the system (\ref{eq:system_sos}) leads us to:
\begin{equation}
k = \dfrac{1}{2} \ln(T_{tot}-1)
\end{equation}
and
\begin{align}
\label{eq:eig_met_k}
E = \pm \left ( \dfrac{T_{tot}}{\sqrt{T_{tot}-1}} \right )v 
\end{align}
These are the same results for the asymptotic regime - and neglecting boundary conditions - calculated in the previous sections; see Eqs.(\ref{eq:l_states_4_t})\footnote{Provided to define $\xi = 2k$ because of different definitions of $k$ and $\xi$.} and (\ref{eq:eigenvalues_4_diff_t}) respectively.\\
Now a question arises: what's more in the approach presented in the previous sections?
First of all, the method described in previous sections allows us to calculate the energies of the localized states for \textit{any finite N} by using Eq.(\ref{eq:polyn_ttot}) and \textit{not only for the asymptotic regime}, as calculated in this section. Therefore, one can compute case by case the gap between the energy corresponding to the finite value of \textit{N} and that given by the asymptotic limit, in order to understand when one can consider the asymptotic regime reached in relation to the accuracy that one wants to achieve.
Secondly, it is also possible to calculate the localization length for any
 finite $N$, by replacing the energy calculated with Eq.(\ref{eq:polyn_ttot}) in Eq.(\ref{eq:ratio_gen}). 
Thirdly, the method described in previous sections takes into account \textit{boundary conditions}, see Eqs.(\ref{eq:ratio_gen2}) and (\ref{eq:boundary}), while the one used in this section does not allows that; 
therefore, the last method introduced considers only the case $\mathcal{B} \equiv 1$ in reference to Equation (\ref{eq:ratio_gen2}). We stress the fact that boundary conditions are crucial for the determination of lifetimes of localized states because the correction provided by $\mathcal{B}$ is very important at the last site that it is the site that must be considered for the lifetime, see Chap.\ref{chap:4}, Eq.(\ref{eq:tau_loc5}).
Finally, in the method described in previous sections we did not make any assumption on the solutions to be found, contrary
to what
we have done in the last section by choosing a suitable ansatz
(not suitable if one takes into account boundary conditions).

\renewcommand\chapterheadstartvskip{\vspace*{2\baselineskip}}





\begin{savequote}[20pc]
\sffamily
I'm personally convinced that computer science has a lot in common with physics. Both are about how the world works at a rather fundamental level. The difference, of course, is that while in physics you're supposed to figure out how the world is made up, in computer science you create the world. Within the confines of the computer, you're the creator. You get to ultimately control everything that happens. If you're good enough, you can be God. On a small scale.
\qauthor{Linus Torvalds - creator of Linux}
\end{savequote}
\chapter{Open tight-binding model:\\ transition to superradiance}
\label{chap:4}
\section{Effective Hamiltonian for the star graph system}
\label{sec:eff_ham_open}
\begin{figure}[h!,b,t]
\vspace{0cm}
\center{\includegraphics[width=10cm,angle=0]{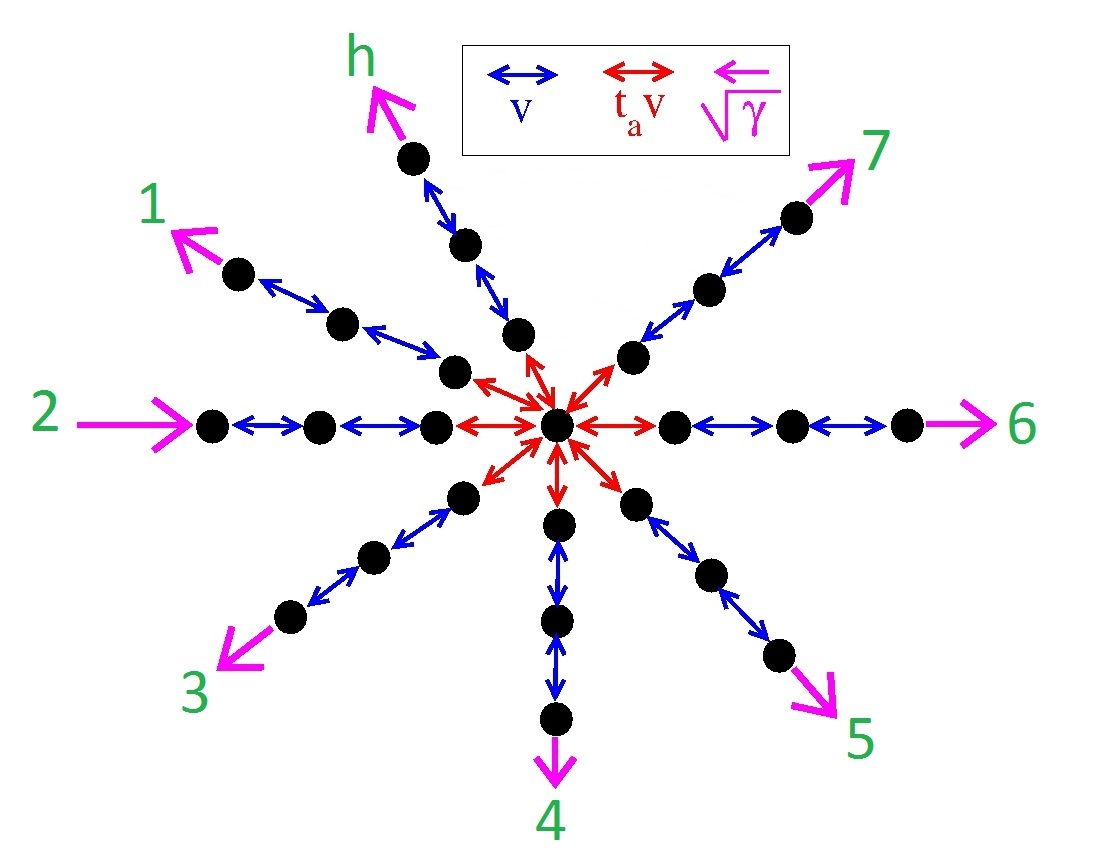}}
\caption{{\footnotesize The star graph system.
}}
\label{fig:4stargraph}
\end{figure}
Let us consider a system composed of $h$ chains of sites of lengths $N_a$, $a=1,\ldots,h$; the chains have a common vertex at the point $O$, the origin. We named this system $S_h(N_1,\ldots,N_a)$ \textit{system}\footnote{Furthermore, if we set $N_a \doteq N$ $\forall a$ we obtain the $S_h(N)$ system.}, see Chap.\ref{chap:3}, Sec.(\ref{sec:3model}).
The total wave function of a particle in such a closed system is
\begin{equation}
\label{eq:psi_form_gen}
\vert \psi\rangle = \sum_{a=1}^h \sum_{n=1}^{N_a}C_n^a \vert a; n\rangle + C_0 \vert 0 \rangle.
\end{equation}
We assume that the numeration of the sites in each chain starts, $n=1$, at the site closest to the origin, while the site $\vert a; N_a\rangle$ will be coupled to the external world. The level energy in each site (also the origin) equals zero, for simplicity.
Thus, the tight-binding Hamiltonian with nearest neighbour interaction of the $S_h(N_1,\ldots,N_a)$ system can be written as
\begin{align}
\label{eq:h_open}
\begin{split}
\mathsf{H} =& \sum_{a=1}^h \sum_{n=1}^{N_a-1} v \vert a; n \rangle \langle a; n + 1 \vert + 
\sum_{a=1}^h \sum_{n=2}^{N_a} v \vert a; n \rangle \langle a; n - 1\vert \\  &+ \sum_{a=1}^h t_av \left( \vert 0 \rangle \langle a; 1 \vert + \vert a; 1 \rangle \langle 0 \vert \right)
\end{split}
\end{align}
This is the Hamiltonian for the \textit{closed system}; now, we want to introduce the coupling with the environment.
The outside world is characterized by a continuum of states $\vert c, E \rangle$ where $c=1,\ldots,M$ is a discrete number labeling $M$ channels and $E$ is a continuum quantum number representing the energy.
In this thesis, we will couple the last site of each chain to a \textit{single} external channel; this is a choice because it is possible to build more complicated couplings with the outside world, that leads to \textit{hierarchical models}, see for example \cite{imry-hier}.
Under these considerations, the channel states can be written as $\vert a, E \rangle$ where $a=1,\ldots,h$, according to Eq.(\ref{eq:h_open}).
Taking into account Eq.(\ref{eq:2sempl_h}), we have 
\begin{align}
\label{eq:4sempl_h}
\mathcal{H} = \mathsf{H} - \dfrac{i}{2}W && W_{mn}=\sum_{a=1}^h A_m^a A_n^{a*}
\end{align}
where $\mathsf{H}$ is the same defined in Eq.(\ref{eq:h_open}) and $A_n^c$ is the \textit{transition amplitude} between the intrinsic state $\vert a; n\rangle$ and the continuum $\vert a,E \rangle$. For system invariant under time reversal, both $\mathsf{H}$ and $W$ are real symmetric matrices, and the coupling amplitudes $A_n^a$ can be taken as real. Furthermore, we set $A_n^a$ all equal to $\sqrt{\gamma}$; so, $\gamma$ is a parameter that controls the coupling strength with the external world.
Due to the coupling to the channel states, the states $\vert a,N_a\rangle$, $a=1,\ldots,h$, acquire a finite width $\gamma$. Now, we can write the full effective Hamiltonian for this system:
\begin{align}
\label{eq:h_open_s}
\begin{split}
\mathcal{H} = \mathsf{H} -  \dfrac{i}{2}\gamma \sum_{a=1}^h \vert a,N_a \rangle \langle a,N_a \vert 
\end{split}
\end{align}
The diagonalization of the non-Hermitian effective Hamiltonian (\ref{eq:h_open_s}) gives us the complex eigenvalues\footnote{These complex eigenvalues of $\mathcal{H}$ coincide with the poles of the S-matrix, see Chap.\ref{chap:2}, Eq.(\ref{eq:2pole}).}
\begin{equation}
\mathcal{E}_i = E_i - \dfrac{i}{2}\Gamma_i
\end{equation}
corresponding to resonances centered at $E_i$ with widths $\Gamma_i$ that determine the lifetime of a
state, $\tau_i \sim \hbar/\Gamma_i$.
\subsubsection{Imaginary part of the eigenvalues outside Bloch band}
\begin{figure}[h!]
\centering
\vspace{0cm}
\includegraphics[width=11cm,angle=-90]{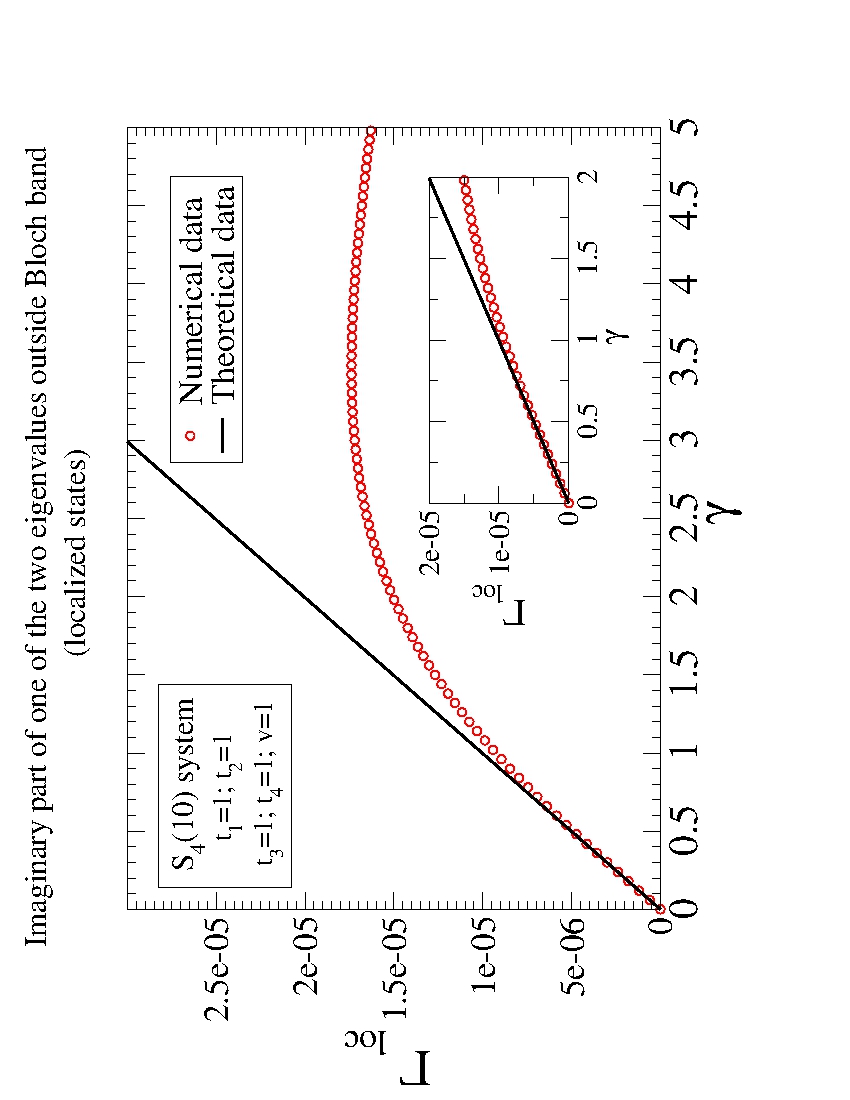}
\caption{{\footnotesize We plot the imaginary part $\Gamma_{loc}$ of one of the two eigenvalues outside Bloch band corresponding to the localized states as a function of the external coupling $\gamma$. We considered the $S_4(10)$ system with $t_a =1$ $\forall a$ and $v=1$. The solid line represents theoretical predictions obtained by using Eq.(\ref{eq:gamma_loc_F}) while circles refer to numerical data; for small $\gamma$, we can see the agreement between numerical and theoretical data. The quantities plotted are dimensionless.
}}
\label{fig:gamma_loc_ml}
\end{figure}

Now, we want to provide an estimation of the imaginary part $\Gamma_{loc}$ of the eigenvalues $\mathcal{E}_{loc}$ outside Bloch band.\\
First of all, we define the wave function for the localized states as
\begin{equation}
\label{eq:psi_loc_gamma}
\vert \psi_{loc}\rangle = \sum_{a=1}^h\sum_{i=1}^N C_{i,loc}^a \vert a; n\rangle + C_{0,loc} \vert 0 \rangle
\end{equation}
and defining $T_{tot} = \sum_{a} t_a^2$, we have for the asymptotic regime\footnote{See Chap.\ref{chap:3}, Eqs.(\ref{eq:ratio_gen2}) and (\ref{eq:boundary}).}:
\begin{align}
\label{eq:ratio_gen_cap4}
\left \vert \dfrac{C^a_{i+1}}{C^a_{i}} \right \vert^2 =
\left ( \dfrac{1}{T_{tot}-1} \right ) \mathcal{B}\left (T_{tot},N,i \right )  
&& i=1,\ldots, N-1
\end{align}
with
\begin{equation}
\label{eq:boundary_cap4}
\mathcal{B}\left (T_{tot},N,i \right ) \doteq \left \vert \dfrac{1 - \left (T_{tot}-1\right )^{-\left ( N-i \right )}}{1-  \left (T_{tot}-1\right )^{-\left ( N+1-i \right )}} \right \vert^2
\end{equation}
The term $\mathcal{B}$ is a boundary effect that it is very important (namely, very different from 1) for $i$ very close to $N$ which is precisely the case that must be considered in this paragraph.
Let us now consider the ratio
\begin{align}
\left \vert \dfrac{C_{N,loc}^a}{C_{0,loc}} \right \vert^2 = \left \vert \dfrac{C_{1,loc}^a}{C_{0,loc}} \right \vert^2
\cdot \left \vert \dfrac{C_{2,loc}^a}{C_{1,loc}^a} \right \vert^2 \cdot \left \vert \dfrac{C_{3,loc}^a}{C_{2,loc}^a} \right \vert^2 \ldots \left \vert \dfrac{C_{N,loc}^a}{C_{N-1,loc}^a} \right \vert^2
\end{align}
that, taking into account Eqs.(\ref{eq:ratio_gen_cap4}), (\ref{eq:boundary_cap4}) and the fact that\footnote{See Chap.\ref{chap:3}, Eq.(\ref{eq:3ratio_t1_t0}).} 
\begin{equation}
\left \vert \dfrac{C_{1,loc}^a}{C_{0,loc}} \right \vert^2 = \dfrac{t_a^2}{T_{tot}-1}
\end{equation}
lead us to the following relation:
\begin{align}
\left \vert \dfrac{C_{N,loc}^a}{C_{0,loc}} \right \vert^2 = \dfrac{t_a^2}{\left (T_{tot}-1\right )^N}\prod_{i=1}^{N-1} \mathcal{B} \left (T_{tot},N,i \right )
\end{align}
which, using the definition of $\xi$, $\xi= \ln(T_{tot}-1)$, can be written as
\begin{align}
\label{eq:cn_c0_b}
\boxed{
\left \vert \dfrac{C_{N,loc}^a}{C_{0,loc}} \right \vert^2 = t_a^2  e^{-\xi N} \prod_{i=1}^{N-1} \mathcal{B} \left (T_{tot},N,i \right )}
\end{align}
Furthermore, we know from the literature\footnote{See for example Refs.\cite{zele1992} and \cite{zele1989}. } that, for small $\gamma$, the width of a state can be written as
\begin{equation}
\label{eq:gam_loc_g}
\Gamma_{loc} = \sum_{a=1}^h \vert \langle a;N_a \vert \psi_{loc}\rangle \vert^2 \gamma
\end{equation}
and it agrees with the intuitive argument that the decay 
executes the decomposition of the state, isolating the components matched to the specific open channels.
For semplicity, we consider $h$ chains of identical length; hence, Eq.(\ref{eq:gam_loc_g}) becomes
\begin{equation}
\label{eq:gamma_loc}
\Gamma_{loc} = \sum_{a=1}^h \vert \langle a;N \vert \psi_{loc}\rangle \vert^2 \gamma
\end{equation}
Taking into account Eqs.(\ref{eq:psi_loc_gamma}) and (\ref{eq:cn_c0_b}), we obtain
\begin{align}
\label{eq:gamma_loc_1}
\Gamma_{loc} = \vert C_{0,loc} \vert^2 e^{-\xi N } \prod_{i=1}^{N-1} \mathcal{B} \left (T_{tot},N,i \right ) \sum_{a=1}^h t_a^2 \gamma 
\end{align}
that gives us
\begin{equation}
\Gamma_{loc} = T_{tot} \vert C_{0,loc} \vert^2 e^{-\xi N } \prod_{i=1}^{N-1} \mathcal{B} \left (T_{tot},N,i \right ) \gamma 
\end{equation}
or, using the definition of $\xi$,
\begin{equation}
\label{eq:gamma_loc_F}
\boxed{\Gamma_{loc} = \vert C_{0,loc}\vert^2 \dfrac{T_{tot}}{\left (T_{tot}-1\right )^N} \prod_{i=1}^{N-1} \mathcal{B} \left (T_{tot},N,i \right )\gamma }
\end{equation}
Moreover, we know that the lifetime of a state is $\tau_i \sim \hbar/\Gamma_i$; so, setting $\hbar = 1$, we get for the localized states:
\begin{equation}
\label{eq:tau_loc5}
\boxed{ \tau_{loc} =  \left [\dfrac{\left (T_{tot}-1\right )^N}{\vert C_{0,loc} \vert^2 T_{tot}} 
\left (\prod_{i=1}^{N-1} \mathcal{B} \left (T_{tot},N,i \right )\right )^{-1} \right ]\dfrac{1}{\gamma}}
\end{equation}
which is valid \textit{only} for small $\gamma$.
We can see from equation (\ref{eq:tau_loc5}) that the larger $N$ 
(the number of sites for each branch),
 the greater the lifetime of the localized states; hence, for very large $N$, we have $\tau_{loc} \simeq 0$ for small values of the external coupling $\gamma$.
Moreover, we note that in Eq.(\ref{eq:tau_loc5}) 
the factor $\vert C_{0,loc} \vert^2$ appears and it
can be calculated in a straightforward manner from the 
knowledge of the relationship between two successive eigenvectors, Eq.(\ref{eq:ratio_gen_cap4}), and by imposing the normalization of the wave function.\\
In Fig.(\ref{fig:gamma_loc_ml}), we show the imaginary part $\Gamma_{loc}$ of one of the two eigenvalues corresponding to the localized states as a function of the external coupling $\gamma$ (the other is the same). We considered the $S_4(10)$ system setting $t_a =1$ $\forall a$ and $v=1$. The solid black line refers to theoretical data obtained by using Eq.(\ref{eq:gamma_loc_F}) while circles refer to numerical data; for small $\gamma$, there is perfect correspondence between theoretical and numerical data.

\section{Superradiance transition}

\subsection{Evolution of the complex eigenvalues}
\begin{figure}[h!]
\centering
\vspace{0cm}
\includegraphics[width=11.5cm,angle=-90]{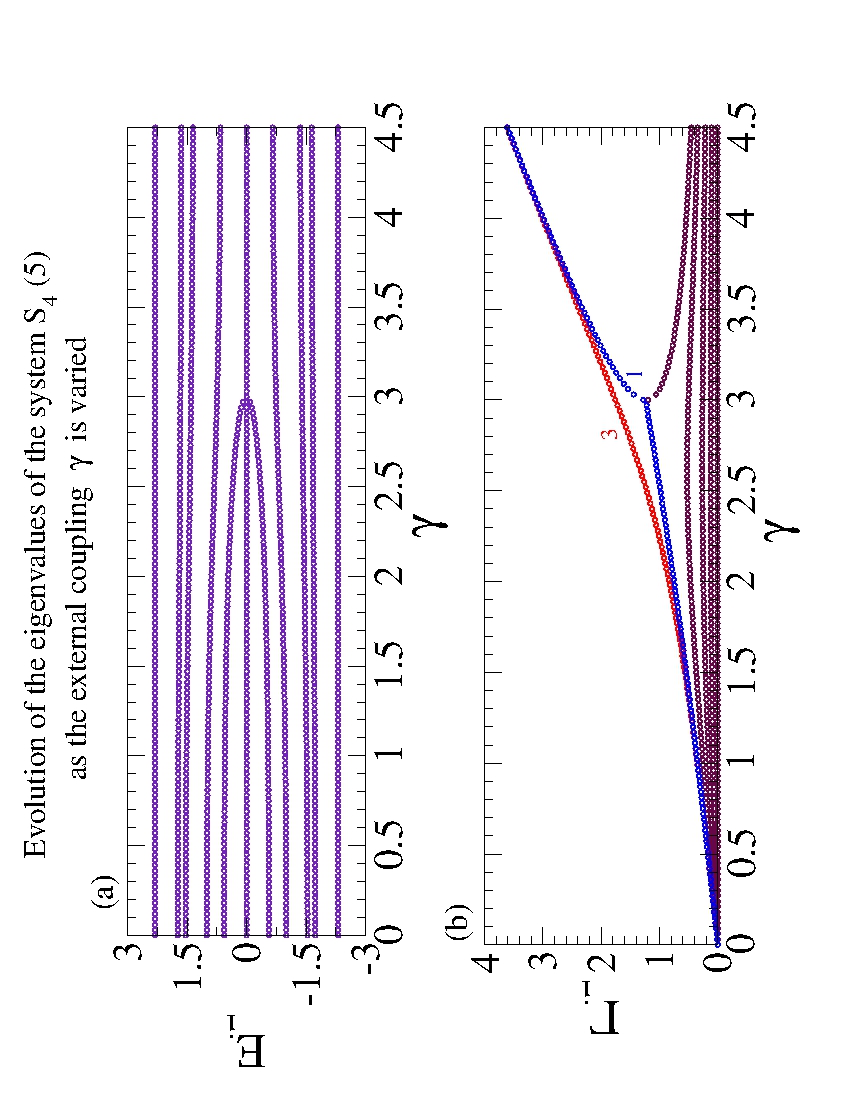}
\caption{{\footnotesize The evolution of the eigenvalues $\mathcal{E}_i = E_i - i/2 \Gamma_i$ of the effective Hamiltonian  is shown as the external coupling $\gamma$ is varied. The $S_4(5)$ system (21 sites) with $t_a=1$ $\forall a$ is considered.  The quantities plotted are dimensionless. (a) The real part of the eigenvalues $\mathcal{E}_i$ as a function of the coupling parameter $\gamma$. (b) The imaginary part of the eigenvalues $\mathcal{E}_i$ as a function of the coupling parameter $\gamma$. The emergence of four superradiance states is clearly visible: one state with multiplicity 3 - red circles - and one state with multiplicity 1 - blue circles; see text for further explanation.
}}
\label{fig:gamma_e_img_s4_5}
\end{figure}
Now, we want to show how and when the superradiance transition occurs. First of all, we know that the eigenvalues of the effective Hamiltonian $\mathcal{H}$ can be written as $\mathcal{E}_i = E_i - i/2\Gamma_i$.
For $\gamma \ll 1$, the real part of the complex eigenvalues, $E_i$, are close to the eigenvalues of the internal Hamiltonian $\mathsf{H}$ and the imaginary part, $\Gamma_i$, gives the width of the isolated resonances.
In the upper 
 panel of Fig.(\ref{fig:gamma_e_img_s4_5}), we show $E_i$ as a function of the coupling constant $\gamma$ for the $S_4(5)$, which stands for $4$ chains of $5$ sites each, with $t_a=1$ $\forall a$; similarly, in the lower panel, we plot $\Gamma_i$ as a function of the coupling $\gamma$.
We see that, as $\gamma$ increases, $h=4$ states - blue and red circles in the lower panel of Fig.(\ref{fig:gamma_e_img_s4_5}) - acquire the most of the width and become short-lived or \textit{superradiant}.
Still looking at the lower panel of Fig.(\ref{fig:gamma_e_img_s4_5}), we see that there are two states that have the same value of $\Gamma$ up to a critical value $\gamma \approx 3$ where a sort of bifurcation occurs and one state becomes superradiant at the cost of the other. At exactly the same critical value of $\gamma$, the real part of these two states both go to zero, as can be seen in the upper panel of the figure. 
The  superradiant state will appear only at  the
 critical  coupling, $\gamma \approx 3$. The other three (degenerate) 
superradiant states (red circles in the lower panel) are 
corresponding to $E_i = 0$.
Summarizing, we can state the following: as the coupling increases and reaches a critical value, the resonances overlap, and a sharp restructuring of the system occurs. Beyond this critical value, a few resonances become short-lived states, leaving all other (long-lived) states effectively decoupled from the environment. This general phenomenon is referred
to as the superradiance transition, due to its analogy with Dicke superradiance in quantum optics, see Ref.\cite{dicke}. At the critical point a transition takes place which is caused by the feedback between environment and system.
\begin{figure}[h!]
\centering
\vspace{0cm}
\includegraphics[width=11cm,angle=-90]{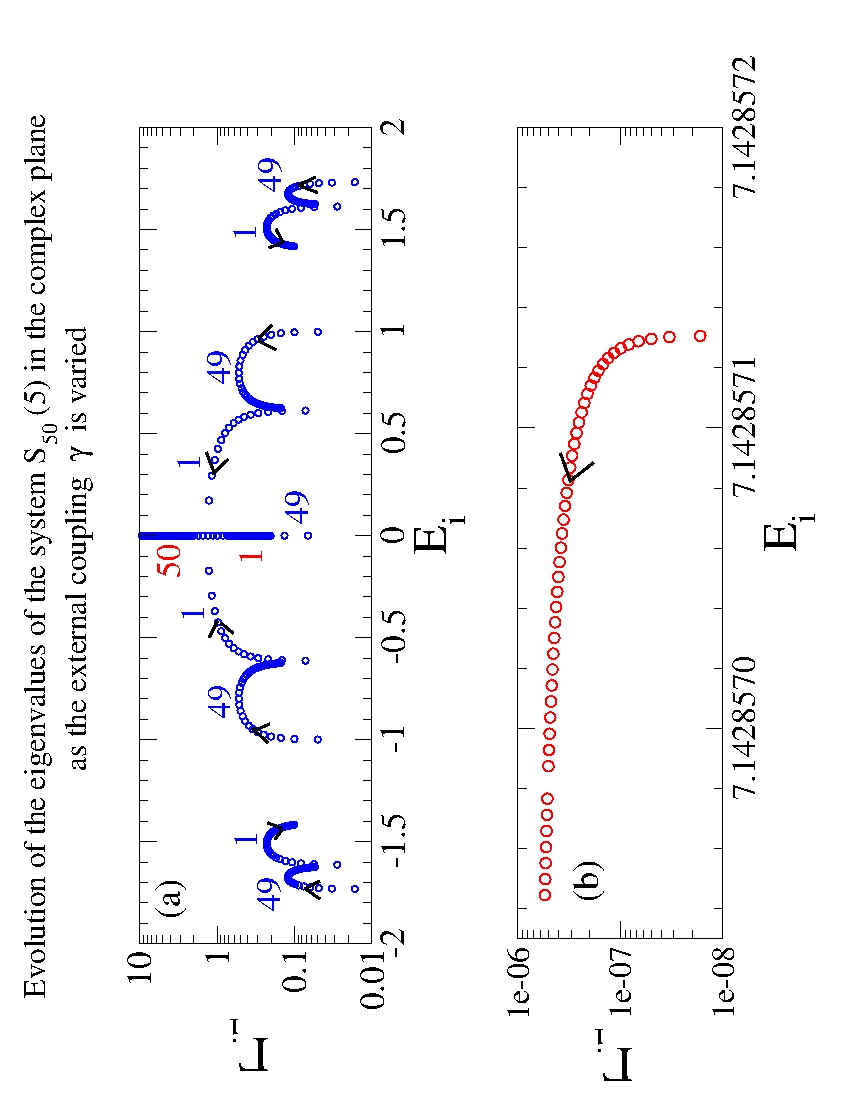}
\caption{{\footnotesize The evolution in the complex plane of the eigenvalues of the effective Hamiltonian is shown as the external coupling $\gamma$ is varied. The $S_{50}(5)$ system (251 sites) with $t_a=1$ $\forall a$ and $v=1$ is considered. Arrows indicate the motion of the poles in the complex plane as the external coupling $\gamma$ is increased ($0\leq \gamma \leq 10$ in the simulation). The quantities plotted are dimensionless. (a) The evolution in the complex plane of the 249 eigenvalues within Bloch band. We reported on the symbol the multiplicity of the eigenvalue (blue). After a critical value of $\gamma$, the superradiance transition occurs. 
After that, we have 51 states with $E_i \simeq 0$; 50 states are superradiants and one state is not superradiant.
 (multiplicity in red). (b) The evolution in the complex plane of one of the two eigenvalues outside Bloch band.}}
\label{fig:ev_eig_s50_5}
\end{figure}
\newline In Fig.(\ref{fig:ev_eig_s50_5}), the evolution in the complex plane of the eigenvalues of the effective Hamiltonian is shown as the external coupling $\gamma$ is varied. Arrows indicate the motion of the poles in the complex plane as $\gamma$ is changed. The $S_{50}(5)$ system (251 sites) with $t_a=1$ $\forall a$ and $v=1$ is considered. According to Chap.\ref{chap:3}-Sec.(\ref{sec:3sol_closed}), for the closed system ($\gamma=0$) we have two eigenvalues outside of a normal Bloch band while the others $249$ are enclosed within the Bloch band.
In the panel (a) of Fig.(\ref{fig:ev_eig_s50_5}), we plot the evolution of the eigenvalues inside Bloch band; we reported on the symbol the multiplicity of the eigenvalue (blue). Generally speaking, we can see that, as $\gamma$ increases, the poles tend to go toward the center of the band; moreover, it is interesting to look at what happens at the two states closest to $E=0$ in more detail. As $\gamma$ increases, these two states approach $E=0$, until a critical value of $\gamma$, when the superradiance transition occurs. 
After that, we have 51 states with $E_i \simeq 0$; 50 states are superradiants and one state is not superradiant.
In the panel (a) of Fig.(\ref{fig:ev_eig_s50_5}), we reported in red the multiplicity of the states with $E_i \simeq 0$ after the superradiance transition.
In the panel (b), we plot the eigenvalue trajectory of one of the two localized states outside Bloch band as a
function of the overall coupling strength $\gamma$. We see that the 
eigenvalues outside the Bloch band
in the complex plane are weakly dependent
on the external coupling. 

\subsection{Evaluation of critical $\gamma$}
\begin{figure}[h!]
\centering
\vspace{0cm}
\includegraphics[width=11cm,angle=-90]{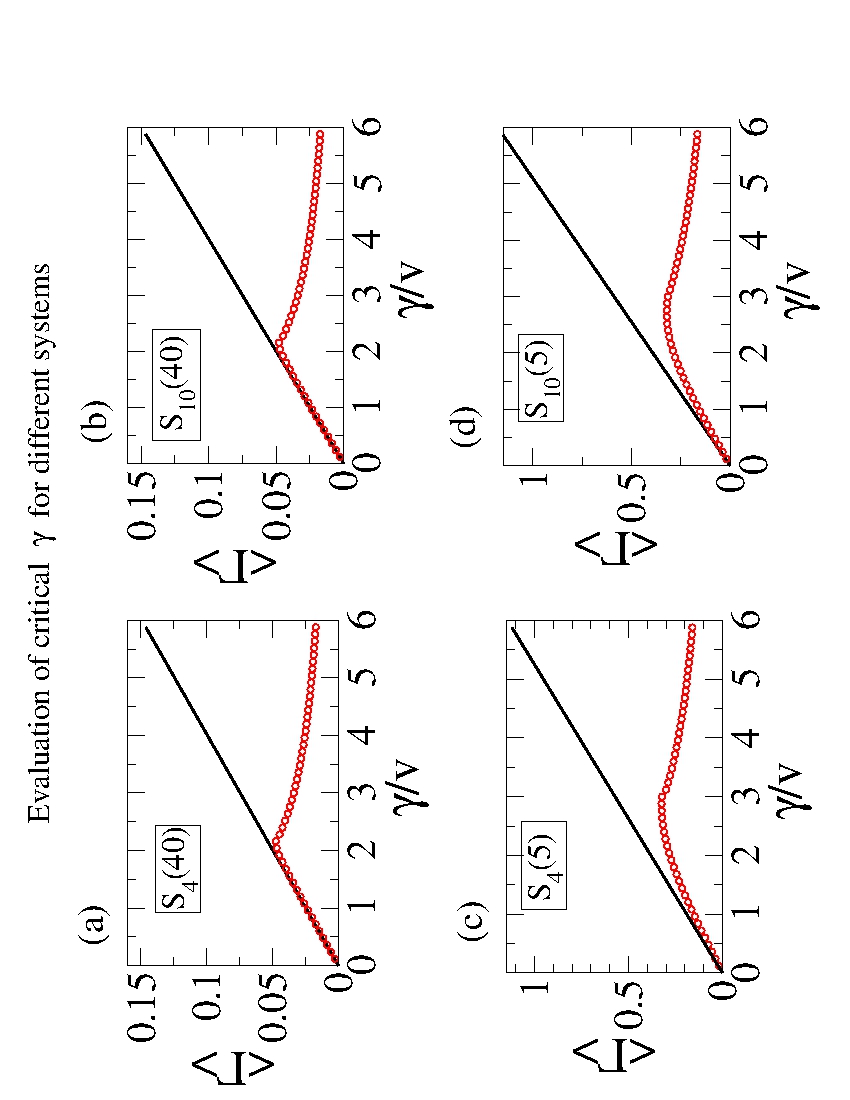}
\caption{{\footnotesize
The average width $\langle \Gamma \rangle$ is shown as a function of $\gamma$. When the number of sites is large, the transition to superradiance is shown to occur at $\gamma_{cr}/v \simeq 2$, regardless of the value of $h$, see text. The solid line corresponds to an average over all $K$ widths, while the symbols are obtained by averaging over the $K-h$ smallest widths. On each panel, we indicated in the box the system under consideration; for all systems we set $t_a=1$ $\forall a$. The quantities plotted are dimensionless.
}}
\label{fig:crit_g_rep}
\end{figure}

As we explained in Chap.\ref{chap:2}, Sec.(\ref{sec:supertrans}), we can distinguish two different regime for increasing $\gamma$. At weak coupling, all internal states are similarly affected by the opening and acquire widths proportionally to $\gamma$. In the opposite limit of large $\gamma$, only $h$ states (where $h$ is the number of channels) will have a width proportional to $\gamma$, while the widths of the remaining states fall of as $1/\gamma$.
In order to find the critical value of the parameter $\gamma$ at which the superradiance transition occurs, we may analyze the average width $\langle \Gamma\rangle$ of the $N-h$ narrowest widths as a function of the coupling $\gamma$, see Fig.(\ref{fig:crit_g_rep}). At the critical value of $\gamma$, the average width $\langle \Gamma\rangle$ peaks and begins to decrease.
We can evaluate this critical value of $\gamma$ using 
the \textit{criticality criterion} 
discussed earlier in Chap.\ref{chap:2}, Eq.(\ref{eq:2crit_crit}). Roughly, the transition occurs when $\langle \Gamma\rangle /D \approx 1$
where $D$ is the mean level spacing of the Hamiltonian for the closed system.\\
In our system, there is lot of degeneracy; for the $S_h(N)$ system (i.e.
 $hN+1$ sites), we numerically find that there are $M$ different energy levels, where $M=2N+1$; so, we define an \textit{effective mean level spacing} $D_{eff}$:
\begin{equation}
D_{eff} = \dfrac{4 v}{M} = \dfrac{4 v}{2N+1}\approx \dfrac{2v}{N}
\end{equation}
Moreover, the average width is
\begin{equation}
\langle \Gamma\rangle = \dfrac{h\gamma}{hN+1}\approx \dfrac{\gamma}{N}
\end{equation}
Therefore, the \textit{criticality criterion} becomes
\begin{equation}
\dfrac{\langle \Gamma\rangle}{D_{eff}}\approx \dfrac{\gamma_{cr}}{2v} = 1
\end{equation}
This implies $\gamma_{cr}=2v$; thus, the critical value of $\gamma$ is \textit{independent} of $h$ and $t_a$ $\forall a$.
In Fig.(\ref{fig:crit_g_rep}), we show the average width $\langle \Gamma \rangle$ as a function of $\gamma$. The solid line corresponds to an average over all \textit{K} widths, while the symbols are obtained by averaging over the $K-h$ smallest widths. 
The numerical computation of the critical $\gamma$ indicates that the above theoretical estimation is in perfect agreement with numerical data; thus, for the systems $S_4(40)$ and $S_{10}(40)$, we can consider the asymptotic regime already achieved, see panel (a) and (b) of Fig.(\ref{fig:crit_g_rep}).
Furthermore, if we look at panel (c) and (d), we notice that the critical $\gamma$ is different from that of the asymptotic regime and also depends on the 
 \textit{h} value; these effects, however, are due to the finite range considered and disappear in the asymptotic regime.
\begin{figure}[h!]
\centering
\vspace{0cm}
\includegraphics[width=10cm,angle=-90]{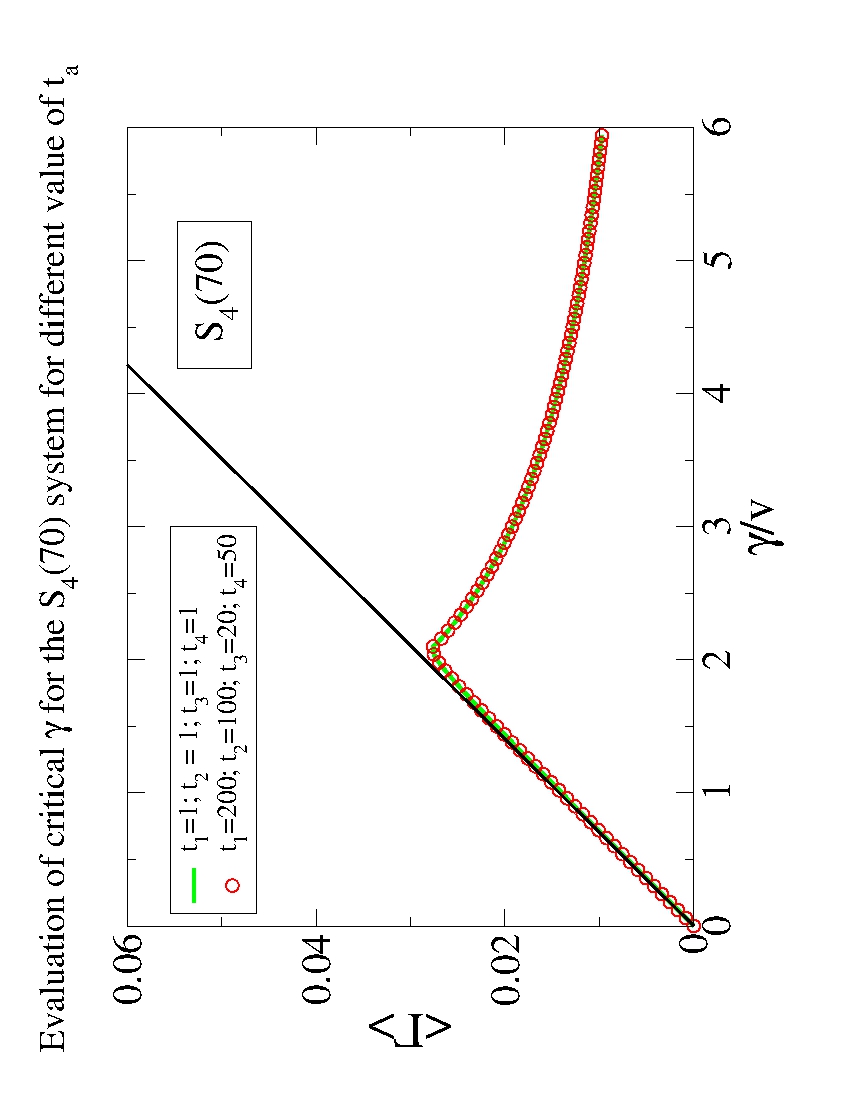}
\caption{{\footnotesize The average width $\langle \Gamma \rangle$ is shown as a function of $\gamma$. When the number of sites is large, the transition to superradiance is shown to occur at $\gamma_{cr}/v \simeq 2$, regardless of the value of $t_a$. In the simulation we considered the $S_4(70)$ system with different values of $t_a$: $t_a = 1$ $\forall a$(green line) and $t_{1}=200$, $t_{2}=100$, $t_{3}=20$, $t_{4}=50$ (red circles). We can see the perfect correspondence between these two sets of data. The solid black line corresponds to an average over all $K$ widths, while the green line and the symbols are obtained by averaging over the $K-4$ smallest widths. 
The quantities plotted are dimensionless.}}
\label{fig:cr_g_t_div}
\end{figure}In order to show that the critical $\gamma$, in the asymptotic regime, is independent of the choice of $t_a$, we consider the $S_4(70)$ system with two different sets
 of $t_a$; in the first case, we choose $t_a=1$ $\forall a$, while in the second we set $t_1= 200; t_2= 100; t_3= 20; t_4= 50$. The result is shown in Fig.(\ref{fig:cr_g_t_div}). The solid black line corresponds to an average over all $K$ widths, while the green line and the symbols are obtained by averaging over the $K-4$ smallest widths; in particular, the green line refers to the first case, $t_a=1$ $\forall a$, while the symbols refer to the second case. The perfect agreement between these two sets of data is evident.
\subsubsection{Chains with different lengths}
\begin{figure}[h!]
\centering
\vspace{0cm}
\includegraphics[width=11.4cm,angle=-90]{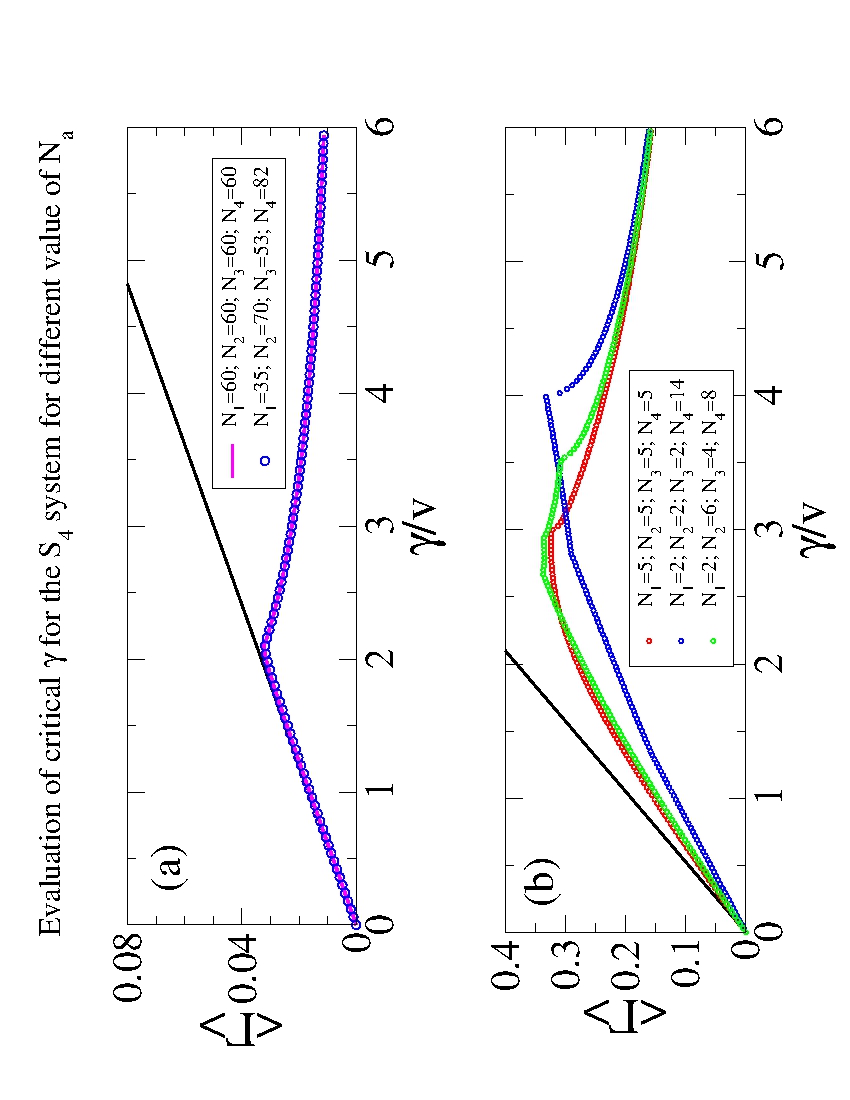}
\caption{{\footnotesize  The average width $\langle \Gamma \rangle$ is shown as a function of $\gamma$. When the total number of sites ($\sum_a N_a$ $+1$) is large, the superradiance transition is shown to occur at $\gamma_{cr}/v \simeq 2$, according to text, regardless of the different values of $N_a$. In the simulation we considered the $S_4(N_1,N_2,N_3,N_4)$ system with different values of $N_a$ and $t_a=1$ $\forall a$. The quantities plotted are dimensionless. (a) $N_a = 60$ $\forall a$(solid magenta line) and $N_{1}=35$, $N_{2}=70$, $N_{3}=53$, $N_{4}=82$ (blue circles). We can see the perfect agreement between these two series of data. The solid black line corresponds to an average over all $K$ widths, while the green line and the symbols are obtained by averaging over the $K-4$ smallest widths. (b) $N_a = 5$ $\forall a$ (red circles), $N_{1}=2$, $N_{2}=6$, $N_{3}=4$, $N_{4}=82$ (green circles) and  $N_{1}=2$, $N_{2}=2$, $N_{3}=2$, $N_{4}=14$ (blue circles).  The solid black line corresponds to an average over all $K$ widths, while the symbols are obtained by averaging over the $K-4$ smallest widths. We can see that, if the total number of sites is small, systems with different $N_a$ exhibit very different behaviors.
}}
\label{fig:n_div_crit_g}
\end{figure}
Now, we want to investigate the behavior of the critical $\gamma$ for a system whose chains have a \textit{different number of sites}, in general $N_a \neq N_{a^\prime}$ if $a \neq a^\prime$.
This question is relevant for \textit{experiments} where the exact number of sites can be hardly controlled; in fact, the sites of our model can represent, for example, biological molecules or ions and it is therefore very common have to deal with chains of different lengths. First of all, we study the \textit{asymptotic regime}, i.e.
 when the total number of sites $\sum_a N_a$ $+1$ is large; after that, we will take a look at systems with a small number of sites.\\
In order to shed light on asymptotic situation, we consider two different systems: the $S_4(60)$ system and the $S_4(35,70,53,82)$ system, both with $t_a=1$ $\forall a$.
These two systems have the same number of sites (241).
In the panel (a) of Fig.(\ref{fig:n_div_crit_g}), we show the average width $\langle \Gamma \rangle$ as a function of $\gamma$. The solid black line corresponds to an average over all \textit{K} widths, the solid magenta line is obtained by averaging over the $K-4$ smallest widths for the $S_4(60)$ system and the blue circles are found by averaging over the $K-4$ smallest widths for the $S_4(35,70,53,82)$ system. We can see the perfect correspondence between these two series of data: this is not due to the particular choice of the parameters $N_a$, but is a general behavior of our system in the asymptotic regime. Thus, we can state that \textit{the critical} $\gamma$, \textit{in the asymptotic regime, is independent of the values of} $N_a$.\\
Now, let us have a look at what happens when the system is composed
of a \textit{small number of sites}. 
We consider three different systems each consisting
 of 21 sites: $S_4(5)$, $S_4(2,6,4,8)$ and $S_4(2,2,2,14)$ all with $t_a=1$ $\forall a$. In the panel (b) of Fig.(\ref{fig:n_div_crit_g}), we plot the average width $\langle \Gamma \rangle$ as a function of $\gamma$. The solid black line corresponds to an average over all \textit{K} widths, while symbols is obtained by averaging over the $K-4$ smallest widths. It is evident that the average over the $K-4$ smallest widths for the $S_4(5)$ system (red circles) shows a single peak: this is the \textit{unique} critical value of $\gamma$. On the other hand, if we look at the two systems with different $N_a$ (blue and green circles), we can easily see
 that the average over the $K-4$ smallest widths exhibits more than one peak, so we have \textit{more than one critical} $\gamma$; this feature is very evident for the data series related to the $S_4(2,6,4,8)$ system (green circles).\\
However, these effects are due to the small number of sites of the systems considered and disappear in the asymptotic limit.

\subsection{Evolution of the localization length}
\begin{figure}[h!]
\centering
\vspace{0cm}
\includegraphics[width=10cm,angle=-90]{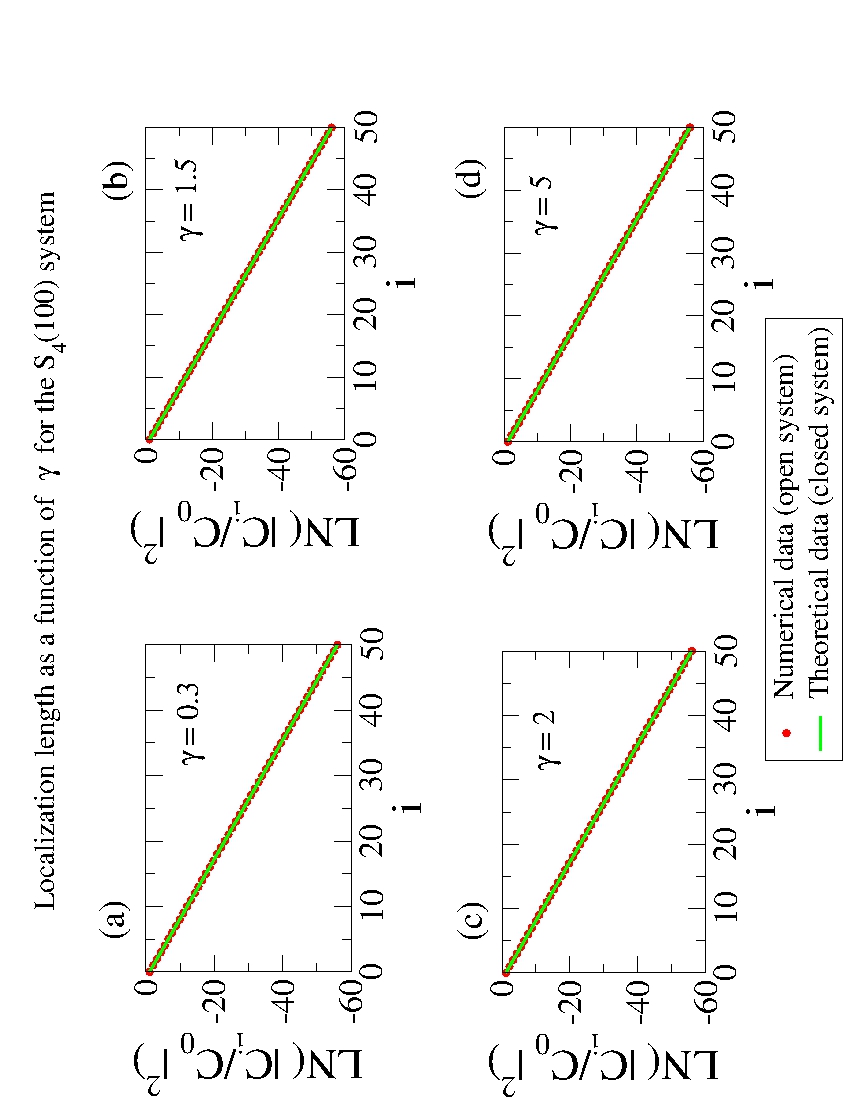}
\caption{{\footnotesize In the simulation, the $S_4(100)$ system with $t_a=1$ $\forall a$ and $v=1$.
Theoretical data (for a closed system) were calculated using equations (\ref{eq:local1}), (\ref{eq:local2}) and (\ref{eq:local2}).
One can see that the localization length is $\gamma$\textit{-independent}.
}}
\label{fig:loc_100}
\end{figure}

In Chapter \ref{chap:3}, Sec.(\ref{sec:3sol_closed}), we showed that for the $S_h(N)$ system we have two localized states corresponding to the two eigenvalues outside of a normal Bloch band. Writing the wave function for the localized states as
\begin{equation}
\vert \psi_{loc}\rangle = \sum_{a=1}^h\sum_{n=1}^N C_{n,loc}^a \vert a; n\rangle + C_{0,loc} \vert 0 \rangle
\end{equation}
neglecting boundary effects (namely, setting $\mathcal{B} \equiv 1$) and defining $T_{tot} = \sum_{a} t_a^2$, we have for the asymptotic regime:
\begin{align}
\label{eq:local1}
\left \vert \dfrac{C_{n,loc}^a}{C_{0,loc}} \right \vert^2 = \mathcal{A}\left (t_a,T_{tot}\right )  e^{-\xi \left (n-1 \right )} && \mbox{with } n=1, \ldots, N; && T_{tot} \geq 2
\end{align}
where
\begin{align}
\label{eq:local2}
\xi= \ln(T_{tot}-1) && T_{tot}\geq 2
\end{align}
and
\begin{align}
\label{eq:local3}
\mathcal{A}\left (t_a,T_{tot}\right ) = \dfrac{t_a^2}{T_{tot}-1}  && T_{tot}\geq 2
\end{align}
Now, we want to study 
 the behavior of the 
localization length as a function of
 the openness of the system.
In Fig(\ref{fig:loc_100}), we plot the evolution of the localization length of one of the two localized states as the external coupling $\gamma$ is varied.
The circles refer to numerical data, while the straight line refers to theoretical data (closed system); theoretical data were calculated using equations (\ref{eq:local1}), (\ref{eq:local2}) and (\ref{eq:local2}). We notice that the localization length is $\gamma$\textit{-independent}.
In the simulation we considered the $S_4(100)$ system with $t_a=1$ $\forall a$ and $v=1$.\\
However, this result is general; it is also valid for different values of $h$, $t_a$ and $v$.
Actually, from Fig.(\ref{fig:ev_eig_s50_5}) on pag.\pageref{fig:ev_eig_s50_5}, we can see that the imaginary part of the eigenvalues outside Bloch band are very small, which means that localized states are very little affected by the opening of the system; in this sense, the results shown in Fig.(\ref{fig:loc_100}) should not be considered as a surprise.

\section{Resonance structure}
In order to show the consequences of the superradiance transition on the transport properties, here we analyze the resonance structure, by considering the transmission $\mathsf{T} \left (E \right )$. 
In Chap.\ref{chap:2}, Eq.(\ref{eq:2t_ab}) and Eq.(\ref{eq:2pole}), we found that the \textit{transmission} for the process from channel \textit{a} to \textit{b} is
\begin{align}
\label{eq:4trasm}
\begin{split}
\mathsf{T}^{ab}(E) = \vert \mathcal{T}^{ab} \left (E \right ) \vert^2= \left \vert \sum_{i=1}^N A_i^{a}\dfrac{1}{E - \mathcal{E}_i} \tilde{A}_i^b \right \vert^2
\end{split}
\end{align}
where $N$ is the number of the internal states and $\mathcal{E}_i = E_i - i/2 \Gamma_i$ are the eigenvalues of the effective Hamiltonian which coincide with the poles of $\mathsf{T}^{ab}(E)$.
Now, we want to compute the trasmission from channel \textit{a} to \textit{b} for the $S_h(N)$ system at the energies corresponding to localized states, $\mathcal{E}_{loc} = E_{loc} - i/2 \Gamma_{loc}$; the last site of each chain is coupled to the external world with coupling amplitudes $\sqrt{\gamma_a} \doteq \sqrt{\gamma}$, $\forall a=1,\ldots,h$.\\ 
We define the wave function\footnote{Please note that the wave function defined in (\ref{eq:psi_loc}) is not normalized; the normalized wave function is the following:
\begin{equation}
\vert \psi_{loc}\rangle = \sum_{a=1}^h\sum_{i=1}^N C_{i,loc}^a \vert a; n\rangle + C_{0,loc} \vert 0 \rangle
\end{equation}
but here we consider $\vert \psi^a_{loc}\rangle$ for convenience.}
for the localized states as follows:
\begin{equation}
\label{eq:psi_loc}
\vert \psi_{loc}^a\rangle = \sum_{i=1}^N C_{i,loc}^a \vert a; n\rangle + C_{0,loc} \vert 0 \rangle
\end{equation}
and we compute the transmission at the energies corresponding to localized states:
\begin{equation}
\mathsf{T}^{ab}(E=E_{loc}) = \left \vert \dfrac{\langle a; N \vert \psi_{loc}^a \rangle \sqrt{\gamma} \langle \psi_{loc}^b \vert b; N \rangle \sqrt{\gamma} }{\frac{i}{2} \Gamma_{loc}} \right \vert^2
\end{equation}
Furthermore, from Eq.(\ref{eq:cn_c0_b}) we know that
\begin{align}
\left \vert \langle a;N \vert \psi_{loc}^a \rangle \right \vert^2 = \vert C_{0,loc} \vert^2 t_a^2  e^{-\xi N} \prod_{i=1}^{N-1} \mathcal{B} \left (T_{tot},N,i \right )
\end{align}
where 
\begin{align}
 T_{tot} = \sum_{a=1}^h t_a^2 &&  \xi = \ln \left ( T_{tot} -1 \right )
\end{align}
and from this chapter, see Sec.(\ref{sec:eff_ham_open}) - Eq.(\ref{eq:gamma_loc}), that
\begin{equation}
\Gamma_{loc} = \sum_{a=1}^h \vert \langle a; N \vert \psi_{loc}^a \rangle \vert^2 \gamma
\end{equation}
Taking into account all these considerations and after some calculation, we get
\begin{align}
\label{eq:trasm_ab}
\boxed{
\mathsf{T}^{ab}(E=E_{loc}) = \dfrac{4 t_a^2 t_b^2}{\vert T_{tot} \vert^2}} &&  T_{tot} = \sum_{a=1}^h t_a^2
\end{align}
This formula gives us the transmission from channel $a$ to $b$ at the energies corresponding to localized states.
If we set, for example, $t_a \doteq 1$ $\forall a$, we have $\mathsf{T}^{ab} = 4/h^2$ $\forall a, b$; so, the transmission is the same in all channels, as one might expect given the symmetry of the system\footnote{See Ref.\cite{pasta}.}.\\
Numerical simulations show that the transmission 
calculated at the energies corresponding to localized states is the same for all the other states; see for example Fig.(\ref{fig:trasm_S4_5}).
\subsubsection{Example 1: identical coupling between the chains and the origin}
\begin{figure}[h!]
\centering
\vspace{0cm}
\includegraphics[width=11cm,angle=-90]{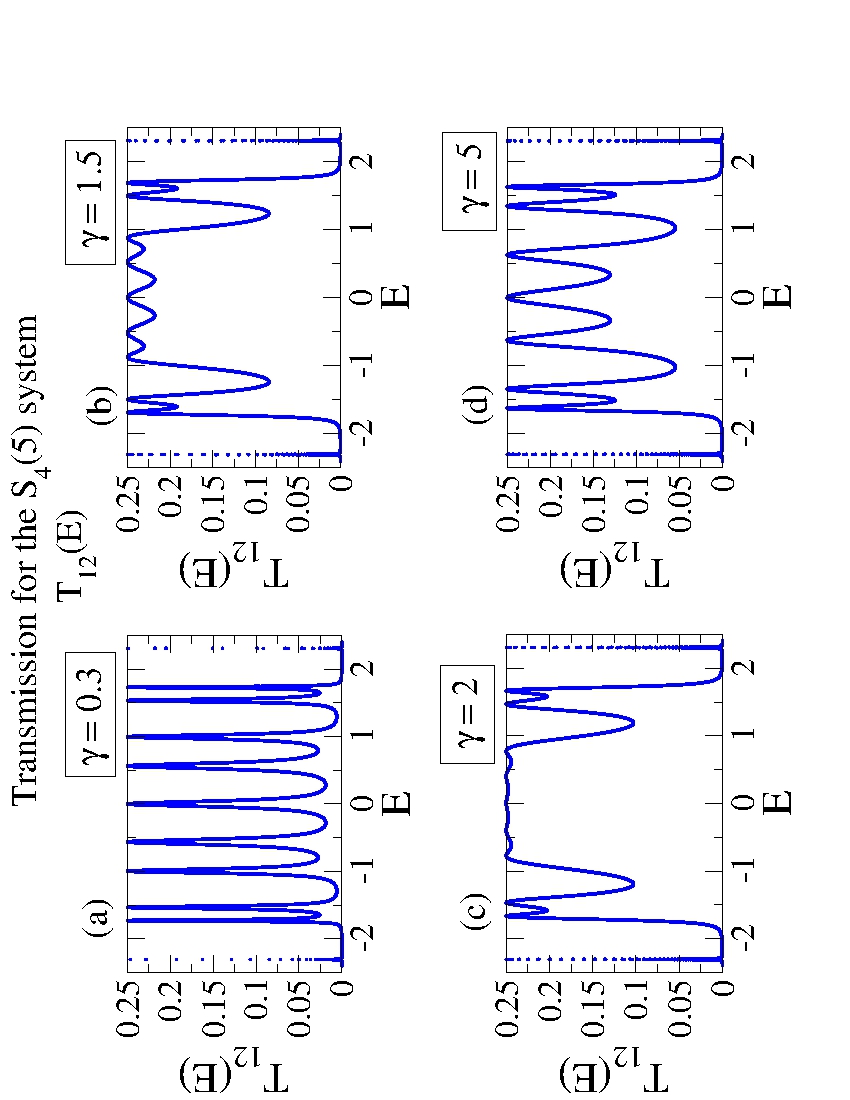}
\caption{{\footnotesize The transmission $\mathsf{T}$ is shown as a function of energy \textit{E} for the $S_4(5)$ system (21 sites) for different values of the external coupling $\gamma$. $\mathsf{T}^{ab}$ is the transmission from channel \textit{a} to \textit{b}. In this figure, we plotted only the results for $a = 1$ and $b = 2$. The results for $a = 2, 3, 4$ and $b = 1, 2, 3, 4$ are the same. We considered $v = 1$. The quantities plotted are dimensionless.}}
\label{fig:trasm_S4_5}
\end{figure}
In this paragraph, we will consider the particular case of $t_a=1$ $\forall a$; this means that all the chains are equally coupled with the origin.
In Fig.(\ref{fig:trasm_S4_5}), the transmission $\mathsf{T}^{12}$ is shown as a function of the energy $E$ for several values of the external coupling $\gamma$ for the $S_4(5)$ system with $t_a=1$ $\forall a$ and $v=1$; from Eq.(\ref{eq:trasm_ab}) we obtain
 that the transmission peaks reach the value of $1/4$, (see 
 Fig.(\ref{fig:trasm_S4_5}).
Moreover, it is clear from Fig.(\ref{fig:trasm_S4_5}) that the \textit{superradiance transition has a clear signature in the resonance structure.}\\
For the $S_h(N)$ system (i.e.
 $hN+1$ sites), we numerically find that there are $2N+1$ different energy levels because of degeneration; so, for the $S_4(5)$ system, we have $11$ different energy levels for the closed system. Hence, the resonances in Fig.(\ref{fig:trasm_S4_5}) are $11$:
 the two external narrow resonances correspond to the localized states, 
outside the normal Bloch band, while the others nine correspond to extended, Bloch-like states.\\
For  weak coupling, $\gamma = 0.3$, see panel (a) of Fig.(\ref{fig:trasm_S4_5}), there are $11$ isolated resonances which corrispond to the 
$11$ classes of different eigenvalues that acquire a small imaginary part. Since $\gamma$ is small, the imaginary part gives the width of the isolated resonances. The width decreases away from the band center ($E = 0$), reaching a minimum with the last two resonance peaks, corresponding to the localized states; indeed, these resonance peaks are the narrowest. As we increase the external coupling, the transmission increases and near $\gamma=1.5$ all resonances (except the two outer corresponding to localized states) start to overlap, see panel (b). \\ At $\gamma = 2$, see panel (c), the five central resonances are almost completely superimposed. 
\begin{figure}[h!]
\centering
\vspace{0cm}
\includegraphics[width=11.5cm,angle=-90]{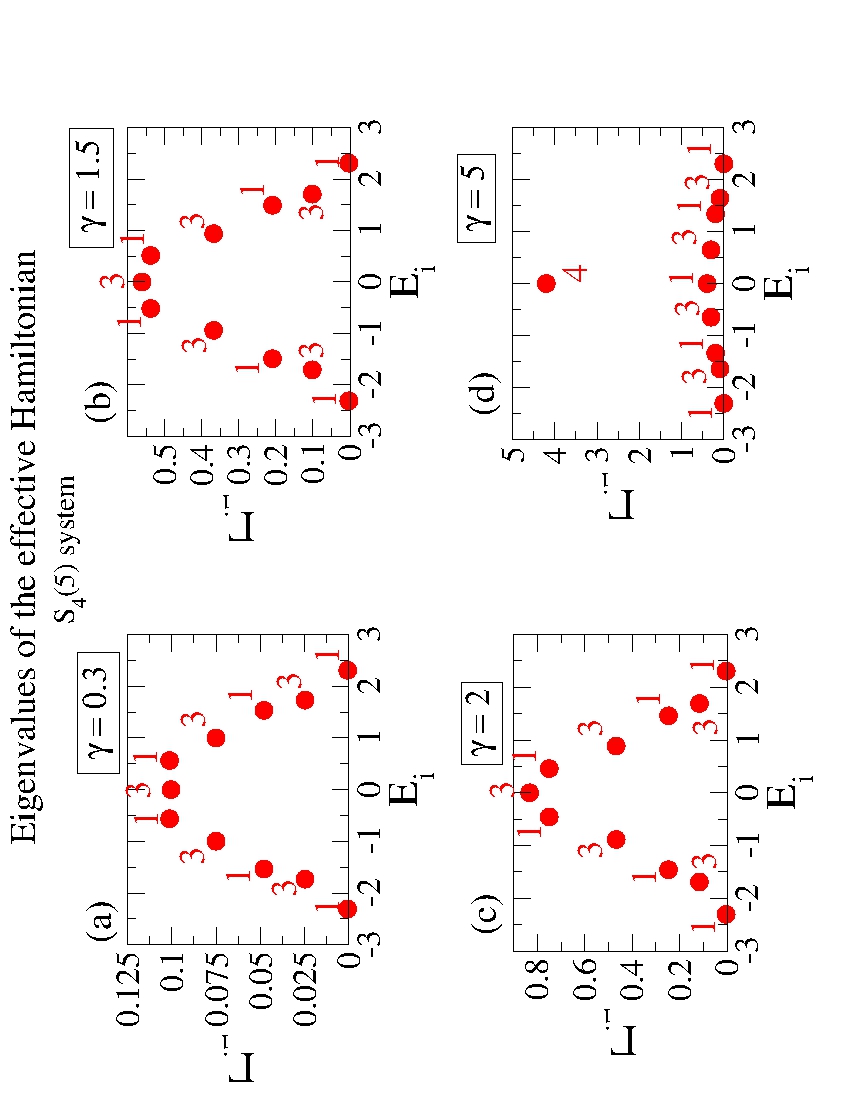}
\caption{{\footnotesize The evolution in the complex plane of the eigenvalues of the effective Hamiltonian is shown as the external coupling $\gamma$ is varied. The $S_4(5)$ system is considered. We reported on the symbol the multiplicity of the eigenvalue. The emergence of the four superradiance states is clearly visible in the panel (d). The quantities plotted are dimensionless.
}}
\label{fig:autov_img_g_deg}
\end{figure}
For $\gamma=5$ (strong coupling), we notice, see panel (d), that we return to the case of isolated resonances but this time, their number is reduced to $11 - 2 = 9$ resonances; therefore
 we lose only two resonance peaks because of the degeneration of the system. Indeed, the peak at $E\simeq0$ is associated with three states and each of the two states closest to $E \simeq 0$ has multiplicity 1; the superradiant states are four, two at $E\simeq0$ and the other two closer to $E\simeq0$, as expected for four channels. Thus, the peak in the middle of the band remains because it
 had (before the 
superradiance transition) multiplicity 3, while the other two neighbors disappear. In Fig.(\ref{fig:autov_img_g_deg}), we show the evolution in the complex plane of the eigenvalues of this system as the external coupling $\gamma$ is varied.\\ Furthermore, we can notice from panel (d) of Fig.(\ref{fig:trasm_S4_5}) that the width of the resonances is smaller for energies closer to the band edge, just as it was for weak coupling.
Finally, we note that the resonance peaks in the figure (\ref{fig:trasm_S4_5}) reach the same value of $1/4$ in each branch, as expected since the $t_a$ are all the same.

\subsubsection{Example 2: different coupling between the chains and the origin}
\begin{figure}[h!]
\centering
\vspace{0cm}
\includegraphics[width=11.8cm,angle=-90]{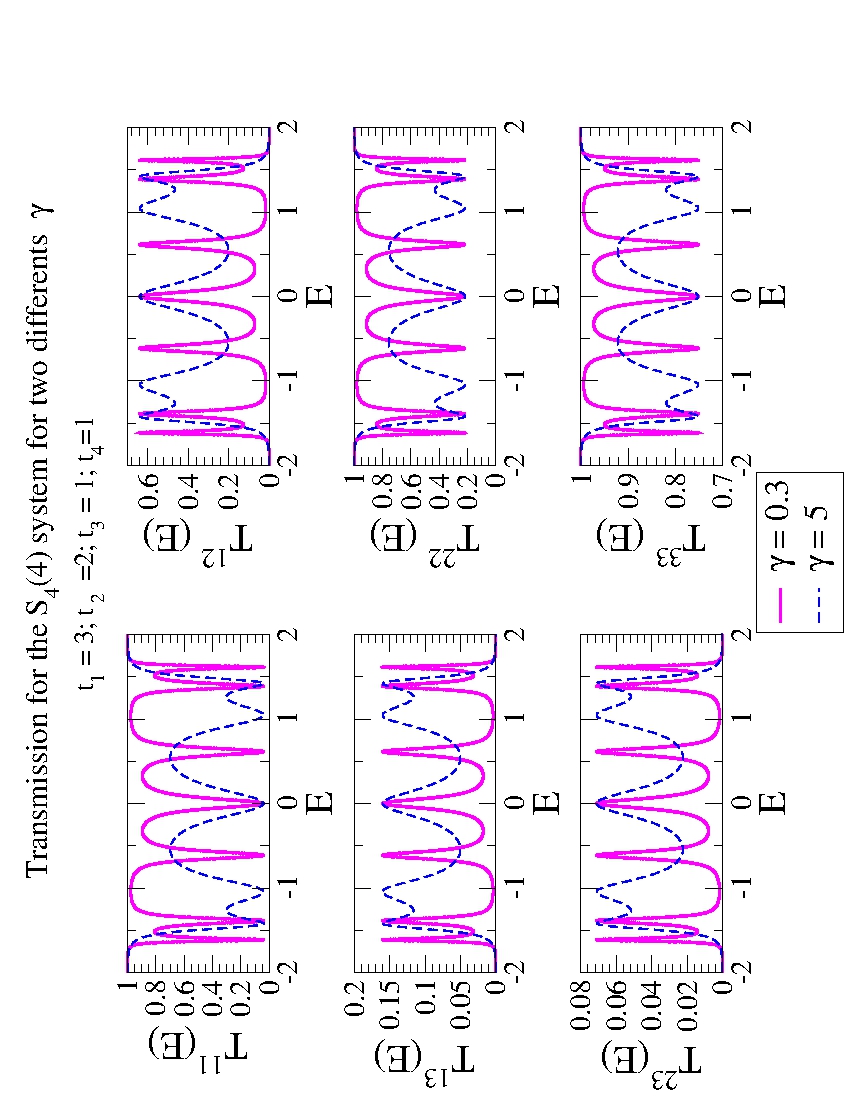}
\caption{{\footnotesize The transmission $\mathsf{T}$ is shown as a function of energy \textit{E} for the $S_4(4)$ system (17 sites) for two different values of the external coupling $\gamma$, $\gamma=0.3$ (solid magenta line), $\gamma=5$ (dashed blue line) before and after the superradiance transition, respectively. We set: $t_1=3;t_2=2;t_3=1;t_4=1$ and $v=1$. Here, we plot only the states within Bloch band. $\mathsf{T}^{ab}$ is the transmission from channel \textit{a} to \textit{b}. It is easy to check that Eq.(\ref{eq:trasm_ab}) is valid also for the states within Bloch band, [-2,+2] in this case.
The quantities plotted are dimensionless.}}
\label{fig:S4_4_diff_g}
\end{figure} 

\begin{figure}[h!]
\centering
\vspace{0cm}
\includegraphics[width=10cm,angle=-90]{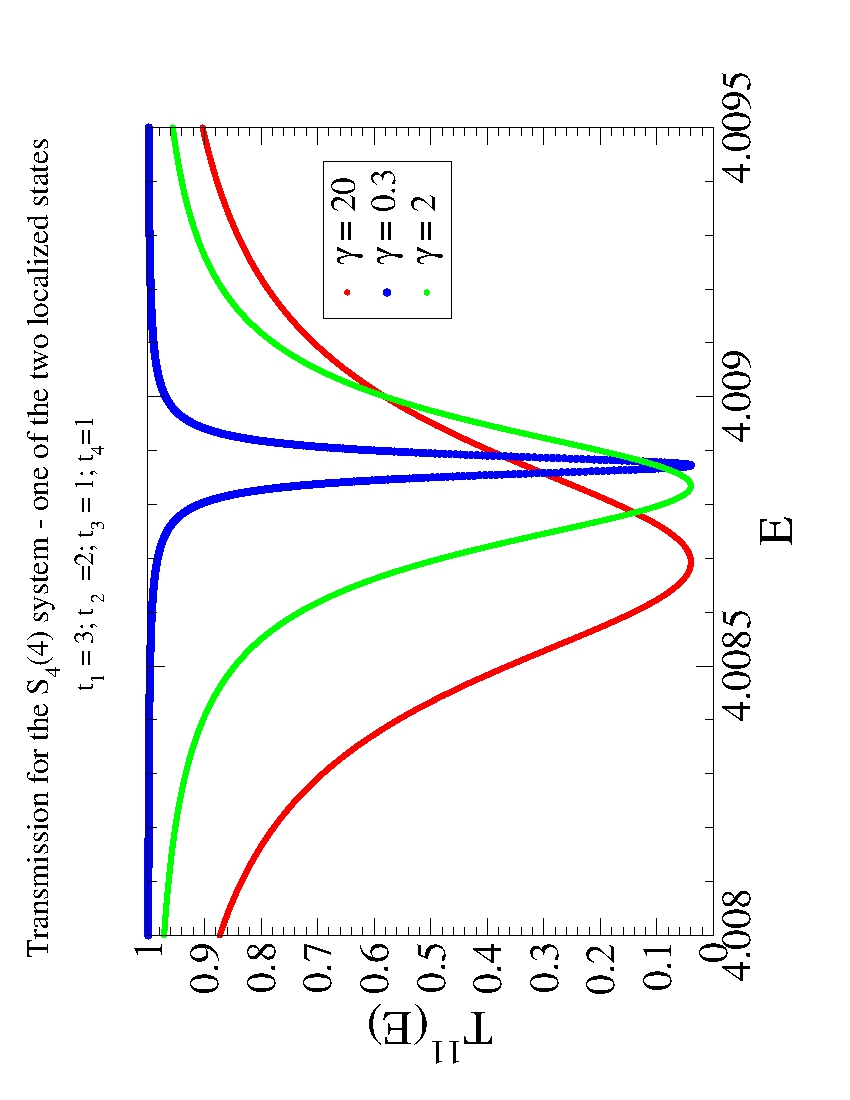}
\caption{{\footnotesize The transmission $\mathsf{T}$ is shown as a function of energy \textit{E} for the $S_4(4)$ system (17 sites) for different values of the external coupling $\gamma$ only for one of the two states outside Bloch band. $\mathsf{T}^{ab}$ is the transmission from channel \textit{a} to \textit{b}.  We set: $t_1=3;t_2=2;t_3=1;t_4=1$ and $v=1$. In this figure, we plotted only the results for $a = 1$ and $b = 1$. The quantities plotted are dimensionless.}}
\label{fig:s4_4_loc}
\end{figure} 
Let us consider an example where the branches have different couplings with the origin; here, we consider the $S_4(4)$ system setting $t_1=3;t_2=2;t_3=1;t_4=1$ and $v=1$. We plot the transmission from several input channels to different output channels for the states within Bloch band in Fig.(\ref{fig:S4_4_diff_g}) and for one of the two localized states in Fig.(\ref{fig:s4_4_loc}); one can easily check that the resonance peaks reach the values found in Eq.(\ref{eq:trasm_ab}) for all states.
Looking at the figure (\ref{fig:S4_4_diff_g}), we see that after the superradiance transition two resonance peaks disappear; actually, the superradiant states are four because there are four open channels. Only two peaks are missing because of degeneration of our system; in other words, the situation is very similar to Fig.(\ref{fig:autov_img_g_deg}) related to the previous example.
Finally, in Fig.(\ref{fig:s4_4_loc}), we plot one of the two eigenvalues outside Bloch band; we notice that the resonances related to the localized states are very narrow compared with those corresponding to extended states and we can also appreciate the energy shift due to the fact that we considered three different external couplings, $\gamma=0.3$ (blue line), $\gamma=2$ (green line), $\gamma=20$ (red line).

\section{Integrated transmission}
\begin{figure}[h!]
\vspace{0cm}
\includegraphics[width=10cm,angle=-90]{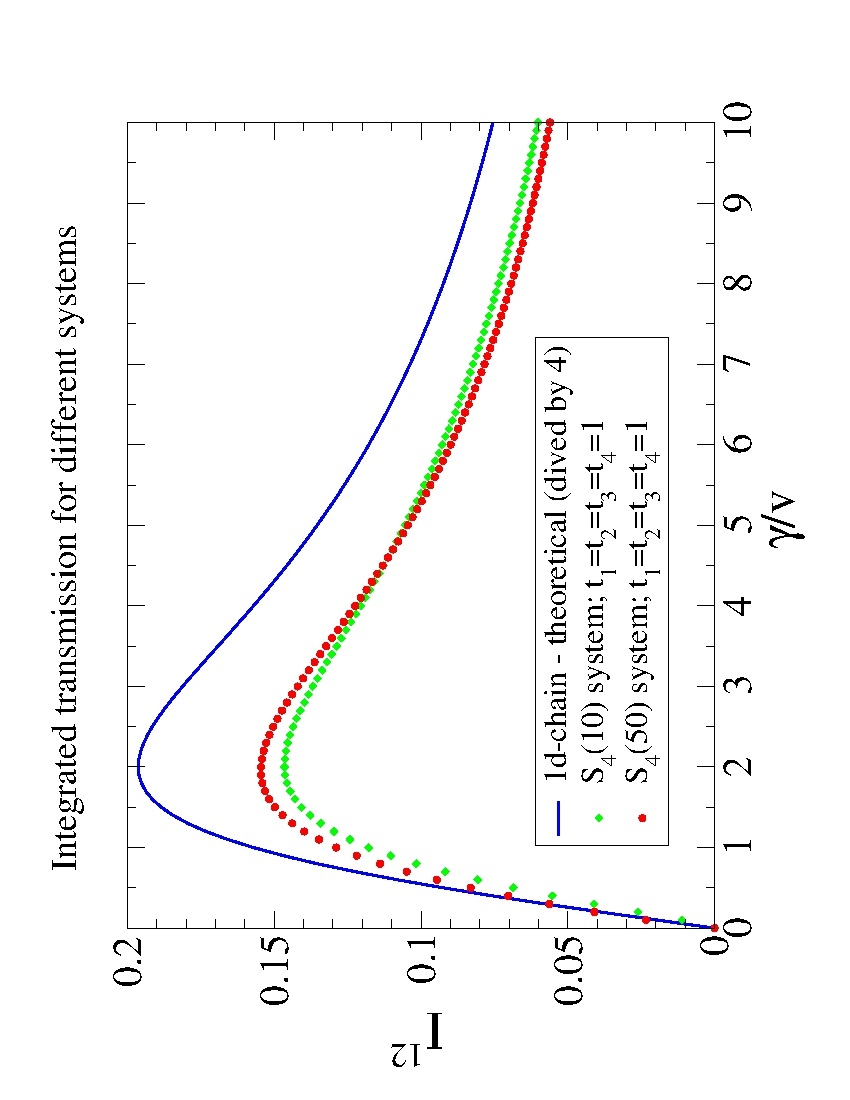}
\caption{{\footnotesize
The integrated transmission $I$ is shown as a function of external coupling $\gamma$ for two different systems: $S_4(10)$ and $S_4(50)$ systems. $S^{ab}$ is the transmission from channel \textit{a} to \textit{b}. In this figure, we plotted only the results for $a=1$ and $b=2$. The results for $a=2,3,4$ and $b=1,2,3,4$ are the same. We considered $v=1$. The quantities plotted are dimensionless.
The integrated transmission $I^{12}$ has a maximum at $\gamma_{tr}/v \simeq 2$: this is the critical value of $\gamma$ for the transmission. This value is almost independent from the number of sites considered as we can see from the figure above.
We plot also the theoretical integrated transmission for a chain of sites, divided by 4.}}
\label{fig:int_plus_th}
\end{figure}
It is also interesting compute the normalized integrated transmission:
\begin{align}
I = \dfrac{1}{\Delta E}\int_{E_{min}}^{E_{max}} \mathsf{T}\left (E \right )  dE && \Delta E = E_{max}-E_{min}
\end{align}
where the interval $\Delta E$ includes all the system eigenvalues.\\
In Fig.(\ref{fig:int_plus_th}), the integrated transmission $I$ is shown as a function of external coupling $\gamma$ for two different systems: $S_4(10)$ and $S_4(50)$ systems.
We plot also the normalized integrated transmission for a one dimensional 
chain \footnote{See Ref.\cite{lucak} for more details.}
for comparison:
\begin{equation}
\mathcal{I} = \dfrac{1}{4v}\int T(E)dE = \dfrac{\pi \gamma / v}{4 + \left( \gamma / v \right)^2 }
\end{equation}
divided by 4. 
Indeed, since the system under consideration consists of four chains coupled in the origin, it is reasonable to compare the integrated transmission with the results already known for the system consisting of one chain, divided by four.\\
The integrated transmission $I^{12}$ has a maximum at $\gamma_{tr}/v \simeq 2$: this is the critical value of $\gamma$ for the transmission. This value is almost independent from the number of sites considered as we can see from figure (\ref{fig:int_plus_th}). Moreover, for large $N$, \textit{the transmission maximum is reached precisely at the superradiance transition}: $\gamma_{cr} \simeq \gamma_{tr} = 2v$, compare Fig.(\ref{fig:crit_g_rep}) and Fig.(\ref{fig:int_plus_th}); so, the maximum of the integrated transmission is strongly related to 
the superradiance transition.
Moreover, we can see from Fig.(\ref{fig:int_plus_th}) that the normalized integrated transmission for a one dimensional chain of sites is very different from the normalized integrated transmission obtained with our crossing configuration system.





\pagestyle{fancy}


\renewcommand\chapterheadstartvskip{\vspace*{0\baselineskip}}





\begin{savequote}[15pc]
\sffamily
Fortunately science, like that nature to which it belongs, is neither limited by time nor by space. It belongs to the world, and is of no country and of no age. The more we know, the more we feel our ignorance; the more we feel how much remains unknown; and in philosophy, the sentiment of the Macedonian hero can never apply, there are always new worlds to conquer.
\qauthor{Sir Humphry Davy -  British chemist}
\end{savequote}
\chapter{Perspective}
\label{chap:5}
\section{Disorder}
\begin{figure}[h!]
\vspace{0cm}
\includegraphics[width=11cm,angle=-90]{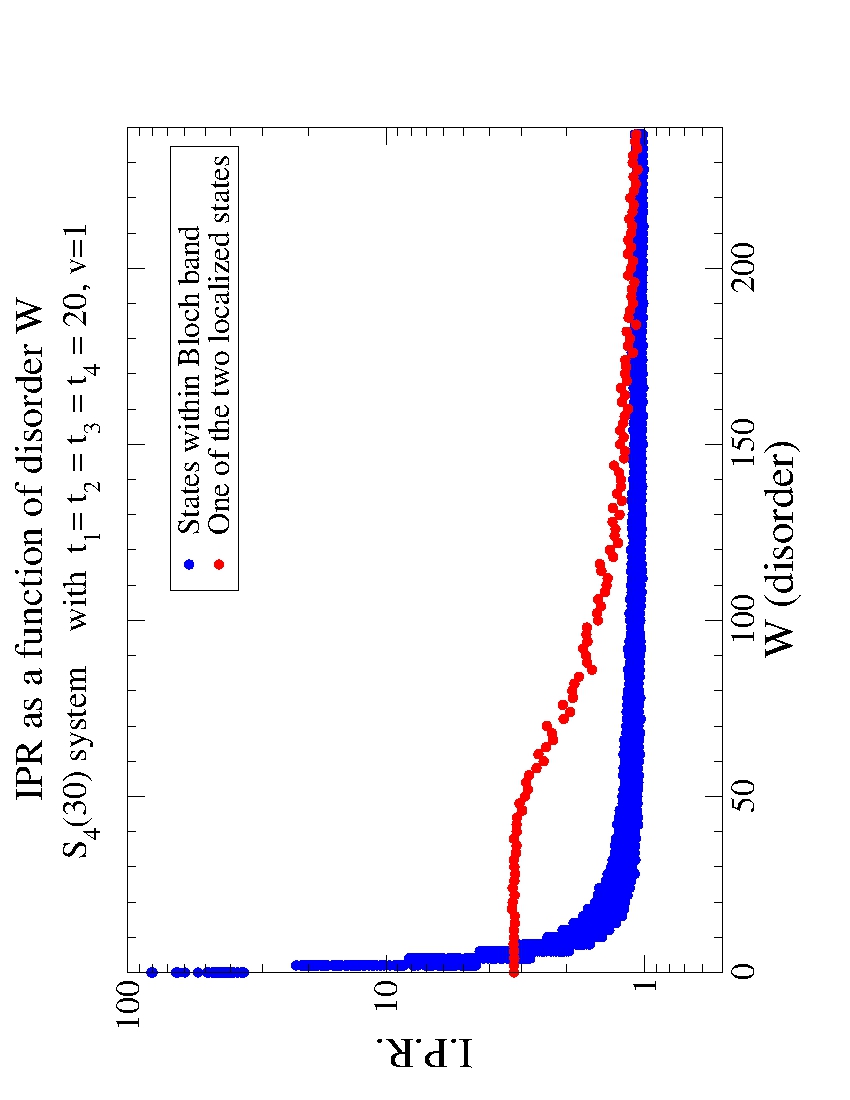}
\caption{The I.P.R. as a function of the disorder parameter $W$ for the $S_4(30)$ system with $t_a=20$ $\forall a$ and $v=1$. In blue, we plot the I.P.R. of the states within Bloch band, [-2,2] in this case, while in green we plot one of the two localized outside Bloch band (the other is the same). It is evident the difference in the I.P.R. between the two types of states. The quantities plotted are dimensionless.
}
\label{fig:loc_len_rum}
\end{figure}

We know from previous chapters, see Eqs.(\ref{eq:3_hamil}), Eq.(\ref{eq:3h_bloch}) and Eq.(\ref{eq:3h_co}), that the tight-binding Hamiltonian $\mathsf{H}$ with nearest neighbour interaction for the star graph system $S_h(N)$ can be written as
\begin{align}
\label{eq:def_h}
\mathsf{H} = \mathsf{H}_{dc} + \mathsf{H}_{c}
\end{align}
where
\begin{align}
\begin{split}
\label{eq:h_bloch}
\mathsf{H}_{dc} &= \varepsilon_0 \vert 0 \rangle \langle 0 \vert + \sum_{a=1}^h \sum_{n=1}^N \varepsilon_a \vert a; n \rangle \langle a; n \vert \\ &+\sum_{a=1}^h \sum_{n=1}^{N-1} v \vert a; n \rangle \langle a; n + 1 \vert + 
\sum_{a=1}^h \sum_{n=2}^{N} v \vert a; n \rangle \langle a; n - 1\vert
\end{split}
\end{align}
and
\begin{equation}
\label{eq:5h_co}
\mathsf{H}_{c} = \sum_{a=1}^h t_av \left( \vert 0 \rangle \langle a; 1 \vert + \vert a; 1 \rangle \langle 0 \vert \right)
\end{equation}
It is clear from Eqs.(\ref{eq:h_bloch}) and (\ref{eq:5h_co}) that $\mathsf{H}_{dc}$ gives us the tight-binding model for $h$ \textit{decoupled chains} of sites (plus the origin) whereas $\mathsf{H}_{c}$ represents the coupling between the origin and the first site of each chain.\\
Up to now, we set the level energy in each site (also the origin) equal to $\varepsilon_0 = \varepsilon_a = 0$ and no disorder was present in this system.
Here, we want to introduce the \textit{disorder} in such a closed system, in order to apply the effective Hamiltonian technique to the case of random variations in the diagonal energies\footnote{Also in this chapter, we set $\varepsilon_0 = \varepsilon_a = 0$, but here $\delta E_0$ is different from zero.}: $\pm$ $\delta E_0$, where $\delta E_0$ is a random variable uniformly distributed in $[- W/2, + W/2]$ and $W$ is a disorder parameter.
For a one-dimensional chain, there is a theoretical model, named \textit{Anderson model} that leads to \textit{Anderson localization}.
Anderson localization, also known as strong localization, is the absence of diffusion of waves in a disordered medium. This phenomenon is named after the American physicist P. W. Anderson, who was the first one to suggest the possibility of electron localization inside a semiconductor, provided that the degree of randomness of the impurities or defects is sufficiently large\footnote{See Ref.\cite{anderson}.}. Anderson localization is a general wave phenomenon that applies to the transport of electromagnetic waves, acoustic waves, quantum waves, spin waves, etc.\\
The eigenstate of the Anderson model are exponentially localized on the system sites, with exponential tails given by $\exp(-x/L_{and})$, where, for \textit{weak disorder}, the localization length $L_{and}$ at the center of the energy band can be written as
\begin{equation}
\label{eq:l_and}
L_{and} \approx 105.2 \left (\dfrac{W}{v} \right )^{-2}
\end{equation}
For $L_{and} \ll N$ ($N$ is the number of the sites in the 1-d chain) we have the localized regime. The condition $L_{and} = N$ defines a critical value of $\left (W/v \right )_{cr}$ for the localized regime, at any given $N$.
Obviously, our system is not a 1-d chain; however, the equation (\ref{eq:l_and}) can provide a rough estimate of the critical value of disorder in which the system moves to the localized regime.
We define the wavefunction $\vert \psi_i \rangle$ of the star graph system as the eigenfunction of the stationary Schr\"{o}dinger:
\begin{equation}
\mathsf{H} \vert \psi_i \rangle = E_i \vert \psi_i \rangle
\end{equation}
where $\mathsf{H}$ is the same defined in Eq.(\ref{eq:def_h}) and $E_i$ are the energy levels of the system.
Writing the eigenstates $\vert \psi_i \rangle$ on the site basis\footnote{The site basis defined here is (for convenience) different from that adopted so far, see for example Eq.(\ref{eq:h_bloch}) and Eq.(\ref{eq:5h_co}).} $\lbrace \vert j \rangle \rbrace$, $j=0,\ldots,hN$ we obtain
\begin{equation}
\vert \psi_i \rangle = \sum_{j=0}^{hN} a_{ij} \vert j \rangle
\end{equation}
We define as a measure of localization the \textit{Inverse Partecipation Ratio} (I.P.R.) of the state $\vert \psi_i \rangle$ which is 
\begin{equation}
\mbox{I.P.R.}^i = \dfrac{1}{hN+1}\left (\sum_{j=0}^{hN} \vert a_{ij} \vert^4 \right) ^{-1}
\end{equation}
The value of $\mbox{I.P.R.}^i$ can be understood as a measure of the localization of the state $\vert \psi_i \rangle$.
In the limiting case of complete delocalization of state $i$ we have $a_{ij}= \left (\sqrt{hN+1} \right )^{-1}$ $\forall j$ that gives us $\mbox{I.P.R.}^i=1$ $\forall i$. In the opposite case (complete localization) we have $\mbox{I.P.R.}^i = \left (hN+1 \right )^{-2}$ $\forall i$. 
In Fig.(\ref{fig:loc_len_rum}) we plot $\mbox{I.P.R.}^i$ as a function of the disorder parameter $W$ for all the states within Bloch band (blue circles) and for one of the two localized states outside Bloch band (red cirles). We considered the $S_4(30)$ system with $t_a = 20$ $\forall a$ and $v=1$. With these 
 parameter values, we have $(121)^{-2} \leq \mbox{I.P.R.} \leq 1$.
We notice that with increasing disorder, the delocalized states within the
Bloch band become localized, while the localized states remain localized, despite the disorder.\\
However, we point out that this is only a \textit{preliminary result}; further study will indicate
 whether this is a general behavior of our system or not and how 
the system behavior varies according to the parameters $t_a$, $v$ and $N$.

\section{Future developments}
There are many possible developments in the study of our star graph system. First of all, we would like to
 study in detail the system's behavior in the presence of disorder in the closed system (changing all the parameters of the system), but also introducing the coupling with the outside world, namely $\gamma \neq 0$.
Secondly, we are going to
 study the system with asymmetric coupling with
the environment and also a system with differents $v$ from chain to chain. Thirdly, it would be very interesting to find possible applications of our model, maybe realizing
an experiment that could support the theoretical and numerical results
 found. Finally, it would be of great interest the study of the superradiance transition in a many-body framework and also seek to apply the superradiance model to the photosynthesis, in order to understand how plants realize the coherent quantum transport at room temperature.






\pagestyle{fancy}


\renewcommand\chapterheadstartvskip{\vspace*{-5\baselineskip}}





\begin{savequote}[15pc]
\sffamily
A great discovery solves a great problem, but there is a grain of discovery in the solution of any problem.
\qauthor{George Polya - Hungarian mathematician}
\end{savequote}

\chapter{Conclusions}
\label{chap:6}
In this thesis, we have studied in detail the star graph system (ie chains of sites coupled with the origin) and we have found several unpublished results, both for the closed and the open system. Regarding to the open system, we used the effective non-Hermitian Hamiltonian approach that enabled us to correctly describe the coupling of the internal system to the outside world. 
Here, we divide the results obtained in two parts: closed and open system.

\subsubsection{Closed system}
Using the tight-binding approach, we found the eigenvalues and the eigenvectors for the star graph system, also with different coupling between the chains and the origin. We analitically showed that there are two states outside of a normal Bloch band and we also demonstrated that these two states are localized; we also provided the localization lengths of these states as a function of the parameters considered. We took in exam chains with a finite number of sites and also chains with a very large numbers of site.

\subsubsection{Open system}
Using the effective non-Hermitian Hamiltonian, we studied the eigenvalues and the localization lengths of the eigenstates by varying the coupling with the environment. We found that the localized states are very little affected by the opening and we also provided an analytical estimation, for small coupling, of the decay rate of these states. 
Furthermore, we investigated how transport properties depend on the coupling with the outside world. We found an analytical estimation of the value of the coupling at which the superradiance transition occurs, which is also valid for chains of different lengths among them and with different couplings with the origin. Moreover, we have shown through numerical simulations that, for large number of sites in each chain, the maximum of the integrated transmission coincides with the value of the coupling at which the transition to superradiance occurs.

\appendix
\chapter{Recurrence relations}
In mathematics, a recurrence relation is an equation that recursively defines a sequence: each term of the sequence is defined as a function of the preceding terms.
In particular, we are interested in \textit{linear homogeneous recurrence relations with constant coefficients}.\\
An order \textit{d} linear homogeneous recurrence relation with constant coefficients is an equation of the form:
\begin{equation}
a_n = c_1 a_{n-1} + c_2 a_{n-2} + \ldots + c_d a_{n-d}
\end{equation}
where the \textit{d} coefficients $c_i$ (for all \textit{i}) are constants.\\
More precisely, this is an infinite list of simultaneous linear equations, one for each $n>d-1$. A sequence which satisfies a relation of this form is called a \textit{linear recursive sequence} or \textit{LRS}. There are \textit{d} degrees of freedom for LRS, the initial values $a_0, \ldots, a_{d-1}$ can be taken to be any values, but then the linear recurrence determines the sequence uniquely.\\
The same coefficients yield the \textit{characteristic polynomial}
\begin{equation}
p(t) = t^d - c_1t^{d-1} - c_2t^{d-2} - \ldots - c_d
\end{equation}
whose d roots play a crucial role in finding and understanding the sequences satisfying the recurrence.\\
Suppose $\lambda$ is a root of $p(t)$ having multiplicity $r$. This is to say that $(t - \lambda)^r$ divides $p(t)$. The following two properties hold:
\begin{enumerate}
\item Each of the $r$ sequences $\lambda^n,n\lambda^n,n^2\lambda^n, \ldots, n^{r-1}\lambda^n$ satisfies the recurrence relation. 
\item Any sequence satisfying the recurrence relation can be written uniquely as a linear combination of solutions constructed in part 1 as $\lambda$ varies over all distinct roots of $p(t)$.
\end{enumerate}
As a result of this statement a linear homogeneous recurrence relation with constant coefficients can be solved in the following manner:
\begin{enumerate}
\item Find the characteristic polynomial $p(t)$.
\item Find the roots of $p(t)$ counting multiplicity.
\item Write $a_n$ as a linear combination of all the roots (counting multiplicity as shown above) with unknown coefficients $b_i$.
This is the general solution to the original recurrence relation.
\begin{equation}
\begin{split}
a_n = & \left(b_1\lambda_1^n + b_2n\lambda_1^n + b_3n^2\lambda_1^n + \ldots +b_rn^{r-1}\lambda_1^n \right)\\
&+\ldots+
\left(b_{d-q+1}\lambda_{\star}^n+\ldots+ b_dn^{q-1}\lambda_{\star}^n \right)
\end{split}
\end{equation}
where $q$ is the multiplicity of $\lambda_{\star}$. 
\item Equate each $a_0, a_1, \ldots, a_d$  from part 3 (plugging in $n=0, \ldots, d$  into the general solution of the recurrence relation) with the known values $a_0, a_1, \ldots, a_d$ from the original recurrence relation. Note, however, that the values $a_n$ from the original recurrence relation used do not have to be contiguous, just $d$ of them are needed (ie, for an original linear homogeneous recurrence relation of order 3 one could use the values $a_0$, $a_1$, $a_4$). This process will produce a linear system of $d$ equations with $d$ unknowns. Solving these equations for the unknown coefficients  of the general solution and plugging these values back into the general solution will produce the particular solution to the original recurrence relation that fits the original recurrence relation's initial conditions (as well as all subsequent values  of the original recurrence relation).
\end{enumerate}
Interestingly, the method for solving linear differential equations is similar to the method above - the ``educated guess" (\textit{ansatz}\footnote{
In physics and mathematics, an ansatz is an educated guess that is verified later by its results. An ansatz is the establishment of the starting equation(s) describing a mathematical or physical problem. It can take into consideration boundary conditions.
}) for linear differential equations with constant coefficients is $e^{\lambda x}$  where $\lambda$ is a complex number that is determined by substituting the guess into the differential equation.
This is not a coincidence. If you consider the Taylor series of the solution to a linear differential equation:
\begin{equation}
\sum_{n=0}^{\infty}\dfrac{f^{(n)}(a)}{n!}\left (x - a \right )^n
\end{equation}
you see that the coefficients of the series are given by the $n^{th}$ derivative of $f(x)$ evaluated at the point $a$. The differential equation provides a linear difference equation relating these coefficients.
This equivalence can be used to quickly solve for the recurrence relationship for the coefficients in the power series solution of a linear differential equation.\\
Let us give a very famous example of homogenous recurrence relation.
\subsubsection*{Example: Fibonacci sequence}
The Fibonacci sequence is named after Leonardo of Pisa, who was known as Fibonacci (a contraction of filius Bonacci, ``son of Bonaccio''). Fibonacci's 1202 book \textit{Liber Abaci} introduced the sequence to Western European mathematics, although the sequence was independently described in Indian mathematics and it is disputed which came first.\\
Fibonacci numbers are used in the analysis of financial markets, in strategies such as Fibonacci retracement, and are used in computer algorithms such as the Fibonacci search technique and the Fibonacci heap data structure. The simple recursion of Fibonacci numbers has also inspired a family of recursive graphs called Fibonacci cubes for interconnecting parallel and distributed systems. They also appear in biological settings, such as branching in trees, arrangement of leaves on a stem, the fruit spouts of a pineapple, the flowering of artichoke, an uncurling fern and the arrangement of a pine cone.\\
In mathematics, the Fibonacci numbers are the numbers in the following integer sequence:
\begin{equation}
0,1,1,2,3,5,8,13,21,34,55,89,144 \ldots
\end{equation}
By definition, the first two Fibonacci numbers are 0 and 1, and each subsequent number is the sum of the previous two. 
In mathematical terms, the sequence $F_n$ of Fibonacci numbers is defined by the recurrence relation
\begin{equation}
F_n = F_{n-1} + F_{n-2}
\end{equation}
with seed values
\begin{align}
F_0=0 && F_1=1
\end{align}
The characteristic equation is
\begin{equation}
p(t) = t^2-t-1 = 0
\end{equation} with roots 
\begin{align}
\alpha = \dfrac{1+\sqrt{5}}{2} && \beta = \dfrac{1-\sqrt{5}}{2}.
\end{align}
We have 
\begin{equation}
F_n = A\alpha^n + B\beta^n
\end{equation}
From the initial conditions $F_0=0$ and $F_1=1$,
\begin{equation}
\label{eq:initial_cond_system}
 \left\{ \begin{split}
F_0 &= A + B = 0 \\
F_1 &= A\alpha + B\beta = 1\\
\end{split} \right.
\end{equation}
Solve for A and B, we have $A=1/\sqrt{5}$ and $B=-1/\sqrt{5}$.\\
Hence
\begin{equation}
F_n = \dfrac{1}{\sqrt{5}}\left[\left(\dfrac{1+\sqrt{5}}{2}\right )^n- \left(\dfrac{1-\sqrt{5}}{2}\right )^n\right ]
\end{equation}
In this way, we obtained a \textit{closed-form solution}, known as \textit{Binet's formula}.

\chapter{Determinant of a tridiagonal matrix}
A tridiagonal matrix is a matrix that has nonzero elements only in the main diagonal, the first diagonal below this, and the first diagonal above the main diagonal.
This is the most general tridiagonal matrix:
\begin{equation}
\begin{split}
\mathcal{T_M} = \left( 
\begin{array}{cccc}
a_1 & b_1 & 0 & 0 \\
c_1 & \ddots & \ddots &0 \\
0 & \ddots & \ddots &b_{n-1} \\
0 & 0 &c_{n-1} & a_n \\
\end{array}
\right)
\end{split}
\end{equation}
For tridiagonal matrices, determinants can be evaluated via multiplication of $2 \times 2$ matrices as follows\footnote{See Ref.\cite{lgmolin}.
}
\begin{equation}
\label{eq:det_trid}
\begin{split}
\det \mathcal{T_M} = \det & \left( 
\begin{array}{cccc}
a_1 & b_1 & 0 & 0 \\
c_1 & \ddots & \ddots &0 \\
0 & \ddots & \ddots &b_{n-1} \\
0 & 0 &c_{n-1} & a_n \\
\end{array}
\right)= \\
= &\left[\left( 
\begin{array}{cc}
a_n & -b_{n-1}c_{n-1} \\
1 & 0 \\
\end{array}
\right) 
\ldots  
 \left(
\begin{array}{cc}
a_2 & -b_{1}c_{1} \\
1 & 0 \\
\end{array}
\right)
\left(
\begin{array}{cc}
a_1 & 0 \\
1 & 0 \\
\end{array} \right)
\right]_{11}
\end{split}
\end{equation}
where the subscript 11 indicates that we have to consider the first element in the first column.






\begin{savequote}[20pc]
\sffamily
Thankfulness is the tune of angels. \\
(La riconoscenza è la melodia degli angeli.)
\qauthor{Edmund Spenser - English poet}
\end{savequote}

\chapter*{Acknowledgments (Ringraziamenti)}
	\addcontentsline{toc}{chapter}{Acknowledgments (Ringraziamenti)\textbf{}}

\vspace{5em}
Per i miei ringraziamenti voglio utilizzare una forma letteraria di rado utilizzata: l'\textit{elenco}.
Ci sono liste che hanno fini pratici e sono finite, come la lista di tutti i libri di una biblioteca; ma ve ne sono altre che vogliono suggerire grandezze innumerabili e che si arrestano incomplete ai confini dell'indefinito. Ed è a questo secondo tipo di lista a cui mi voglio riferire nei miei ringraziamenti, perché tutti coloro con cui ho avuto a che fare fino ad ora, nel bene o nel male, mi hanno fatto crescere. Ed è a tutti loro che sono grato.\\
Permettetemi però di fare alcuni ringraziamenti in particolare.\\
\newline
Ringrazio i miei genitori, per avermi fatto diventare la persona che sono fiero di essere. \\
\newline
Ringrazio mia sorella, per me un modello da seguire.\\
\newline
Ringrazio la mia nipotina Camilla, per avermi sempre regalato un sorriso, anche nei momenti più bui.\\
\newline
Ringrazio la mia ragazza Laura, per avermi sempre sostenuto, in ogni momento.\\
\newline
Ringrazio gli zii e la nonna di Alfianello, per avermi dato ospitalità quando ne avevo bisogno.\\
\newline
Ringrazio i miei compagni di università, per i fantastici cinque anni passati insieme.\\
\newline
Ringrazio Laura e Michela, per la loro particolare vicinanza.\\
\newline
Ringrazio gli amici della compagnia, per le esilaranti serate trascorse insieme.\\
\newline
Ringrazio il Booz, per avermi permesso di assaporare il gusto del perdono.\\
\newline
Ringrazio i miei amici di Bettegno, perché sono per me come una seconda famiglia.\\
\newline
Ringrazio Renato per i suoi fantastici programmi.\\
\newline
Ringrazio Maura, per avermi trasmesso il suo amore per la scienza.\\
\newline
Ringrazio il Prof. Fausto Borgonovi, per avermi insegnato moltissime cose, per aver saputo capire quali erano i miei interessi e per i suoi precisi e puntuali consigli.\\
\newline
Ringrazio il Dott. Luca Celardo, per gli innumerevoli pomeriggi passati a discutere e per avermi seguito passo passo in questo mio lavoro. \\
\newline
Thanks to Prof. L. Kaplan, Prof. V.G. Zelevinsky, Prof. F.M. Izrailev for fruitful discussions.\\
\newline
Ringrazio Dio, per così tante ragioni che non basterebbe tutta la carta del mondo per scriverle.


\appendix

\end{document}